\RequirePackage{lineno} 

\documentclass[twocolumn,showpacs,preprintnumbers,amsmath,amssymb,nofootinbib]{revtex4}

\usepackage{graphicx}
\usepackage{dcolumn}
\usepackage{bm}

\usepackage{epstopdf}
\def\bea{\begin{eqnarray}}
\def\eea{\end{eqnarray}}

\def\pp{\mbox{$p$-$p$}}
\def\pa{\mbox{$p$-$A$}}

\def\auau{\mbox{Au-Au}}

\def\pbpb{\mbox{Pb-Pb}}
\def\aa{\mbox{$A$-$A$}}
\def\nn{\mbox{$N$-$N$}}

\def\dau{\mbox{$d$-Au}}

\def\pt{$p_t$}
\def\titlept{$\bf p_t$}
\def\titlepp{$\bf p$-$\bf p$}
\def\mt{$m_t$}
\def\yt{$y_t$}

\def\nch{$n_{ch}$}
\def\mmpt{$\bar p_t$}
\def\titlempt{$\bf \bar p_t$}
\def\ppb{\mbox{$p$-Pb}}

\begin{document} 

\setlength{\pdfpagewidth}{8.5in}
\setlength{\pdfpageheight}{11in}

\setpagewiselinenumbers
\modulolinenumbers[5]

\preprint{Version 2.0}

\title{
A two-component model of hadron production applied to $\bf p_t$ spectra\\ from 5 TeV and 13 TeV $\bf p$-$\bf p$ collisions at the large hadron collider
}

\author{Thomas A.\ Trainor}\affiliation{CENPA 354290, University of Washington, Seattle, Washington 98195}


\date{\today}

\begin{abstract}

The ALICE collaboration at the large hadron collider (LHC) recently reported high-statistics $p_t$ spectrum data from 5 TeV and 13 TeV $p$-$p$ collisions. Particle data for each energy were partitioned into event classes based on the total yields within two disjoint pseudorapidity $\eta$ intervals denoted by acronyms V0M and SPD. For each energy the spectra resulting from the two selection methods were then compared to a minimum-bias INEL $> 0$ average over the entire event population. The nominal goal was determination of the role of jets in high-multiplicity $p$-$p$ collisions and especially the jet contribution to the low-$p_t$ parts of spectra. A related motivation was response to recent claims of ``collective'' behavior and other nominal indicators of quark-gluon plasma (QGP) formation in small collision systems. In the present study a two-component (soft + hard) model (TCM) of hadron production in $p$-$p$ collisions is applied to the ALICE spectrum data. As in previous TCM studies of a variety of A-B collision systems the jet and nonjet contributions to $p$-$p$ spectra are accurately separated over the entire $p_t$ acceptance. Distinction is maintained among spectrum normalizations, jet contributions to spectra and systematic biases resulting from V0M and SPD event selection. The statistical significance of data-model differences is established. The effect of {\em spherocity} (azimuthal asymmetry measure nominally sensitive to jet production) on ensemble-mean $p_t$ vs event multiplicity $n_{ch}$ is investigated and found to have little relation to jet production. The general results of the TCM analysis are as expected from a conventional QCD description of jet production in $p$-$p$ collisions.

\end{abstract}

\pacs{12.38.Qk, 13.87.Fh, 25.75.Ag, 25.75.Bh, 25.75.Ld, 25.75.Nq}

\maketitle

\section{Introduction} \label{intro}


Claims of quark-gluon plasma (QGP) formation in \auau\ collisions at the relativistic heavy ion collider (RHIC) and subsequently in \pbpb\ collisions at the large hadron collider (LHC) were based on certain data features initially seen as unique to more-central A-A collisions and not appearing in small asymmetric A-B control systems such as \dau\ or \ppb\ where QCD theory suggested that QGP formation should be unlikely. However, in recent years similar features have been observed in LHC data for \ppb\ collisions and high-charge-multiplicity \pp\ collisions and have been interpreted as evidence for QGP formation in small collision systems~\cite{ppcms,dusling}. But  the recent LHC results could also be interpreted to indicate that data features conventionally associated with QGP formation may result from unexceptional QCD processes.


The ALICE collaboration recently published a comprehensive high-statistics study of \pt\ spectra from 5 TeV and 13 TeV \pp\ collisions~\cite{alicenewspec}. The analysis employs two methods to sort collision events into ten multiplicity classes each and application of {\em spherocity} $S_0$, a measure of the azimuthal nonuniformity of distributed $\vec p_t(\phi)$, to estimate the ``jettiness'' of events. Several methods are applied to determine variation of spectrum shape with charge multiplicity, event-selection method and spherocity. 

The study reported in Ref.~\cite{alicenewspec} is motivated in part by claimed observation in \pp\ and \ppb\ collisions of evidence for radial and elliptic flow (``collectivity'')~\cite{aliflows1,aliflows2} as well as strangeness enhancement~\cite{alistrange} similar to that observed in more-central \aa\ collisions and attributed there to QGP formation. Observation of such effects in low-density systems occupying small space-time volumes runs counter to initial theoretical expectations concerning QGP formation. The study seeks to understand hadron production associated with jets in relation to soft particle production: ``The aim of this study is to investigate the importance of jets in high-multiplicity pp collisions and their contribution to charged-particle production at low $p_T$.''


The phenomenology of high-energy \pp\ collision data serves as an essential reference for high-energy \pa\ and \aa\ collisions, specifically regarding claims of novel physical mechanisms such as QGP formation~\cite{perfect} or possible manifestations of hydrodynamic flows even in small collision systems~\cite{ppflow,moreppflow}. One can formulate a set of critical questions addressed to available \pp\ data: (a) What is the evidence for or against azimuthally symmetric radial flow, and for or against elliptic flow and ``higher harmonics'' as manifestations of azimuthal asymmetry? (b) Are nominally flow-related azimuthal features certainly disjoint from jet production? (c) Are jet contributions to spectra significantly modified by a dense medium (i.e.\ QGP)? (d) Does  the charge-multiplicity (\nch) dependence of certain data features reflect the onset of such a medium with increasing particle density? If the QGP scenario is valid then systematic behavior of various data features should be synchronized with emergence of a {\em common} underlying dense medium. Are comprehensive data trends consistent with such expected synchronization?


A two-component (soft + hard) model (TCM) of hadron production near mid-rapidity in A-B collisions was initially derived from the charge-multiplicity \nch\ dependence of \pt\ spectra from 200 GeV \pp\ collisions~\cite{ppprd}. The \nch\ dependence of yields, spectra and two-particle correlations has played a key role in establishing (a) the nature of hadron production mechanisms in \pp\ collisions and (b) that the TCM hard component of \pt\ spectra is {\em quantitatively} consistent with predictions based on event-wise reconstructed jets~\cite{fragevo,jetspec,jetspec2}. The question of recently-claimed collectivity or flows in small (\pp\ and \pa) systems has been addressed in terms of the resolved TCM soft and hard components and evidence (or not) for radial flow in differential studies of \pt\ spectra~\cite{hardspec,ppquad,ppbpid}. 

Reference~\cite{alicetomspec} reported previous TCM analysis of 13 TeV \pp\ spectra that serves as a precursor to the present study (see App.~\ref{old13}). Then-available spectrum data~\cite{alicespec} were presented only as ratios to minimum-bias spectra and only for a limited \nch\ range. The present study extends those results utilizing both 5 TeV and 13 TeV data. Spectra for the two energies and for two methods of event selection, V0M and SPD, are decomposed into soft and hard components for ten event multiplicity classes each. Spectrum biases resulting from event selection methods are evaluated and compared. The quality of the TCM description is determined via data-model {\em differences} (not ratios) compared to statistical uncertainties (Z-scores) as a significance test. 
A TCM for ensemble-mean \pt\ or \mmpt\ is defined and applied to \mmpt\ vs \nch\ data from Ref.~\cite{alicenewspec}. Evolution of the \mmpt(\nch) trend with spherocity $S_0$ is examined in detail -- especially the relation of $S_0$ to dijet production. An ironic result emerges.

The TCM serves as an accurate reference for A-B collision systems that is not derived from fits to individual spectra. The TCM is required to describe diverse data formats applied to a broad array of collision systems self-consistently. It precisely separates jet and nonjet data contributions, greatly facilitating and simplifying data interpretation. Data-TCM deviations may reveal systematic data biases as in the present study or identify new physics beyond conventional models as in Ref.~\cite{ppquad}.

This article is arranged as follows:
Section~\ref{specdata} summarizes \pp\ spectrum data, methods and conclusions reported in Ref.~\cite{alicenewspec}.
Section~\ref{pptcm} describes the TCM for \pp\ \pt\ spectra.
Section~\ref{biases} summarizes selection biases resulting from two event-selection criteria and their evolution with event multiplicity \nch.
Section~\ref{shapes} reviews some \pt\ spectrum shape measures.
Section~\ref{meanpt} describes the TCM for ensemble \mmpt\ and reviews results from Ref.~\cite{alicenewspec} for evolution of \mmpt(\nch) trends with varying spherocity $S_0$.
Section~\ref{syserr} discusses systematic uncertainties.
Sections~\ref{disc} and~\ref{summ}  present discussion and summary.

\section{5 and 13 T$\bf e$V \titlepp\ \titlept\ spectrum data} \label{specdata}

Reference~\cite{alicenewspec} reports \pt\ spectra from 5 and 13 TeV \pp\ collisions for two event selection methods (V0M and SPD) and for ten charge-multiplicity classes each. For each energy the same minimum-bias (INEL $> 0$) event ensemble is effectively sorted into multiplicity classes in two ways. The \pt\ acceptance is $p_t \in [0.15,20]$ GeV/c. The angular acceptance is $2\pi$ on azimuth and $|\eta | \leq 0.8$ on pseudorapidity. The total event numbers are 105 and 60 million for 5 and 13 TeV respectively. The basic \pt\ spectra are further analyzed via several methods. As noted, the stated main goal of the study is to understand the role of jets in high-multiplicity \pp\ collisions. 

\subsection{Motivation and strategy}

The context presented for the spectrum analysis reported in Ref.~\cite{alicenewspec} is claimed observation at the LHC of collectivity -- i.e.\ radial~\cite{aliflows1,aliceppbpid} and anisotropic (e.g.\ elliptic, higher harmonic)~\cite{aliflows2} flows -- and strangeness enhancement~\cite{alistrange} in \pp\ and \ppb\ collisions whereas those phenomena had been designated as indicators of QGP formation only in high-density \aa\ collision systems. A variety of models based on hydrodynamics, string percolation, multiparton interactions or fragmentation of saturated gluon states are ``...able to describe...qualitatively well...some features of data.'' However, concerns have been expressed about interpretations of data from small collision systems in terms of QGP formation without a more-rigorous examination of data and models~\cite{thoughts}.

It is asserted that a \pt\ spectrum ``carries information of the dynamics of soft and hard interactions.'' Reference is made to three \pt\ intervals: $p_t > 10$ GeV/c is said to be ``quantitatively well described by perturbative QCD (pQCD) calculations.''%
\footnote{A pQCD (i.e.\ jet) description is quantitatively consistent with the \pp\ \pt\ spectrum {\em hard component} down to 0.5 GeV/c~\cite{fragevo}.} Below that limit one must ``resort to phenomenological QCD inspired models [i.e.\ Monte Carlo models].'' Novel effects claimed for \pp\ and \ppb\ collisions are said to appear in $p_t < 2$ GeV/c and $p_t \in [2,10]$ GeV/c. Reference~\cite{alicenewspec} asserts that ``The present paper reports a novel multi-differential analysis aimed at understanding charged-particle production associated to partonic scatterings with large momentum transfer and their possible correlations with soft particle production.'' In essence, Ref.~\cite{alicenewspec} poses the question: what is the jet contribution to hadron spectra at low \pt?

\subsection{\titlepp\ \titlept\ spectrum data} \label{alicedata}

\begin{figure}[h]
	\includegraphics[width=3.3in,height=1.6in]{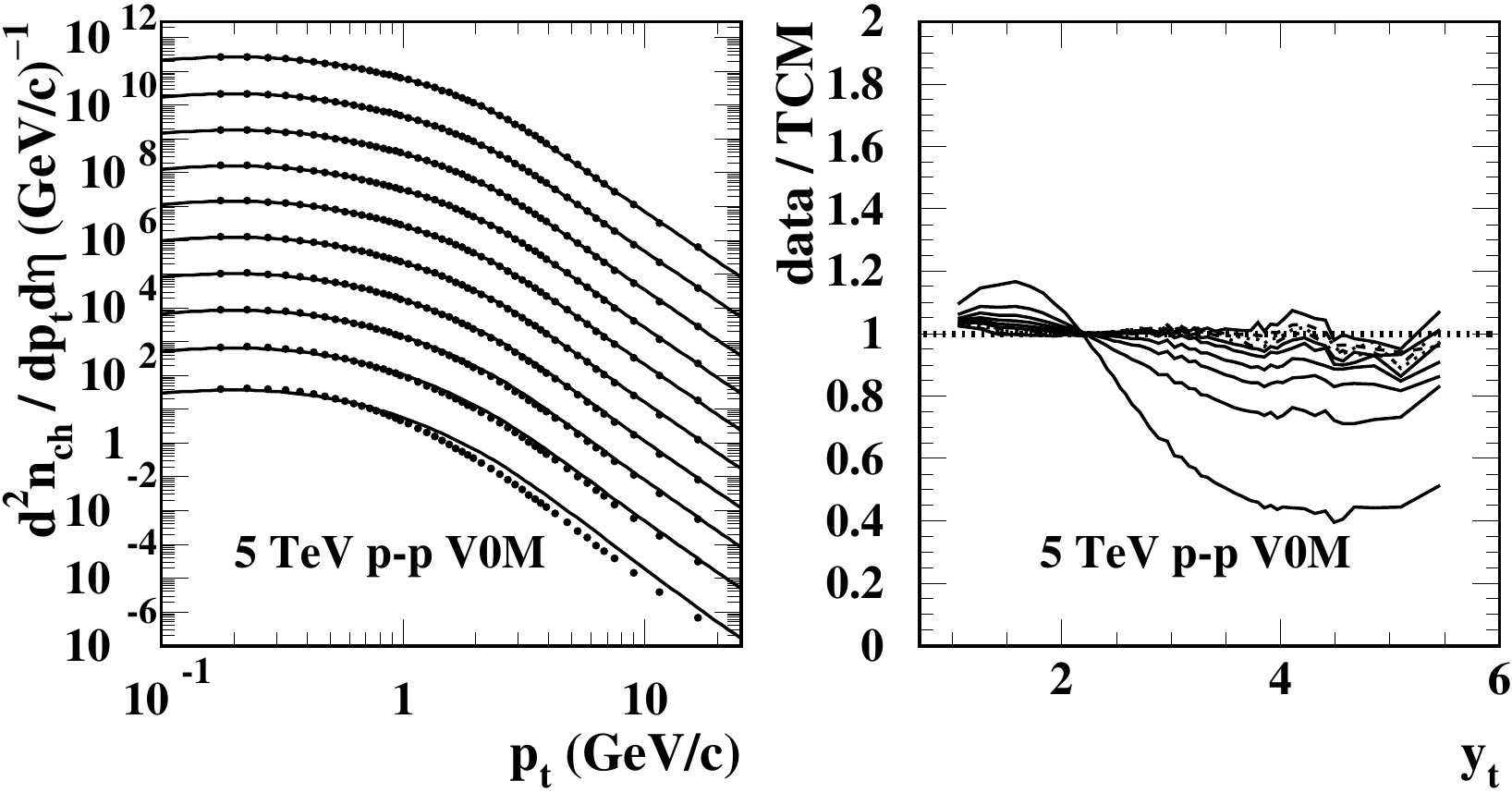}
\put(-140,99) {\bf (a)}
\put(-84,99) {\bf (b)}\\
 	\includegraphics[width=3.3in,height=1.6in]{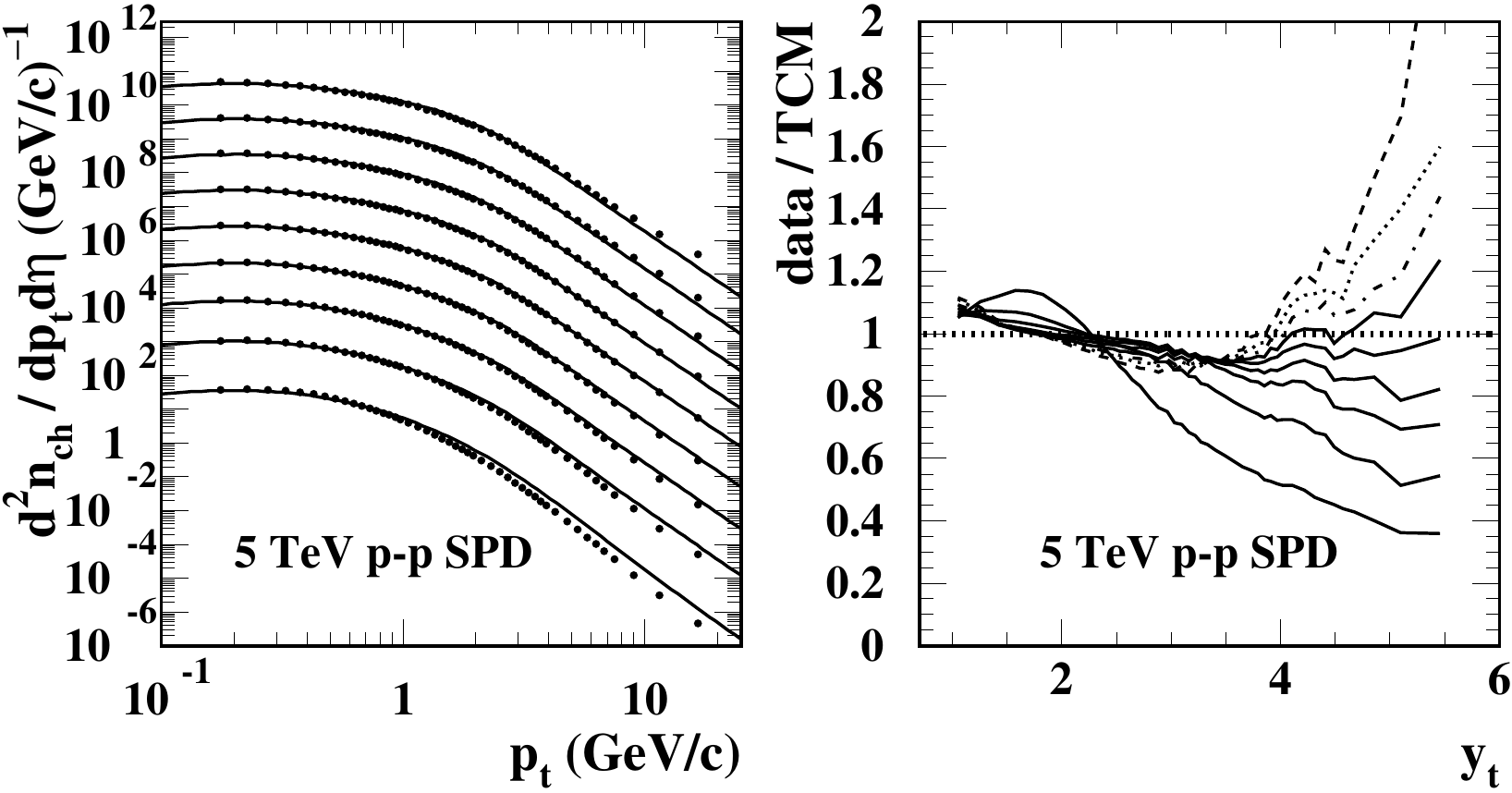}
\put(-140,99) {\bf (c)}
\put(-84,99) {\bf (d)}
	\caption{\label{ptspec1}
Left: \pt\ spectra from ten (V0M) or nine (SPD) multiplicity classes of 5 TeV \pp\ collisions for V0M (a) and SPD (c) event selection.
Right: data/TCM spectrum ratios for data in the left panels for  V0M (b) and SPD (d) event selection.
	} 
\end{figure}

\begin{figure}[h]
	\includegraphics[width=3.3in,height=1.6in]{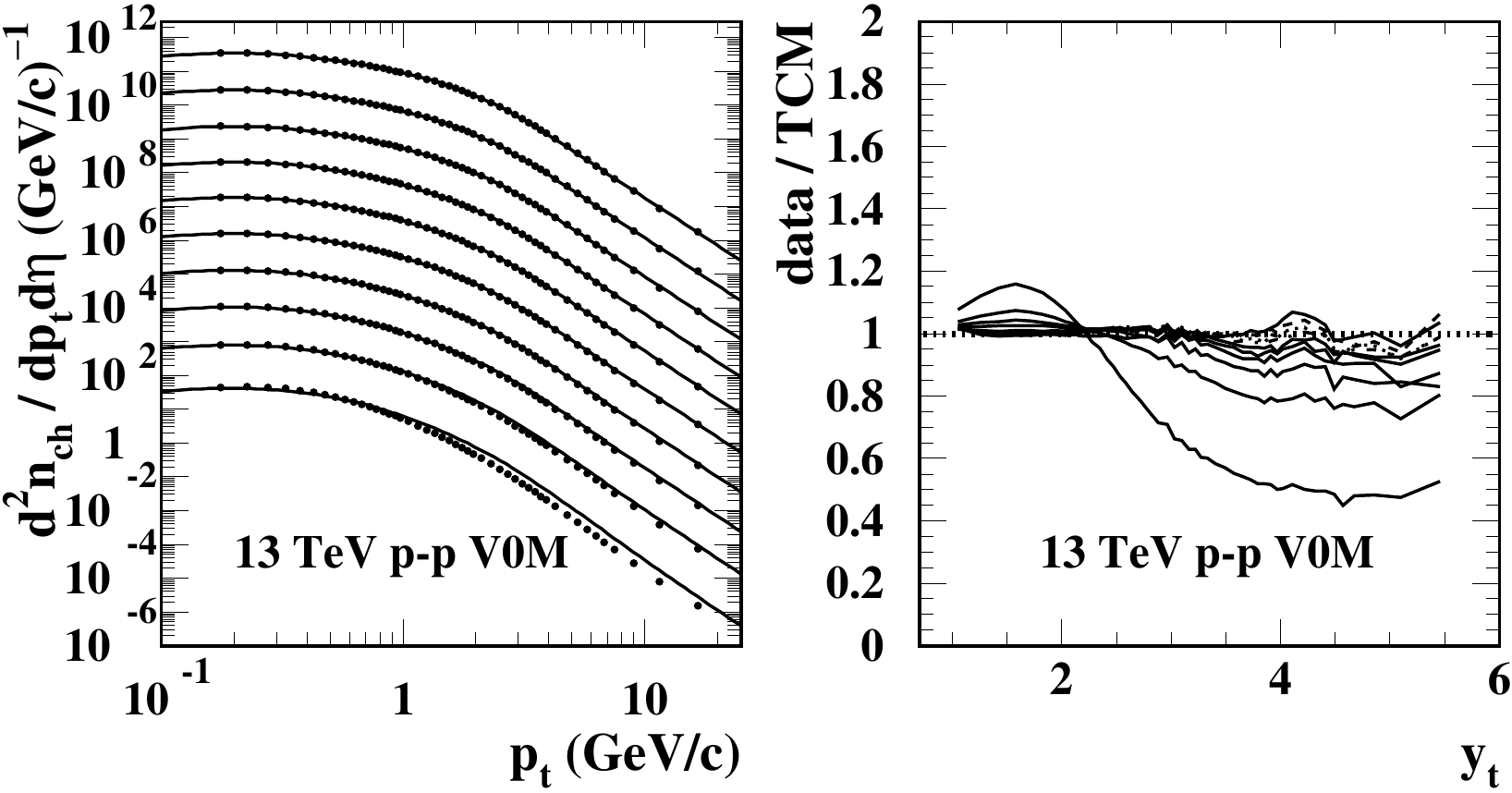}
\put(-140,99) {\bf (a)}
\put(-84,99) {\bf (b)}\\
	\includegraphics[width=3.3in,height=1.6in]{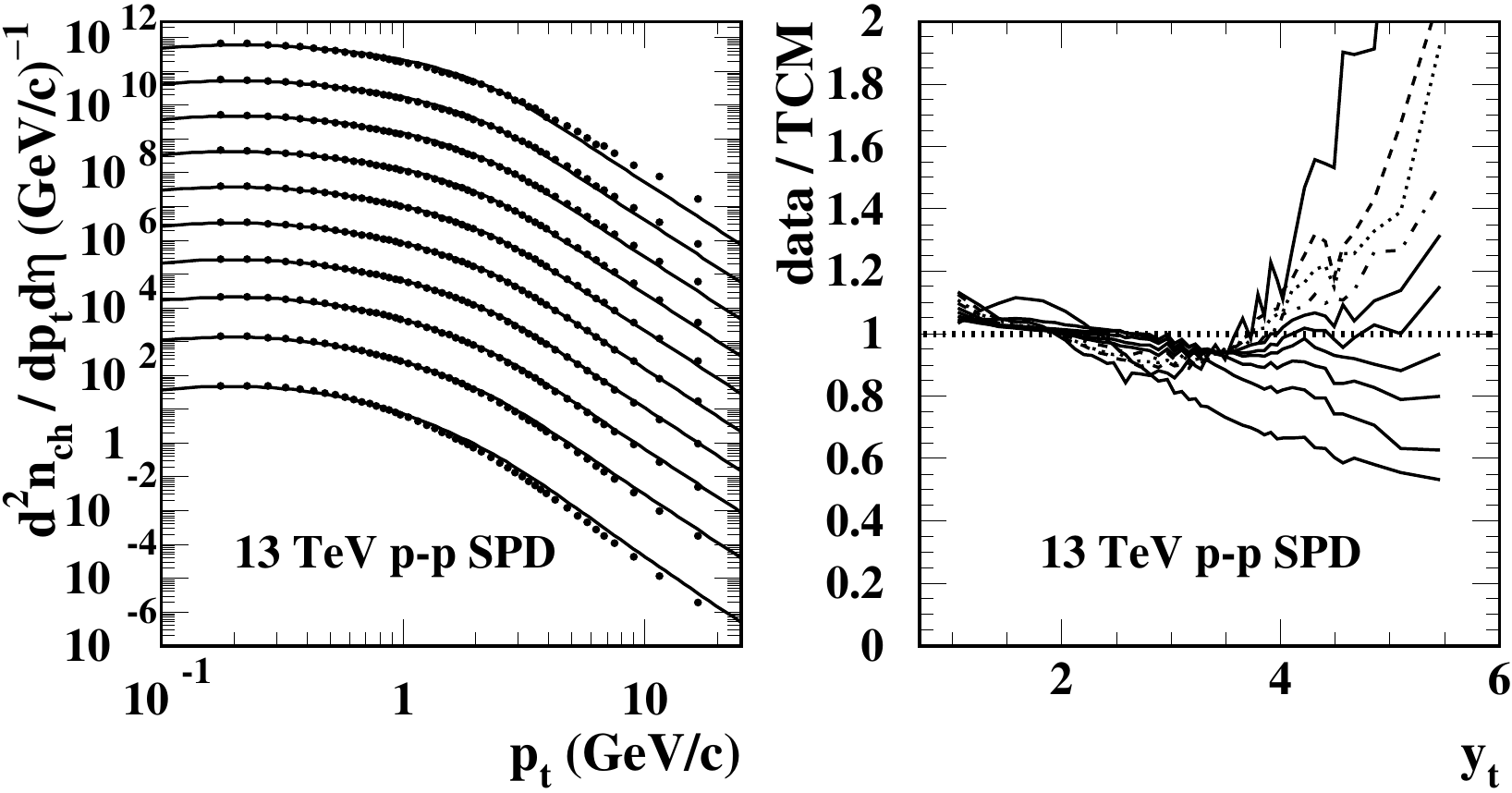}
\put(-140,99) {\bf (c)}
\put(-84,99) {\bf (d)}
	\caption{\label{ptspec2}
Left: \pt\ spectra from ten multiplicity classes of 13 TeV \pp\ collisions for V0M (a) and SPD (c) event selection.
Right: data/TCM spectrum ratios for data in the left panels for  V0M (b) and SPD (d) event selection.
	} 
\end{figure}

Figures~\ref{ptspec1} and \ref{ptspec2} show spectra (points) for 5 TeV and 13 TeV respectively. Panels (a) and (c) show spectra for sorting criteria V0M and SPD respectively multiplied by successive powers of ten from lowest to highest multiplicity in a conventional log-log plot format. The solid curves represent the TCM described in Sec.~\ref{pptcm}. The TCM is not derived from fits to individual spectra.  Panels (b) and (d) show data/model ratios based on the TCM. Line types for the four highest-\nch\ classes vary as solid, dashed, dotted and dash-dotted. That convention is applied consistently in what follows unless explicitly noted. In those ratios certain ``noise'' components appear as common to multiple spectra. It is possible that those features arise from efficiency corrections generated by a Monte Carlo simulation with a more-limited number of events.
Power-law fits to spectra above 6 GeV/c are used to infer a power-law exponent $n$ which is observed to decrease in magnitude with increasing \nch\ (but see Sec.~\ref{powerlaw}). Note that spectra as plotted in Ref.~\cite{alicenewspec} and Figs.~\ref{ptspec1} and \ref{ptspec2} (a,c) below are in the form $d^2 n_{ch} / dp_t d\eta$ whereas $\bar \rho_0(p_t;n_{ch})$ corresponding to $\bar \rho_0(y_t;n_{ch})$ as defined in Eq.~(\ref{rhotcm}) includes an additional factor \pt\ in its denominator.

\subsection{Ensemble-mean $\bf \bar p_t$ vs $\bf n_{ch}$ and spherocity}

Spectra are also sorted according to spherocity $S_0 \in [0,1]$ or $S_0(\%) \in [0\%,100\%]$ [see Eq.~(\ref{spher})], said to be a measure of the ``jettiness'' of the event-wise $\vec p_t(\phi)$ azimuth distribution, where $S_0 \approx 1$ or 100\% is associated with near-isotropic events. For ten classes of spherocity $S_0$ ensemble-mean \mmpt\ vs  multiplicity \nch\ trends for 13 TeV data are evaluated.  Event multiplicity classes are defined within $|\eta| < 0.8$ (i.e. SPD selection).
Reference~\cite{alicenewspec} states: ``Since the goal of the present study is to separate jet events from isotropic ones, we study different spherocity classes for a given multiplicity value.''  ``...the most jet-like and [most] isotropic events will be referred to as 0-10\% and 90-100\% spherocity classes, respectively.''

\subsection{Summary and conclusions}

The abstract of Ref.~\cite{alicenewspec} asserts that \pt\ spectra exhibit little energy dependence (between 5 and 13 TeV), and the high-\pt\ tails of spectra increase faster than linearly with event multiplicity \nch.  Regarding the spherocity study ``For low- (high-) spherocity events, corresponding to jet-like (isotropic) events, the average $p_T$ is higher (smaller) than that measured in INEL $> 0$ pp collisions.'' 

Although it is reported that ensemble-mean \mmpt\ increases with decreasing spherocity (expected to indicate increased jettiness) the observed {\em shape} of \mmpt\ vs \nch\ doesn't change significantly with spherocity (contradicting the expectation, see Sec.~\ref{alicempt}).  Spherocity results are said to ``illustrate the difficulties for the [Monte Carlo] models to describe different observables once they are differentially analyzed as a function of several variables.''

The summary states that ``For a fixed center-of-mass energy, particle production above $p_T = 0.5$ GeV/c exhibits a remarkable multiplicity dependence. Namely, for transverse momenta below 0.5 GeV/c, the ratio of the multiplicity dependent spectra to those for INEL$> 0$ pp collisions is rather constant, and for higher momenta, it shows a significant $p_t$ dependence. The behavior observed for each of the two multiplicity estimators are consistent within the $\langle dN_{ch}/d\eta \rangle$ interval defined by the V0M multiplicity estimator, which gives a $\langle dN_{ch}/d\eta \rangle$ reach of $\sim 25$. For the highest V0M multiplicity class, the ratio increases going from $p_T = 0.5$ GeV/c up to $p_T \approx 4$ GeV/c, then for higher $p_T$, it shows a smaller increase.'' Those qualitative observations  contrast with highly differential and quantitative TCM results from the present study as presented below in Secs.~\ref{pptcm} -  \ref{meanpt}.

\section{\titlepp\ \titlept\ spectrum TCM} \label{pptcm}

The \pt\ spectrum TCM, first reported for 200 GeV \pp\ collisions in Ref.~\cite{ppprd}, is basically consistent with the TCM first reported in Ref.~\cite{pancheri} in response to UA1 ``minijets'' from the CERN S$p \bar p$S. The TCM provides an accurate description of yields, spectra and two-particle correlations for A-B collision systems based on linear superposition of \nn\ or parton-parton collisions. In general, the TCM serves as a {\em predictive reference} for any collision system. Deviations from the TCM then provide systematic and quantitative information on details of collision mechanisms. In  this section the spectrum TCM is reviewed and then applied to 5 TeV and 13 TeV \pt\ spectra for two event selection methods from Ref.~\cite{alicenewspec}.

\subsection{\titlept\ spectrum TCM for unidentified hadrons} \label{unidspec}

The \pt\ or \yt\ spectrum TCM is by definition the sum of soft and hard components with details inferred from data (e.g.\ Ref.~\cite{ppprd}). For \pp\ collisions
\bea  \label{rhotcm}
\bar \rho_{0}(y_t;n_{ch}) &\approx& \frac{d^2n_{ch}}{y_t dy_t d\eta}
\\ \nonumber
&\approx& \bar \rho_{s}(n_{ch}) \hat S_{0}(y_t) + \bar \rho_{h}(n_{ch}) \hat H_{0}(y_t),
\eea
where \nch\ is an event-class index and factorization of the dependences on \yt\ and \nch\ is a central feature of the spectrum TCM inferred from 200 GeV \pp\ spectrum data in Ref.~\cite{ppprd}. The motivation for transverse rapidity $y_{ti} \equiv \ln[(p_t + m_{ti})/m_i]$ (applied to hadron species $i$) is explained in Sec.~\ref{tcmmodel}. The \yt\ integral of Eq.~(\ref{rhotcm}) is $\bar \rho_0 \equiv n_{ch} / \Delta \eta = \bar \rho_s + \bar \rho_h$, a sum of soft and hard charge densities. $\hat S_{0}(y_t)$ and $\hat H_{0}(y_t)$ are unit-normal model functions independent of \nch. The centrally-important relation $\bar \rho_{h} \approx \alpha \bar \rho_{s}^2$ with $\alpha = O(0.01)$ is inferred from \pp\ spectrum data~\cite{ppprd,ppquad,alicetomspec}. $\bar \rho_s$ is then obtained from measured $\bar \rho_0$ as the root of the quadratic equation $\bar \rho_0  = \bar \rho_s + \alpha \bar \rho_s^2$ with  $\alpha$ determined by an energy trend derived from \pp\ spectrum data covering a large energy interval~\cite{alicetomspec}. It is important to distinguish TCM model elements from spectrum data soft and hard components. It is useful to recall that \yt\ values 1, 2, 3, 4 and 5 are approximately equivalent to \pt\ values 0.15, 0.5, 1.4, 3.8 and 10 GeV/c.

\subsection{\titlept\ spectrum TCM model functions} \label{tcmmodel}

The \pp\ \pt\ spectrum soft component is most efficiently described on transverse mass \mt\ whereas the spectrum hard component is most efficiently described on transverse rapidity \yt. The spectrum TCM thus requires a heterogeneous set of variables for its simplest definition. The components can be easily transformed from one variable to the other by Jacobian factors defined below.

Given spectrum data in the form of Eq.~(\ref{rhotcm}) the unit-normal spectrum soft-component model $\hat S_0(y_t)$ is defined as the asymptotic limit of data spectra normalized in the form $X(y_t) \equiv \bar \rho_{0}(y_t;n_{ch}) / \bar \rho_s$ as \nch\ goes to zero. Hard components of data spectra are then defined as complementary to soft components, with the explicit form
\bea \label{yyt}
Y(y_t) \equiv \frac{1}{\alpha \bar \rho_s} \left[ X(y_t) - \hat S_0(y_t) \right],
\eea
directly comparable with TCM model function $\hat H_0(y_t)$.

The data soft component for a specific hadron species $i$ is typically well described by a L\'evy distribution on $m_{ti}  = \sqrt{p_t^2 + m_i^2}$. The unit-integral soft-component model is 
\bea \label{s00}
\hat S_{0i}(m_{ti}) &=& \frac{A_i}{[1 + (m_{ti} - m_i) / n_i T_i]^{n_i}},
\eea
where $m_{ti}$ is the transverse mass-energy for hadron species $i$ with mass $m_i$, $n_i$ is the L\'evy exponent, $T_i$ is the slope parameter and coefficient $A_i$ is determined by the unit-normal condition. Model parameters $(T_i,n_i)$ for several species of identified hadrons have been inferred from 5 TeV \ppb\ spectrum data as described in Ref.~\cite{ppbpid}. A soft-component model function for unidentified hadrons can be defined as the weighted sum
\bea \label{s0m}
\hat S_{0}(m_{t}) &=& \sum_i z_{0i} \hat S_{0i}(m_{ti}),
\eea
where the weights for charged hadrons follow $\sum_i z_{0i} = 1$. In the present context model function $\hat S_0(m_t)$ should not be confused with spherocity $S_0$ introduced in Ref.~\cite{alicenewspec}.

The unit-normal hard-component model  is a Gaussian on $y_{t\pi} \equiv \ln[(p_t + m_{t\pi})/m_\pi]$ (as explained below) with exponential (on $y_t$) or power-law (on $p_t$) tail for larger \yt\
\bea \label{h00}
\hat H_{0}(y_t) &\approx & A \exp\left\{ - \frac{(y_t - \bar y_t)^2}{2 \sigma^2_{y_t}}\right\}~~~\text{near mode $\bar y_t$}
\\ \nonumber
&\propto &  \exp(- q y_t)~~~\text{for larger $y_t$ -- the tail},
\eea
where the transition from Gaussian to exponential on \yt\ is determined by slope matching~\cite{fragevo}. The $\hat H_0$ tail density on \pt\ varies approximately as power law $1/p_t^{q + 1.8} \approx  1/p_t^n$. Coefficient $A~(\approx 0.3)$ is determined by the unit-normal condition. Model parameters $(\bar y_t,\sigma_{y_t},q)$ are derived as described in App.~\ref{old13} except as noted in the main text.

All spectra are plotted vs pion rapidity $y_{t\pi}$ with pion mass assumed. The motivation is comparison of spectrum hard components demonstrated to arise from a common underlying jet spectrum on \pt~\cite{fragevo}, in which case $y_{t\pi}$ serves simply as a logarithmic measure of hadron \pt\ with well-defined zero. $\hat S_0(m_{t})$ in Eq.~(\ref{s0m}) is transformed to $y_{t\pi}$ via the Jacobian factor $m_{t\pi} p_t / y_{t\pi}$ to form  $\hat S_0(y_{t\pi})$ for unidentified hadrons. $\hat H_0(y_{t})$ in Eq.~(\ref{h00}) is always defined on $y_{t\pi}$ as noted. In general, plotting spectra on a logarithmic rapidity variable provides improved access to important spectrum structure in the low-\pt\ interval {\em where the majority of jet fragments appear}.

\subsection{\titlepp\ \titlept\ spectrum data} \label{tcmspecdat}

Figures~\ref{ptspec1} and \ref{ptspec2} (a,c) show the general relation between the TCM (solid) and ALICE data (points). The TCM is not the result of fits to individual spectra. The curves actually represent predictions derived from a self-consistent description of \pp\ spectra covering the energy interval 17 GeV to 13 TeV (Ref.~\cite{alicetomspec} and App.~\ref{old13}). The data/TCM ratios in (b,d)  provide important information on biases resulting from V0M and SPD event sorting methods.

\begin{figure}[h]
	\includegraphics[width=3.3in,height=1.6in]{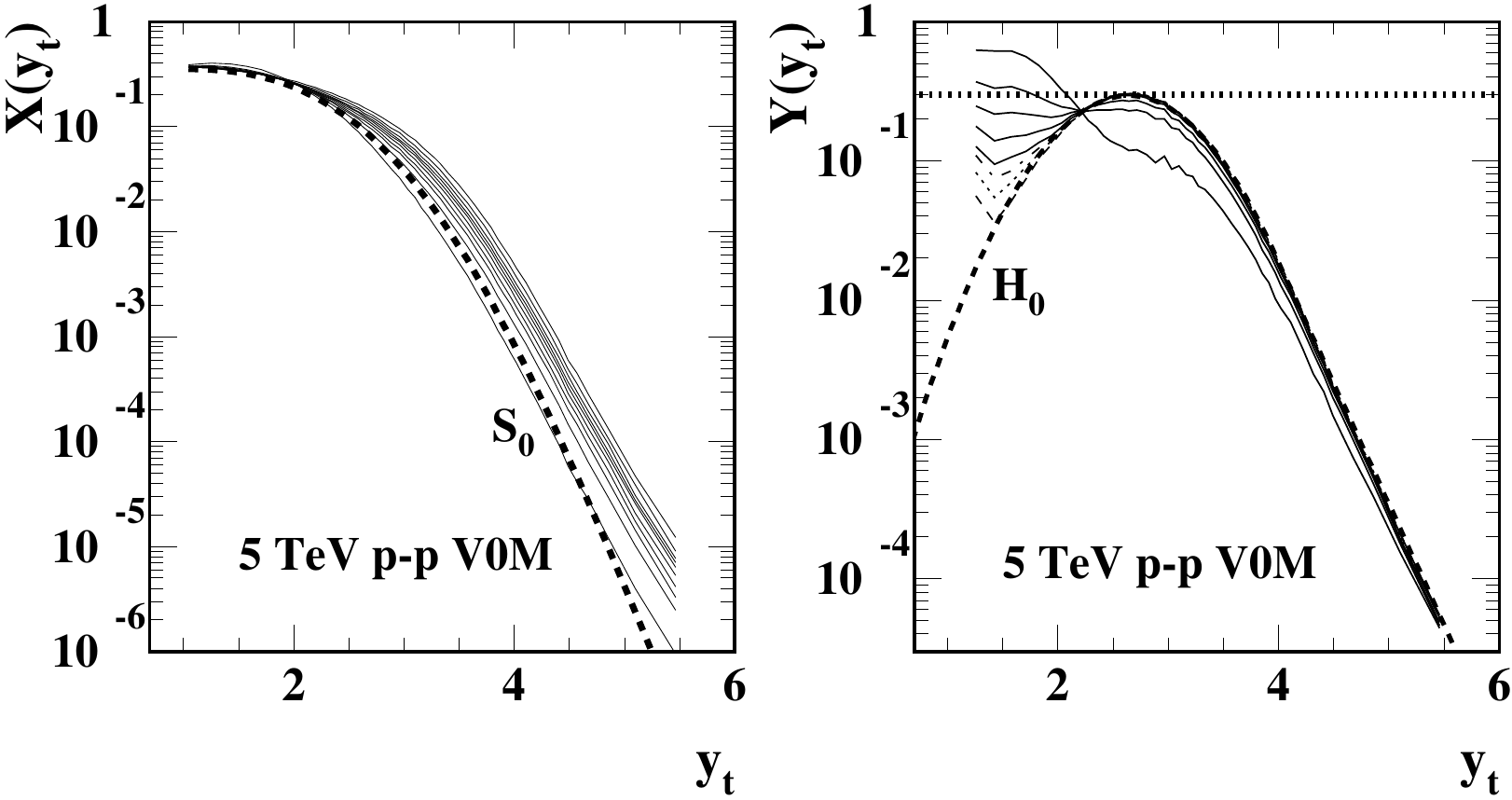}
\put(-140,89) {\bf (a)}
\put(-24,89) {\bf (b)}\\
	\includegraphics[width=3.3in,height=1.6in]{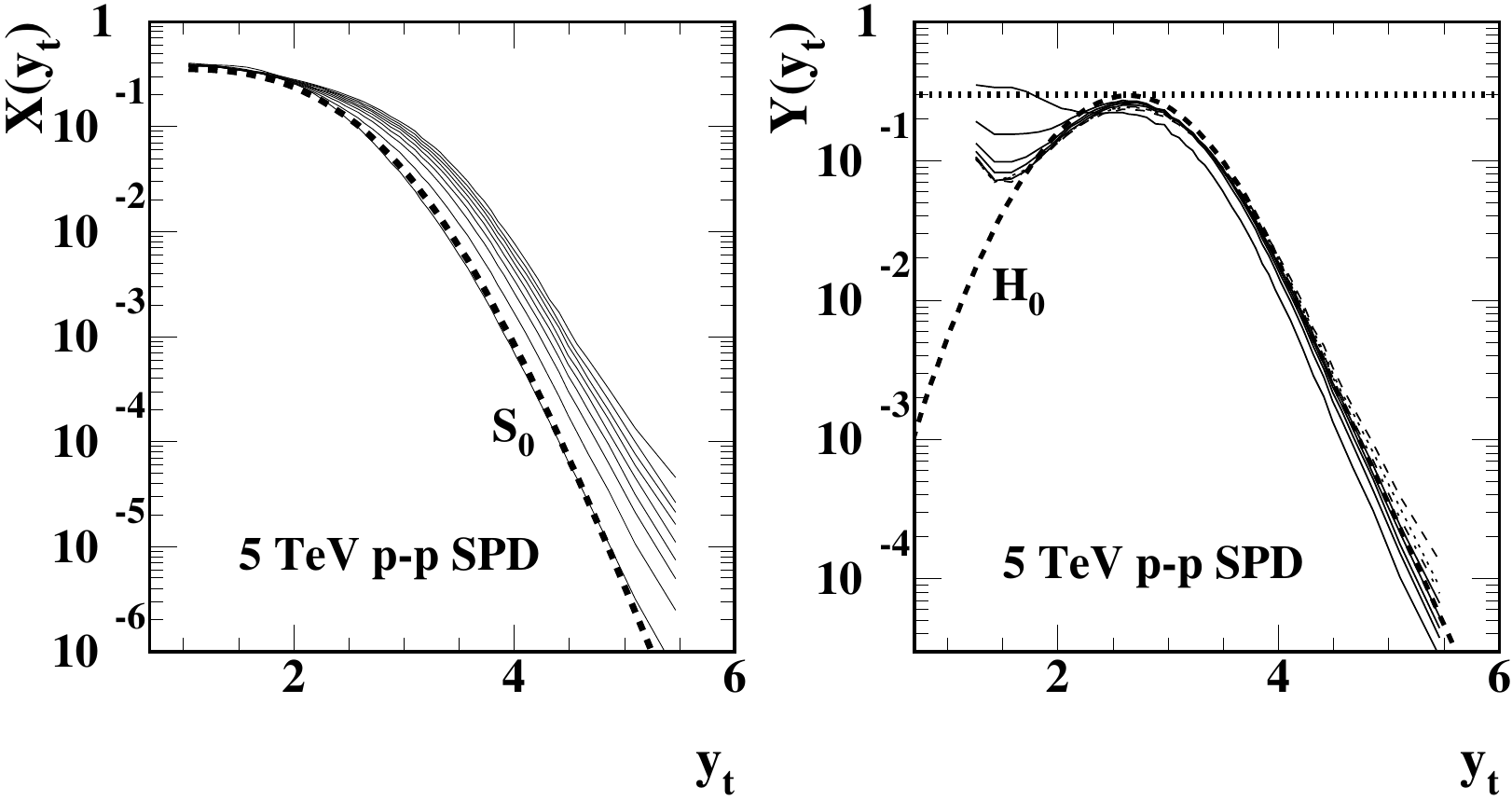}
\put(-140,89) {\bf (c)}
\put(-24,89) {\bf (d)}
	\caption{\label{tcm5}
Left: Normalized \yt\ spectra in the form $X(y_t)$ from ten (V0M) or nine (SPD) multiplicity classes of 5 TeV \pp\ collisions for V0M (a) and SPD (c) event selection.
Right: Normalized spectrum hard components in the form $Y(y_t)$ for data in the left panels for  V0M (b) and SPD (d) event selection. The bold dashed curves are TCM model functions.
	}  
\end{figure}

\begin{figure}[h]
	\includegraphics[width=3.3in,height=1.6in]{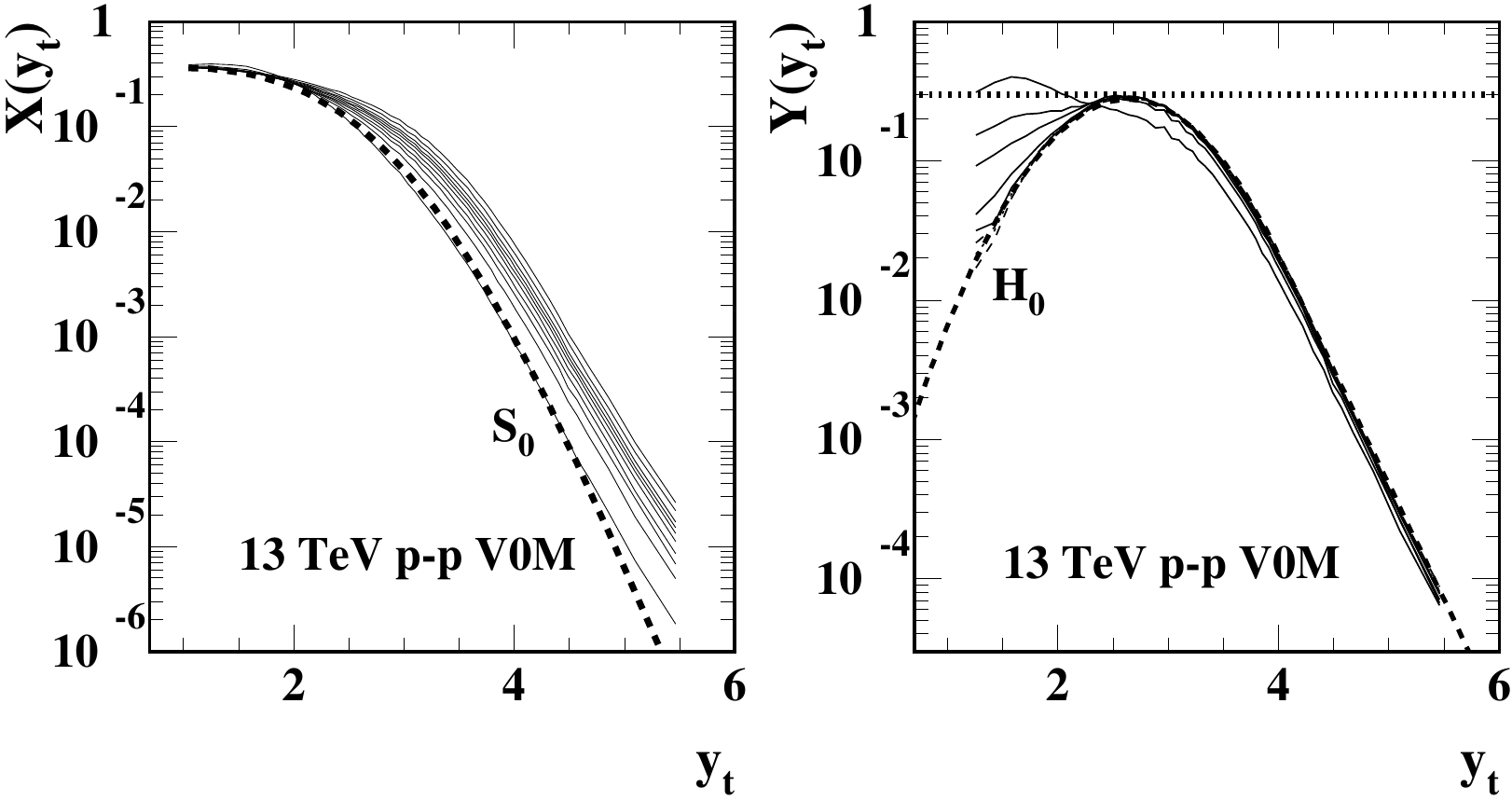}
\put(-140,89) {\bf (a)}
\put(-24,89) {\bf (b)}\\
	\includegraphics[width=3.3in,height=1.6in]{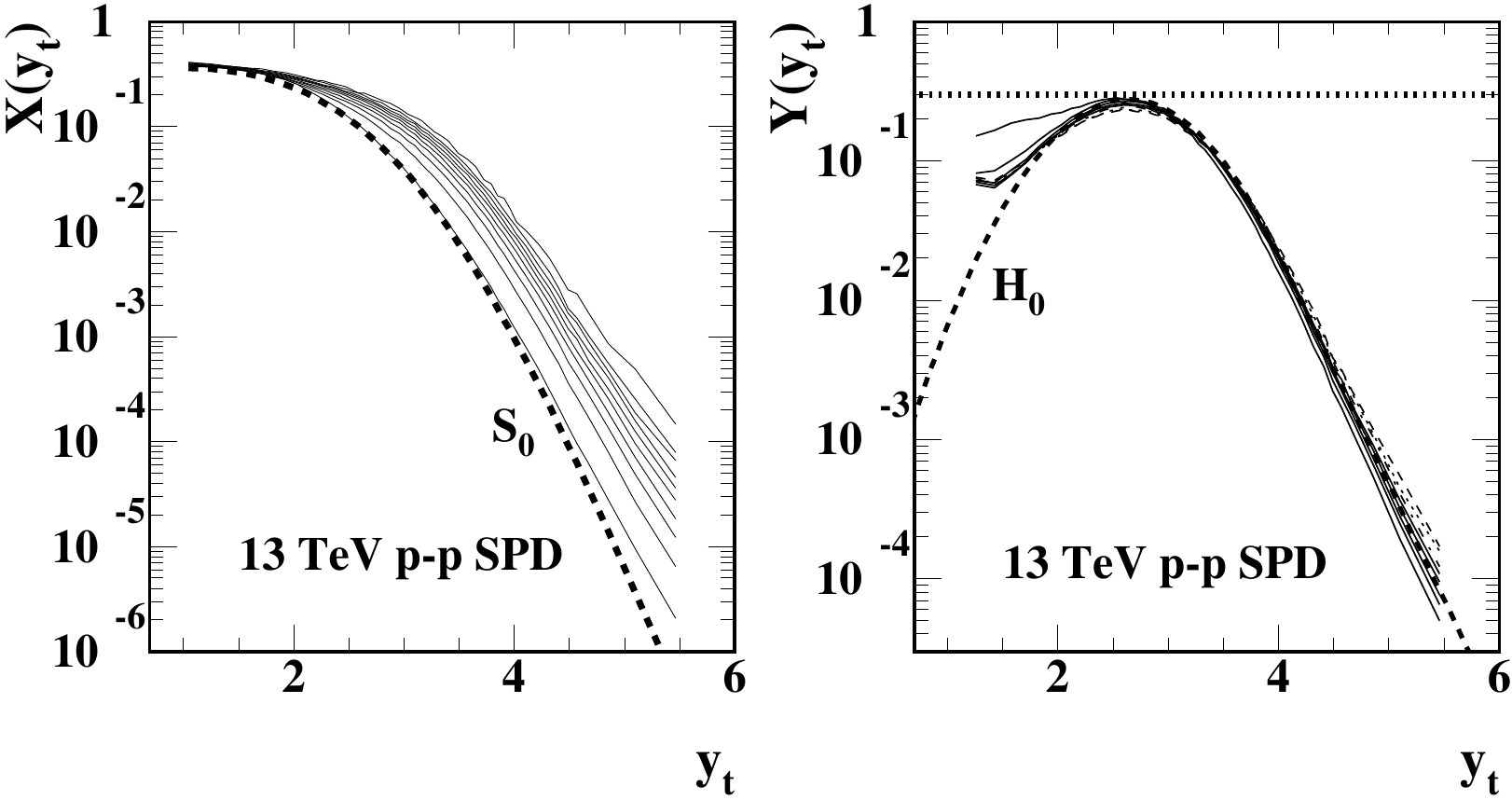}
\put(-140,89) {\bf (c)}
\put(-24,89) {\bf (d)}
	\caption{\label{tcm13}
Left: Normalized \yt\ spectra in the form $X(y_t)$ from ten multiplicity classes of 13 TeV \pp\ collisions for V0M (a) and SPD (c) event selection.
Right: Normalized spectrum hard components in the form $Y(y_t)$ for data in the left panels for  V0M (b) and SPD (d) event selection.  The bold dashed curves are TCM model functions.
	} 
\end{figure}

The TCM format of Figs.~\ref{tcm5} and $\ref{tcm13}$ then provides a more-differential decomposition of spectrum data into soft and hard components. Panels (a,c) show full data spectra (thin solid) in the normalized form $X(y_t)$ defined above that are directly comparable with soft-component model $\hat S_0(y_t)$ (bold dashed). Below 0.5 GeV/c ($y_t \approx 2$) the data curves closely follow the model.  The same $\hat S_0(y_t)$ model is used for both event-selection methods.

Panels (b,d) show inferred data hard components $Y(y_t)$ defined by Eq.~(\ref{yyt}) (thin, several line styles) compared to TCM hard-component model $\hat H_0(y_t)$ (bold dashed). Deviations from $\hat H_0(y_t)$ below $y_t = 2$ appear in every \pp\ collision system (e.g.\ 200 GeV as reported in Ref.~\cite{ppprd}). The horizontal dotted lines provide a check on proper normalization of hard-component model $\hat H_0(y_t)$ The data hard component for the lowest multiplicity class is not shown because there is in effect very little jet contribution to those events due to strong selection bias. Note that full spectra in panels (a,c) for the lowest \nch\ class are approximately consistent with $\hat S_0(y_t)$. The same $\hat H_0(y_t)$ model is used for both event-selection methods.

Although Ref.~\cite{alicenewspec} states that \pp\ \pt\ spectra show ``little energy dependence'' both soft and hard components exhibit significant energy dependence as previously reported in Ref.~\cite{alicetomspec}. The TCM parameters $n$ (soft-component exponent) and $q$ (hard-component exponent) vary systematically with $\log(\sqrt{s})$. The variation is apparent as shifts of $\hat S_0(y_t)$ and $\hat H_0(y_t)$ intercepts on \yt\ (at plot lower bounds) to larger values with increasing energy. The former may be related to  increasing depth of the longitudinal splitting cascade on proton momentum fraction $x$ with increasing energy~\cite{alicetomspec}, and the latter is certainly related to expected evolution of the underlying jet spectrum with increasing \pp\ collision energy~\cite{jetspec2}.

\subsection{Spectrum TCM parameter summary} \label{parsum}

Table~\ref{engparamy} presents TCM parameters for 5 TeV and 13 TeV \pp\ collisions. Entries are grouped as soft-component parameters $(T,n)$, hard-component parameters $(\bar y_t,\sigma_{y_t},q)$, hard/soft ratio parameter $\alpha$ and NSD (non-single-diffractive) soft density $\bar \rho_{sNSD}$. For unidentified hadrons soft component $\hat S_0(m_t)$ may be approximated by Eq.(\ref{s00}) for pions only with parameters as in Table~\ref{engparamy}. Slope parameter $T = 145$ MeV is held fixed for all cases consistent with data.  Its value is determined  within a low-\yt\ interval where the hard component is negligible. For the present analysis  Eq.~(\ref{s00}) was evaluated separately for pions, kaons and protons with $T_i = 140$, 200 and 210 MeV respectively. Those expressions were then combined to form $\hat S_0(m_t)$ via Eq.~(\ref{s0m}) with $z_{0i} = 0.82$, 0.12 and 0.06 respectively. For each energy the same exponent $n$ was applied to three hadron species. L\'evy exponent $n$ values and hard-component $q$ values are as reported in Table~\ref{engparamy}. 

$\bar \rho_{sNSD}$ values are derived from the universal NSD trend $\bar \rho_{sNSD} \approx 0.81 \ln(\sqrt{s} / \text{10 GeV})$ inferred from spectrum and yield data. $\bar \rho_{0NSD}$ values are derived from the TCM relation $\bar \rho_0 \approx \bar \rho_s + \alpha \bar \rho_s^2$. The NSD $\bar \rho_0$ values can be compared with $\bar \rho_0 = 5.91$  and 7.60 for 5 and 13 TeV INEL $> 0$ events from Ref.~\cite{alicenewspec} [above its Eq.~(1)]. The 13 TeV number contrasts with $6.46 \pm 0.19$ in its Fig.~5 (center panel). Appendix~\ref{old13} describes previous TCM analysis of 13 TeV \pp\ \pt\ spectrum data from Ref.~\cite{alicespec}.

\begin{table}[h]
	\caption{\pt\ spectrum TCM parameters for 5 TeV and 13 TeV NSD \pp\ collisions within $\Delta \eta \approx 2$.
	}
	\label{engparamy}
	\begin{center}
		\begin{tabular}{|c|c|c|c|c|c|c|c|c|} \hline
			$\sqrt{s}$ (TeV) & T\. (MeV) & $n$ & $\bar y_t$ & $\sigma_{y_t}$ & $q$ & $100\alpha$ & $\bar \rho_{s\text{NSD}}$ &  $\bar \rho_{0\text{NSD}}$ \\ \hline
			5.0 & 145  & 8.5 & 2.63 & 0.58  &  4.0  & 1.45 & 5.0 & 5.3 \\ \hline
			13.0  & 145  & 7.8 & 2.66 & 0.60  & 3.8  & 1.70 & 5.8 & 6.3 \\ \hline
		\end{tabular}
	\end{center}
\end{table}

It should be noted that the $\alpha$ values for 5 TeV and 13 TeV in Table~\ref{engparamy} are 0.0145 and 0.0170 whereas Table~\ref{engparamx} includes values reported in Ref.~\cite{alicetomspec} for alpha as 0.013 and 0.015. The earlier values were based on limited 13 TeV \pp\ spectrum data from Ref.~\cite{alicespec}. The updated values better accommodate the much more complete spectrum data reported in Ref.~\cite{alicenewspec}. The $\approx 12$\% increase in $\alpha$ values is compatible with estimated uncertainties in Fig.~16 (left) of Ref.~\cite{alicetomspec} but also favors a prediction of the $\alpha(\sqrt{s})$ trend based on measured \pp\ jet cross sections and fragmentation functions (dashed curve in that panel).

\section{\titlept\ spectrum biases and evolution} \label{biases}

V0M and SPD event-selection methods bias event structure (e.g.\ \pt\ spectra) in different ways. Biases are examined here relative to a TCM reference. The TCM for \pp\ collisions assumes linear superposition of parton-parton interactions consistent with basic QCD (e.g.\ published jet measurements). The TCM is not fitted to individual spectra; it serves as a fixed reference for comparison of biases from different event-selection methods.

\subsection{Spectrum ratios vs $\bf n_{ch}$}

In its Figs.~2 and 3 Ref.~\cite{alicenewspec} presents ratios of \pt\ spectra for ten \nch\ classes to those for minimum-bias INEL $> 0$ ensemble averages for each of two event selection criteria. It is noted that ``While at low $p_T$ ($< 0.5$ GeV/c) the ratios exhibit a modest $p_T$ dependence, for $p_T > 0.5$ GeV/c they strongly depend on multiplicity and $p_T$.'' It is possible to arrive at more-detailed quantitative conclusions using TCM-based techniques first reported in Ref.~\cite{ppprd}.

\begin{figure}[h]
	\includegraphics[width=1.65in,height=1.6in]{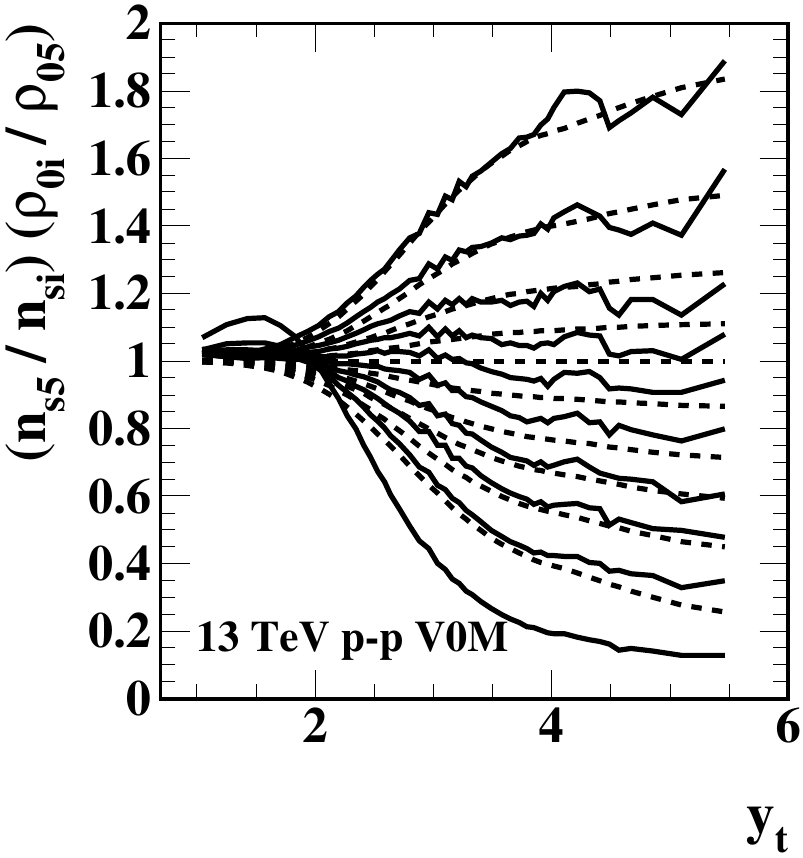}
	\includegraphics[width=1.65in,height=1.6in]{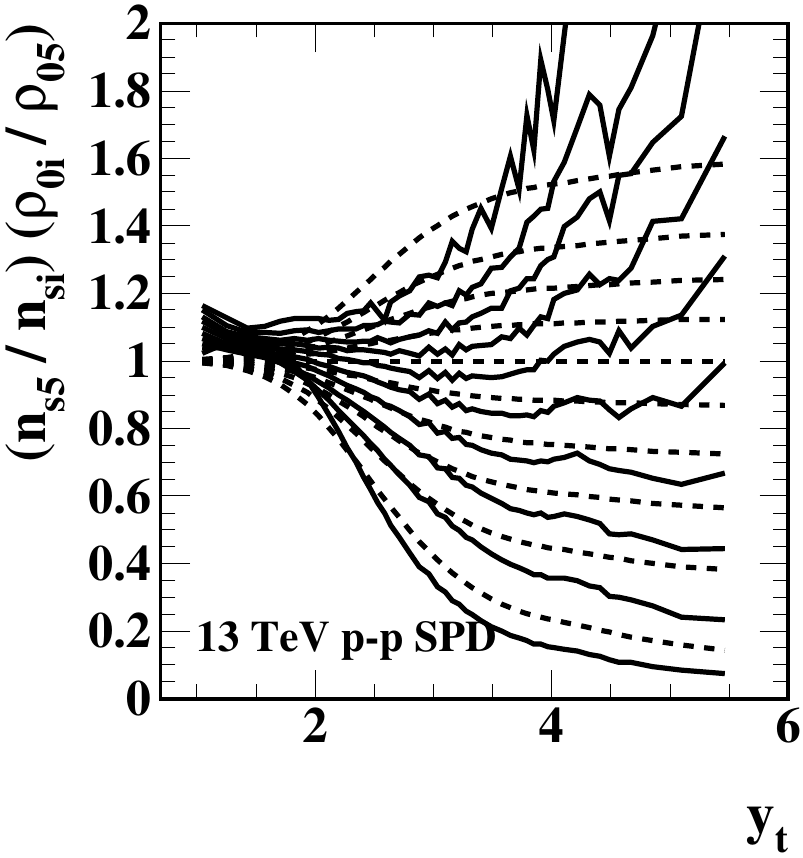}
	\caption{\label{specrat}
		Spectrum ratios for data (solid) and TCM (dashed) from 13 TeV \pp\ collisions in ratio to the TCM spectrum for event class V for V0M (left) and SPD (right) event selection.
	}  
\end{figure}

Figure~\ref{specrat} shows spectra from 13 TeV \pp\ collisions for V0M (left) and SPD (right) event classes and for data from Ref.~\cite{alicenewspec} (solid) and corresponding TCM (dashed). The various spectra are in ratio to the TCM spectra for \nch\ class 5 where 1-10 goes from higher to lower \nch. The spectra have first been normalized by the corresponding soft-component density $\bar \rho_s$ as they appear in Fig.~\ref{tcm13} (a,c).

The corresponding Figs.~2 and 3 in Ref.~\cite{alicenewspec} show ratios of \pt\ spectra to a single minimum-bias INEL $>0$ reference spectrum. As such the spectrum ratios include three sources of systematic variation: (a) mean event multiplicity varying from class to class, (b) varying jet contribution relative to charge multiplicity and (c) bias effects that are of major interest. It is then essentially impossible to sort out what cause produces which effect. 

Figure~\ref{specrat} removes variation due to mean event multiplicity, but strong variation of jet contributions relative to total yield is still confused with bias effects.  As a consequence of that plotting format, below \yt\ = 2 (0.5 GeV/c) spectra nearly coincide as a result of the chosen normalization and are nearly constant on \yt, qualitatively consistent with the observation in Ref.~\cite{alicenewspec} (modulo distortions from selection bias discussed in the next subsection). Above that point the ratios vary strongly with \nch\ and \pt, also qualitatively consistent with Ref.~\cite{alicenewspec}. However, such variation is to be expected given that jet production varies approximately {\em quadratically} with \nch. 

Close examination of the left panel reveals that the data spectrum for the highest V0M \nch\ class corresponds to the TCM (dashed) within statistics. Then with decreasing \nch\  spectra are suppressed relative to the TCM at higher \pt\ but  are enhanced at lower \pt, the enhancement {\em mode} moving to lower \pt. Correlated suppression and enhancement lead to trends in Fig.~\ref{endpoint} where spectrum {\em integrals} adhere to the TCM trend $\bar \rho_h \propto \bar \rho_s^2$ for all event classes even as the spectrum hard components are significantly modified in shape with decreasing \nch.

\subsection{V0M and SPD biases relative to the TCM}  \label{bias}

Event sorting or selection in Ref.~\cite{alicenewspec} is based on different angular acceptances denoted by acronym. SPD denotes tracklets (two hits plus vertex) within $|\eta|< 0.8$, the same angular acceptance as for the \pt\ spectra. V0M denotes a ``forward estimator'' with combined acceptances $\eta \in [2.8,5.1]$ and  $\eta \in [-3.7,-1.7]$ that is said to ``minimize the possible autocorrelations introduced by the use of the mid-pseudorapidity estimator.'' The term ``autocorrelations'' is here misused. The autocorrelation function (special case of cross-correlation function) is an established statistical method for analyzing time series~\cite{autocorr}. A better term is {\em selection bias} wherein event selection is based on the same particle sample (e.g.\ mid-rapidity hadrons) as the object of study (mid-rapidity \pt\ spectra).

\begin{figure}[h]
	\includegraphics[width=1.65in,height=1.6in]{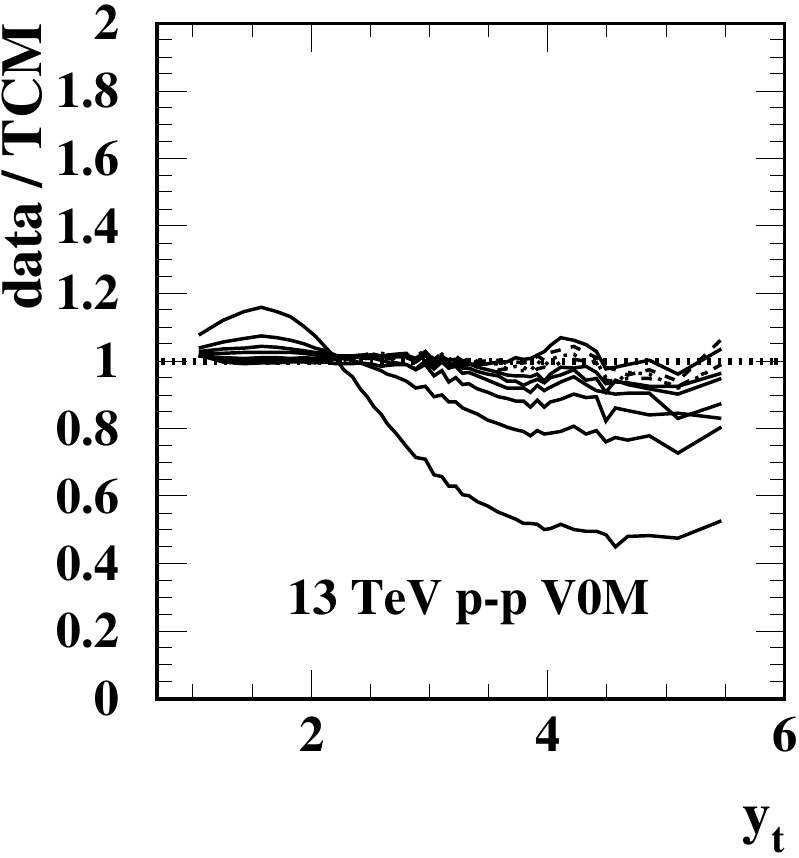}
	\includegraphics[width=1.65in,height=1.6in]{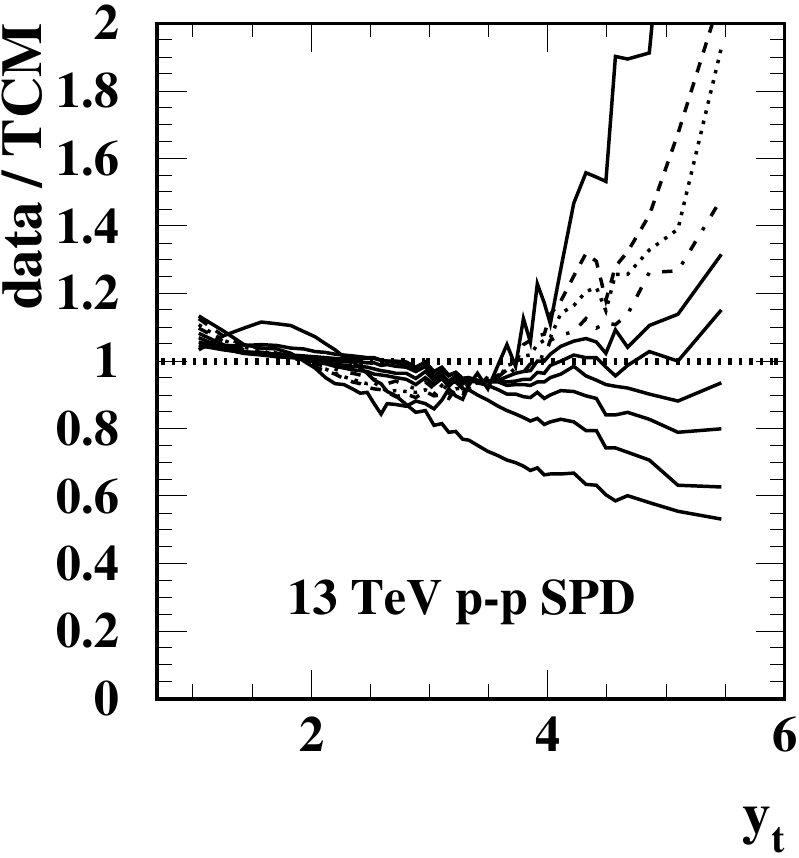}
	\caption{\label{v0mspd}
		Data/TCM spectrum ratios from 13 TeV \pp\ collisions for V0M (left) and SPD (right) event selection.
	} 
\end{figure}

Figure~\ref{v0mspd} shows data/TCM ratios for 13 TeV \pp\ collisions and for V0M (left) and SPD (right) event classes. In contrast to Fig.~\ref{specrat} the systematic variation of jet yield relative to total yield is largely canceled in the data/TCM ratio. The same TCM reference is used for both selection methods. What remains is the bias effects of interest. The 5 TeV results are similar. The two methods bias spectra substantially but in {\em apparently} different ways. V0M for smaller multiplicities suppresses spectra at higher \yt\ but produces complementary enhancement at lower \yt.  SPD for smaller multiplicities also suppresses higher \yt\ and enhances lower \yt, but for greater \nch\ there is {\em apparently} strong enhancement for  higher \yt\ whereas V0M produces no significant corresponding effect. However, spectrum {\em ratios} exaggerate structure at higher \yt\ while concealing important structure at lower \yt~\cite{alicetomspec}. For example, compare these results with data-model {\em differences} in ratio to statistical errors (Z-scores) in Fig.~\ref{comparex}.

As can be seen in  Figs.~\ref{tcm5} and \ref{tcm13} (b,d) the physical process biased by event selection is jet production represented by the TCM spectrum hard component. The TCM hard-component reference has fixed exponent $q$ corresponding to \pt\ power-law exponent $n$. The constant trend for the data/TCM ratio above \yt\ = 4 in Fig.~\ref{v0mspd} (left) is consistent with $n$ or $q$ independent of \nch, whereas the result in Fig.~\ref{v0mspd} (right) is consistent with strong decrease of those parameters with increasing \nch. See Figs.~\ref{logderiv} (right) and \ref{checkx} (left). The different forms of spectrum bias relating to V0M and SPD as indicated by Fig.~\ref{v0mspd} can thus be understood in terms of jet production.

Minimum-bias jet production near midrapidity arises from three processes: (a) separate parton splitting cascades within projectile protons resulting from inelastic scattering, (b) occasional large-angle scattering of cascade (participant) partons and (c) fragmentation of scattered participant partons to dijets. A proton splitting cascade ({\em event-wise} parton distribution function or PDF) is sensitive to initial conditions and fluctuates strongly from event to event~\cite{gosta}. Likewise, the fragment distribution within a jet (also a splitting cascade) fluctuates strongly from jet to jet. The biases indicated in Fig.~\ref{v0mspd} result from sorting the {\em same} minimum-bias INEL $> 0$ event population according to two different criteria.

Since V0M selection is derived from particle yields at higher $\eta$ outside the acceptance for spectra it cannot significantly bias the jet formation process itself since most jets, near the lower bound of the jet spectrum, are derived from low-$x$ partons appearing near midrapidity within a longitudinal cascade. However, V0M selection for low \nch\ may influence the underlying jet spectrum resulting from the event-wise PDF, i.e.\ softening the jet spectrum. The result is a shift of the data hard component to lower \yt\ without changing its shape, consistent with Fig.~\ref{v0mspd} (left).

Since SPD selection is based on particle yields within the same acceptance as for \pt\ spectra it relates primarily to low-$x$ partons {\em and} low-energy jets. It cannot significantly bias the event-wise PDF at larger $x$ (or $\eta$) but can strongly bias the parton scattering and fragmentation process near midrapidity. Figure~\ref{v0mspd} (right) suggests that whereas lower SPD values produce a bias similar to V0M (softened jet spectrum), for higher SPD values the effective jet spectrum {\em and} mean fragmentation function are biased to harder distributions leading to evolution of the hard-component tail. The manifestation in the data/TCM {\em ratio} seems dramatic but a very small fraction of all particles is actually involved. Figures~\ref{tcm5}, \ref{tcm13} and \ref{comparex} provide a more transparent picture of hard-component biases arising from V0M and SPD event selection.

The role of fluctuations warrants further consideration. It would be informative to have a 2D plot of event density on SPD vs V0M. Given the jet production scenario described above one may conjecture qualitatively that for large V0M yields (and hence event-wise PDF) the SPD event multiplicity and especially jet contribution is free to fluctuate strongly over a large range whereas the multiplicity {\em mean value} for fixed V0M remains modest as observed. In contrast, for large SPD yields the largest jet fluctuations are singled out, V0M is pinned to its highest value (V0M fluctuations are thus limited) and the multiplicity mean value for fixed SPD is large as observed. In Fig.~\ref{comparex} the most significant bias structure (the bipolar excursion on \yt) has maximum amplitude for the {\em lowest} values of V0M and SPD. The trend is consistent with a biased underlying jet spectrum (via the event-wise PDF). 

\subsection{Significance of data-model differences}

Data/model spectrum ratios may exaggerate deviations at higher \pt\ compared to deviations at lower \pt. A more transparent representation is based on statistical-significance measures. The Z-score~\cite{zscore} compares data-model differences to their statistical uncertainties
\bea \label{zscore}
Z_i &=& \frac{O_i - E_i}{\sigma_i},
\eea
where $O_i$ is an observation (e.g.\ spectrum data), $E_i$ is an expectation (e.g.\ predictions derived from a model) and $\sigma_i$ is the statistical uncertainty of the observation.

\begin{figure}[h]
	\includegraphics[width=1.65in,height=1.6in]{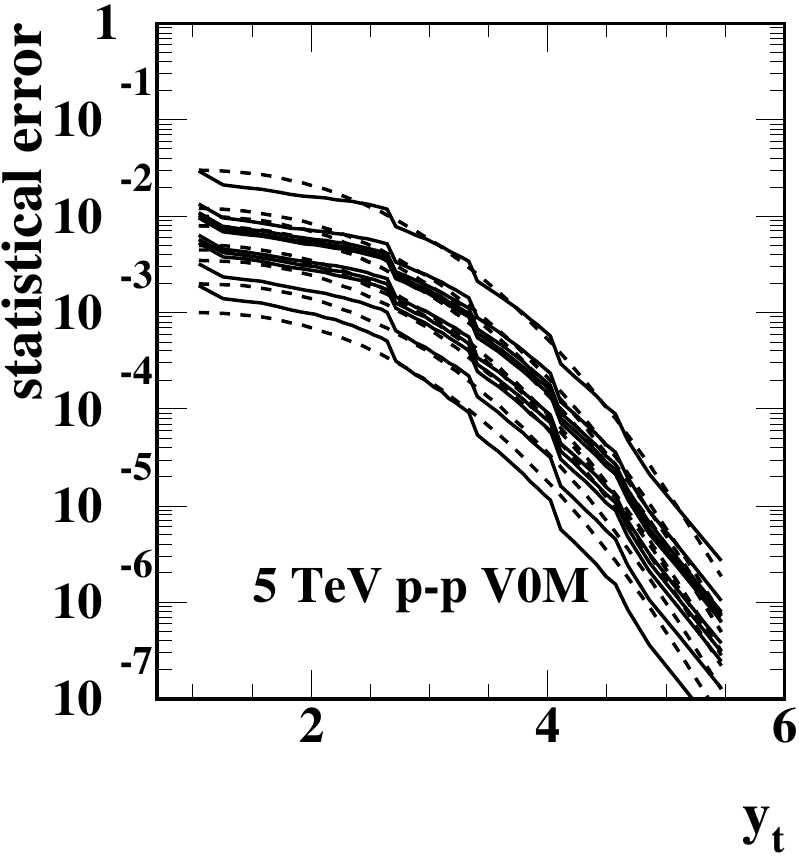}
	\includegraphics[width=1.65in,height=1.6in]{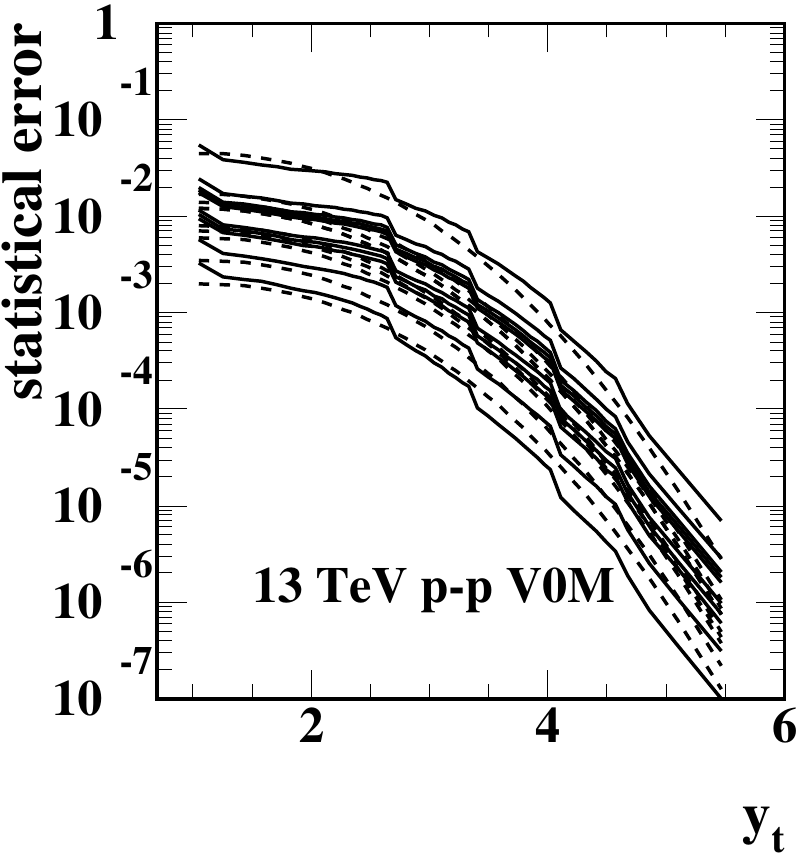}
\put(-147,89) {\bf (a)}
\put(-24,89) {\bf (b)}\\
	\includegraphics[width=1.65in,height=1.6in]{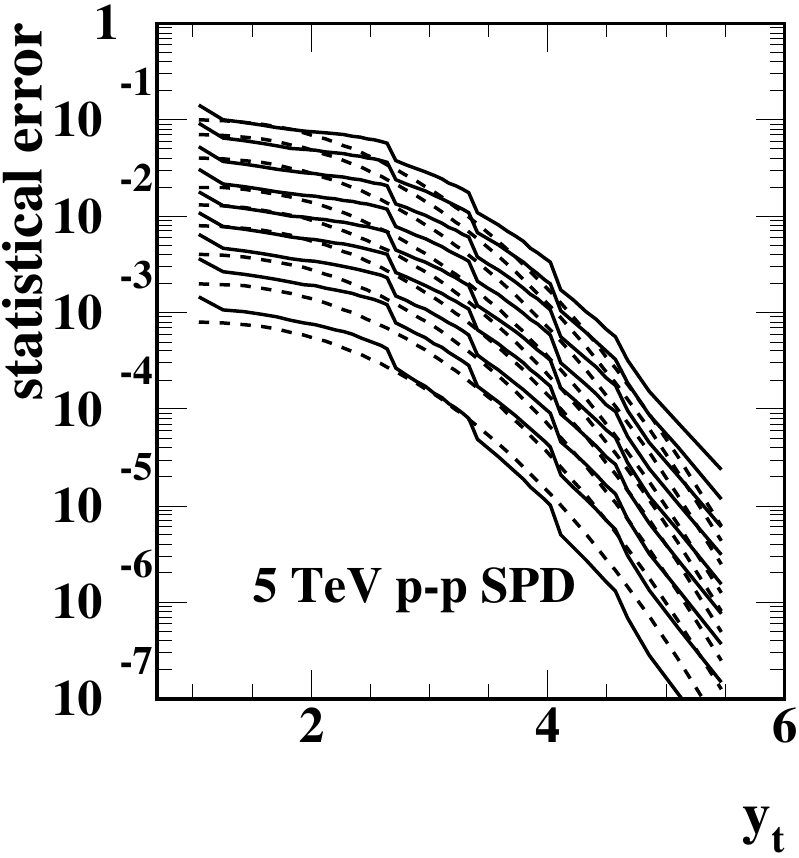}
\includegraphics[width=1.65in,height=1.6in]{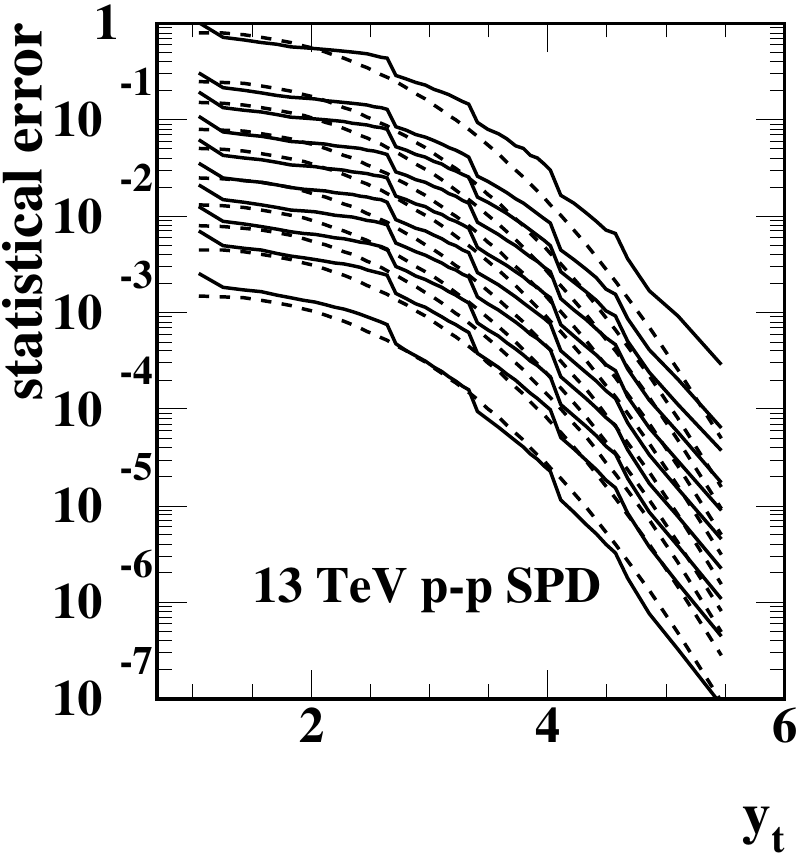}
\put(-147,89) {\bf (c)}
\put(-24,89) {\bf (d)}
	\caption{\label{staterr}
		Statistical errors (solid) as reported in Ref.~\cite{alicenewspec} for 5 TeV (left) and 13 TeV (right) \pp\ collisions. To avoid the step-like structures in the data errors simple models (dashed, Gaussians on \yt) are used to approximate the error trends.
	}  
\end{figure}

Figure~\ref{staterr} shows statistical errors $\sigma_i$ (solid) accompanying published VOM and SPD spectrum data for 5 and 13 TeV \pp\ collisions which exhibit step-like structures. If used to process data within ratios those structures are injected into the result. Smooth approximations (dashed) are introduced to represent statistical errors without step-like structures. The approximated error curves are used for data-model comparisons below.

\begin{figure}[h]
	\includegraphics[width=1.65in,height=1.6in]{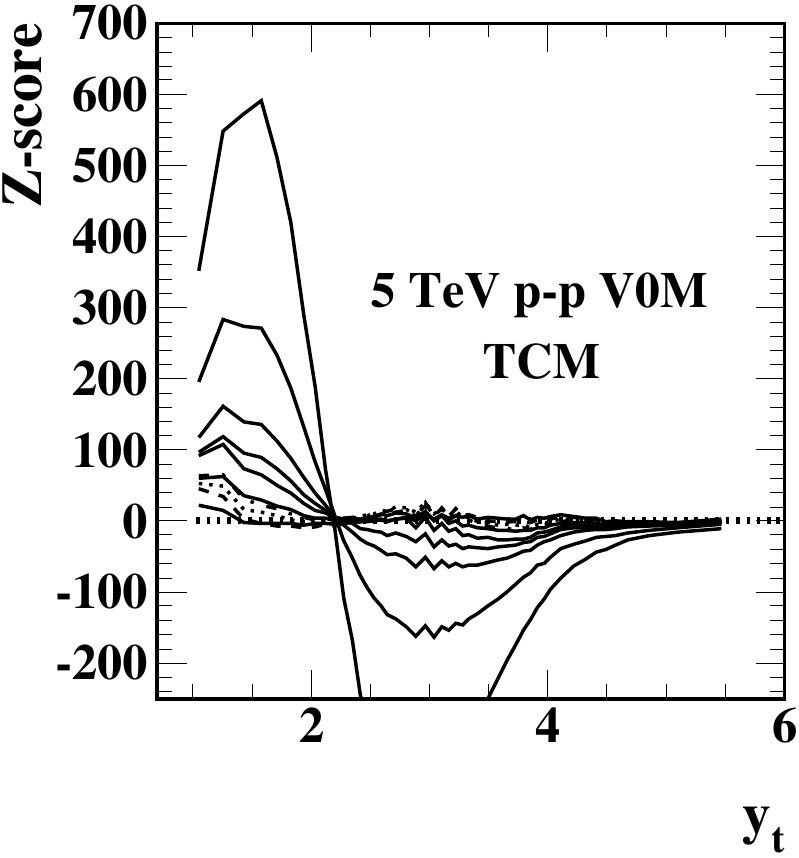}
	\includegraphics[width=1.65in,height=1.6in]{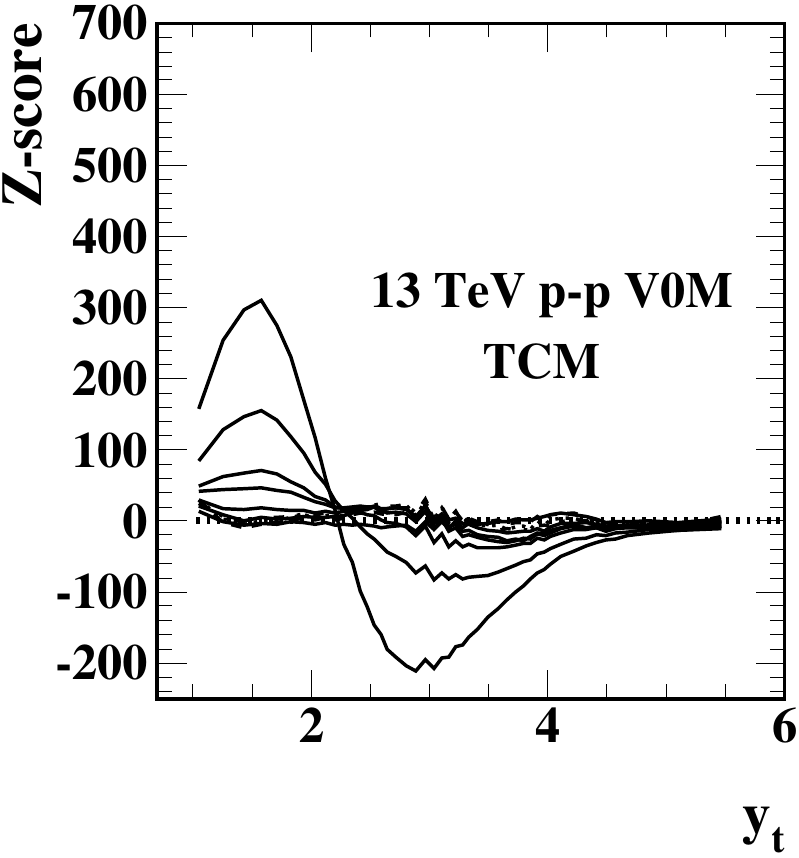}
	\put(-145,99) {\bf (a)}
	\put(-24,99) {\bf (b)}\\
	\includegraphics[width=1.65in,height=1.6in]{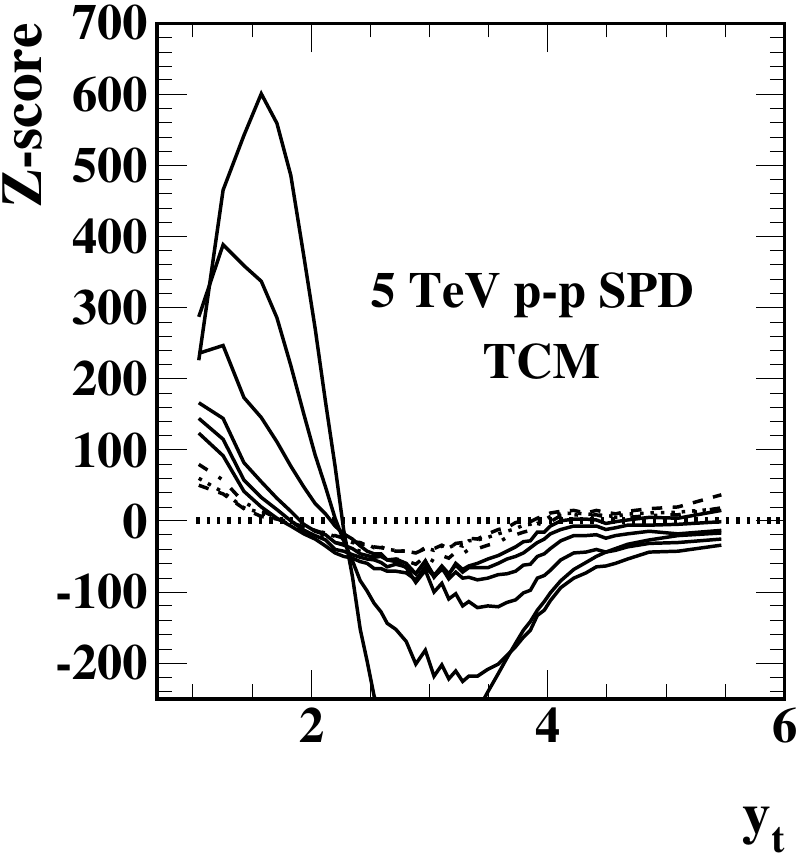}
	\includegraphics[width=1.65in,height=1.6in]{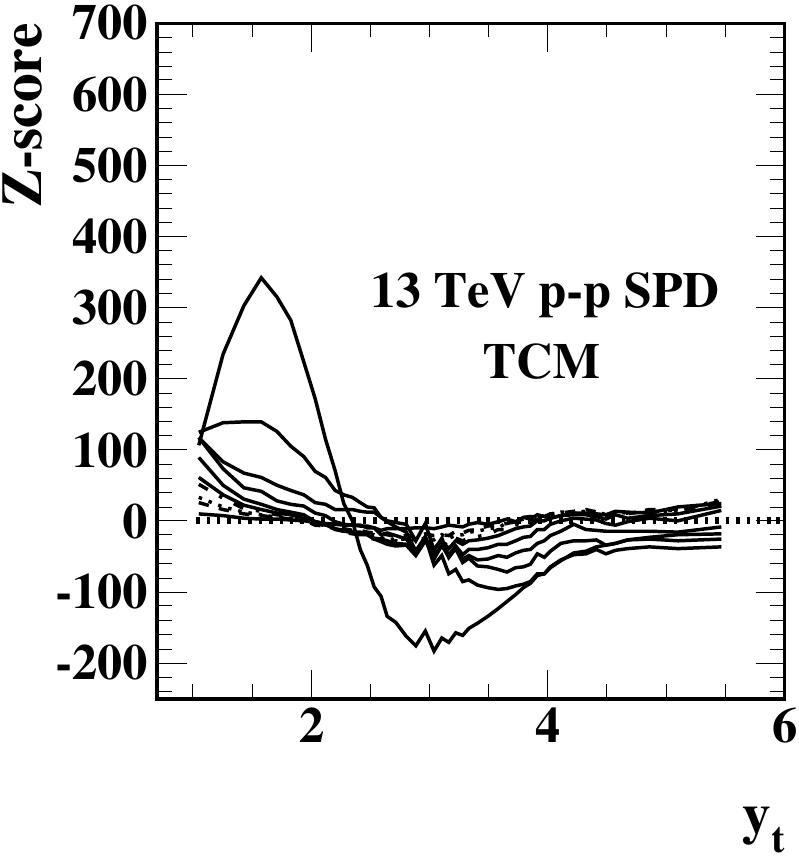}
	\put(-145,99) {\bf (c)}
	\put(-24,99) {\bf (d)}\\
	\caption{\label{comparex}
		Data-TCM differences in ratio to statistical errors (Z-scores) for 5 TeV (left) and 13 TeV (right) and for V0M event selection (upper) and  SPD selection (lower). The highest multiplicity classes in each case are represented by solid, dashed, dotted and dash-dotted curves in descending order.
	} 
\end{figure}

Figure~\ref{comparex} shows data-TCM {differences} in ratio to statistical uncertainties for 5 TeV (left) and 13 TeV (right) \pp\ collisions and for V0M (upper) and SPD (lower) event selection. Such Z-score results can be contrasted with data/TCM ratios as in Fig.~\ref{v0mspd}. Whereas in the ratio format biases for V0M and SPD appear quite different, the Z-scores in Fig.~\ref{comparex} reveal the {\em statistical significance} of data-model deviations. 
Despite noticeable differences at higher \pt\ the most significant bias effects are similar. The bias amplitude in terms of Z-scores seems to track with fraction of total cross section rather than with \nch.

\section{\titlept\ spectrum shape measures} \label{shapes}

Reference~\cite{alicenewspec} analyzed the evolution of \pt\ spectrum shapes with varying charge multiplicity \nch\ in two ways:  (a) power-law model fits to spectra to infer trends for power-law exponent $n$ and (b) variation of integrated yields within three \pt\ intervals compared to a minimum-bias reference as a function of \nch. This section considers such shape-measure results in the context of the TCM.

\subsection{Power-law fits to high-$\bf p_t$ intervals} \label{powerlaw}

Reference~\cite{alicenewspec} fitted a power-law function to 13 TeV \pp\ spectra above 6 GeV/c ($y_t \approx 4.5$) to estimate exponents $n$ vs \nch\ for V0M and SPD events. A related result can be obtained without curve fitting via a logarithmic derivative applied directly to data hard components $Y(y_t)$ as in Figs.~\ref{tcm5} and \ref{tcm13} (b,d) or $Y(p_t) = y_tY(y_t) / m_t p_t$:
\bea  \label{logg}
-\frac{d\ln[Y(y_t)]}{dy_t} \rightarrow q ~~\text{for}~~ y_t > 4.1
\\ \nonumber
-\frac{d\ln[Y(p_t)]}{d\ln(p_t)} \rightarrow n ~~\text{for}~~ y_t > 4.1.
\eea
 If $dn_{ch}/y_t dy_t \propto \exp(-q y_t)$ [high-\yt\ tail of $\hat H_0(y_t)$] then $dn_{ch}/p_t dp_t = (y_t / m_t p_t)dn_{ch}/y_t dy_t \propto 1/ p_t ^{q + 2} \approx 1/p_t^n$. $n \approx q +2$ is a rough estimate, but $n \approx q +1.8$ is established by direct comparison of results from Eq.~(\ref{logg})  (upper and lower) for the same spectra. Power-law exponent $n$ invoked by Ref.~\cite{alicenewspec} should not be confused with TCM L\'evy exponent $n$ for soft component $\hat S_0(m_t)$.
 
\begin{figure}[h]
	\includegraphics[width=3.3in,height=1.6in]{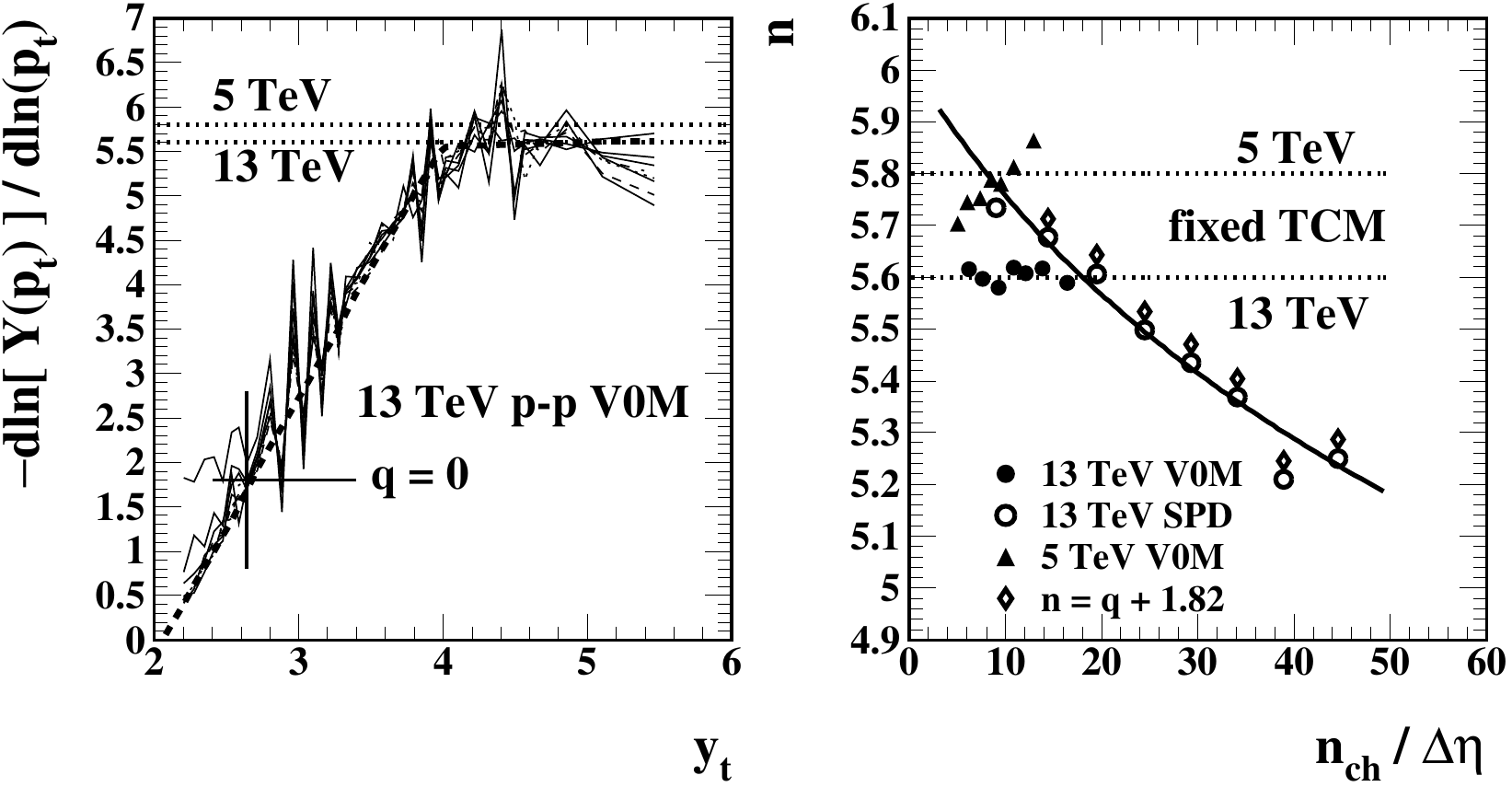}
	\caption{\label{logderiv}
		Left: Logarithmic derivatives defined by Eq.~(\ref{logg}) (second line, thin solid) for ten \nch\ classes of 13 TeV \pp\ collisions and V0M events. The dashed curve is the same operation applied to TCM hard component $\hat H_0(p_t)$. The crossed solid lines are explained in the text.
		Right: Power-law exponents $n$ obtained by averaging the last ten points in the left panel (i.e.\ $y_t > 4.1$, $p_t > 4.2$ GeV/c). Points corresponding to $n = q + 1.8$ with $q$ from Fig.~\ref{checkx} (left) are raised by 0.02 for visibility. The curve is derived from Eq.~(\ref{2qxx}). Power-law exponent $n$ should not be confused with TCM  $\hat S_0$ exponent $n$.
	} 
\end{figure}

Figure~\ref{logderiv} (left) shows logarithmic derivatives vs \yt\ (thin solid) for eight \nch\ classes of 13 TeV V0M \pp\ collisions. The bold dashed curve results from applying the same technique to TCM hard-component model $\hat H_0(p_t)$. Horizontal dotted lines represent values $n \approx q + 1.8$ expected for 5 (upper) and 13 (lower) TeV \pp\ collisions from TCM $q$ energy trends related to Ref.~\cite{alicetomspec} (see Table~\ref{engparamy}). The crossed solid lines for $y_t = 2.65$ and $n = 1.8$ remind that while the mode of the hard component on \yt\ is near 2.65 (where the upper line of Eq.~(\ref{logg}) would pass through zero) the mode on \pt\ is near $p_t = 0.5$ GeV/c ($y_t \approx 2$ where the lower line of Eq.~(\ref{logg}) would pass through zero). See Fig.~\ref{enrat3x} (right) and associated text. 

In this data format the spike artifacts common to all spectrum classes are most evident since differential measures are sensitive to short-wavelength structure.  As noted, it may be that those artifacts arise from efficiency corrections derived from Monte Carlo data with more-limited statistics than the spectrum data themselves. 

Figure~\ref{logderiv} (right) shows exponents $n$ vs \nch\ for three spectrum data types. The $n$ values are in each case averages over the ten highest-\yt\ data points ($y_t > 4.1$ or $p_t > 4.2$ GeV/c) in the left panel. That limit is lower than the fit interval $p_t > 6$ GeV/c applied in Ref.~\cite{alicenewspec}. However, the V0M results in the left panel indicate that the exponent trend is approximately constant down to $y_t = 4.1$, and the added points provide more-stable $n$ values. The values for highest and lowest multiplicity classes are not plotted because of excessive noise in the log-derivative results. The V0M data trends on \nch\ are approximately constant, with values close to 5.60 (13 TeV) and 5.80 (5 TeV). The dotted lines correspond to TCM $\hat H_0(y_t)$ $q =$  3.8 for 13 TeV and 4.0 for 5 TeV related to Ref.~\cite{alicetomspec} (see Table \ref{engparamy}).
The SPD data trend varies strongly, decreasing (i.e.\ to harder spectra) with increasing \nch. Those trends are consistent with spectrum data trends in Figs.~\ref{ptspec1} and \ref{ptspec2} (b,d) showing deviations from fixed TCM references. The solid curve for 13 TeV SPD data is derived from Eq.~(\ref{2qxx}) with $n = q + 1.8$. The 5 TeV SPD data are more scattered than other data trends and are thus omitted to improve visual access to the 13 TeV SPD trend relevant to Fig.~5 of Ref.~\cite{alicenewspec}.

In its summary Ref.~\cite{alicenewspec} concludes ``...within uncertainties, the functional form of $n$ as a function of $\langle dN_{ch}/d\eta \rangle$ is the same for the two multiplicity estimators [V0M and SPD] used in this analysis. Moreover, $n$ is found to decrease with $\langle dN_{ch}/d\eta \rangle$.'' The results in Fig.~\ref{logderiv}  contradict that conclusion: The power-law trend is dramatically different for V0M and SPD, and $n$ does not decrease with charge density for V0M event selection. The contrast between V0M and SPD event selection is most evident in Fig.~\ref{v0mspd}. In the left panel the effective exponent at high \pt\ relative to the fixed TCM value is itself fixed (the ratio is approximately constant above \yt\ = 4). The right panel shows dramatic variation of the effective SPD exponent from high (soft) to low (hard) with increasing \nch.

\subsection{Spectrum response to selection bias} \label{biasresp}

Figures 4 and 6 of Ref.~\cite{alicenewspec} deal with other manifestations of selection bias in \pt\ spectrum structure. Figure~6 compares yields integrated within three specific \pt\ intervals $\Delta p_t$ for ten multiplicity classes in ratio to yields in the same intervals from the INEL $> 0$ minimum-bias class. The resulting data trends are compared to a linear $y = x$ trend corresponding to {\em no change in spectrum shape} with \nch. Figure~4 compares V0M and SPD spectra for nearly-equal charge densities $\bar \rho_0 \approx 20$. Ratio SPD/V0M drops to 0.85 near 4 GeV/c for both 5 and 13 TeV spectra. What follows is an effort to understand systematics details and provide a physical interpretation.

\begin{figure}[h]
	\includegraphics[width=1.65in,height=1.6in]{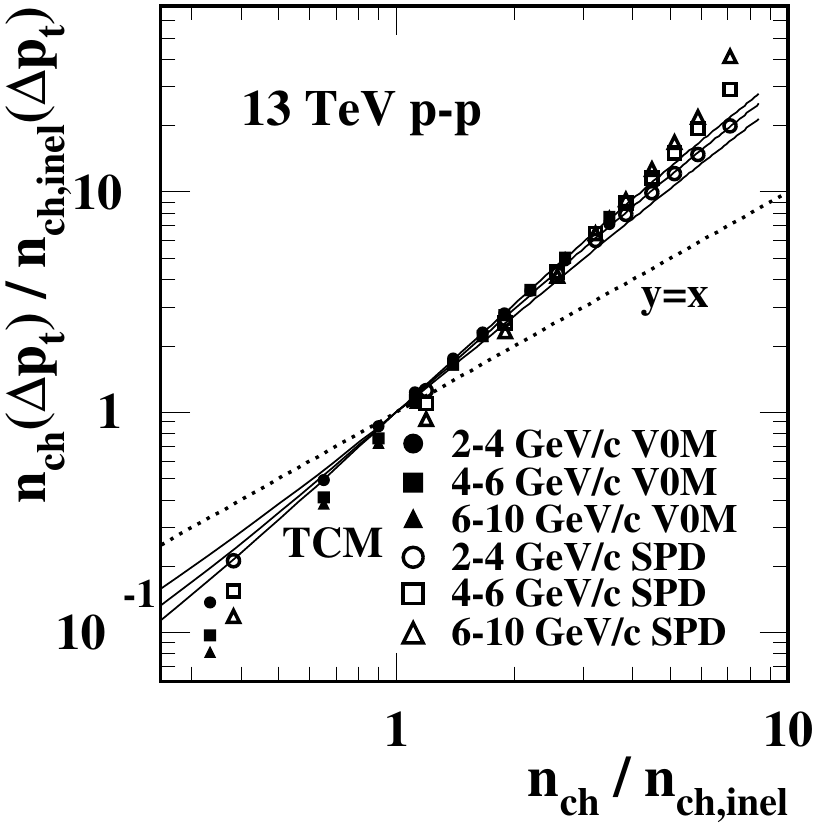}
	\includegraphics[width=1.65in,height=1.63in]{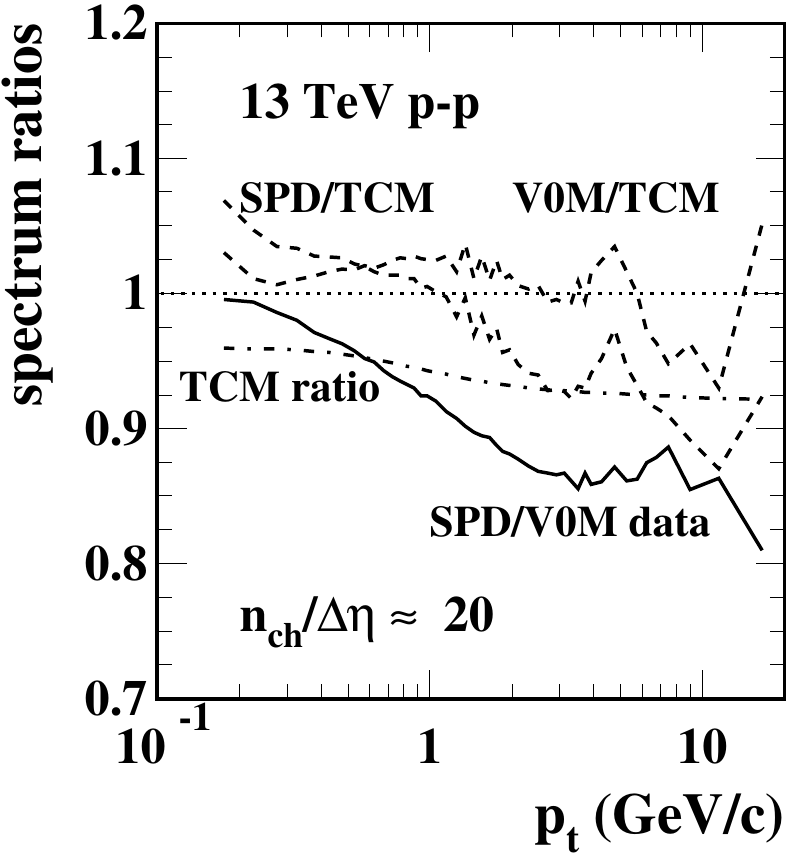}
	\caption{\label{6ax}
		Left: Ratios of yields (points) from \pt\ spectra within three $\Delta p_t$ intervals for ten \pp\ V0M or SPD event classes to yields from the same intervals and from INEL $> 0$ events. The independent variable is mean event multiplicity \nch\ in ratio to INEL $> 0$ mean multiplicity $n_{ch,\text{INEL}}$. The solid reference curves result from the same method applied to TCM spectra defined on a \yt\ continuum.
		Right: Ratios of data spectra to corresponding TCM references for class II of V0M data and class VII of SPD data (dashed), which both approximately correspond to $\bar \rho_0 = 20$. The solid curve is SPD/V0M data spectra compared directly in ratio, reportedly demonstrating that SPD is softer than V0M for the same charge multiplicity. However, the corresponding TCM ratio (dash-dotted) demonstrates that most of the deviation from unity is simply due to the V0M vs SPD multiplicity difference. The two panels relate to Figs.\ 6 and 4 respectively of Ref.~\cite{alicenewspec}.
	}  
\end{figure}

Figure~\ref{6ax} (left) shows integrated yields $n_{ch}(\Delta p_t)$ for three \pt\ intervals from spectra for ten \nch\ classes of SPD and V0M events in ratio to yields $n_{ch,\text{INEL}}(\Delta p_t)$ in the same intervals  (points) from a 13 TeV TCM \pt\ spectrum defined on data \pt\ values with $\bar \rho_0 = 7.60$ as for INEL $> 0$ events in Ref.~\cite{alicenewspec}. The solid lines are TCM references resulting from the same method applied to TCM spectra defined ``on a continuum'' (i.e.\ on 100 equal-spaced \yt\ points extending down to \yt\ = 0). Soft and hard TCM model functions do not vary with \nch. The different slopes of the TCM lines result only from the different fractions of soft and hard components in each \pt\ interval.  The log-log plot format ensures that suppression at smaller \nch\ is as visible as enhancement at larger \nch. The effects of selection bias are indicated by deviations of data from the TCM trends, {\em not} from the $y = x$ dotted line (that assumes no change in jet production). As for Fig.~\ref{specrat} above or Figs.~2 and 3 in Ref.~\cite{alicenewspec} distinction should be maintained between variations due to generic jet trends and bias effects relative to those variations.

Several features are apparent. For V0M events points for different $\Delta p_t$ at higher \nch\ approximately coincide, consistent with Fig.~\ref{v0mspd} (left) where the ratios for higher \nch\ are nearly flat on \pt. In contrast, higher-\nch\ points for SPD vary strongly with \pt\ interval $\Delta p_t$, consistent with Fig.~\ref{v0mspd} (right). For  lowest \nch\ classes there is strong suppression below the TCM references for both V0M and SPD, also consistent with spectrum ratios in Fig.~\ref{v0mspd}, but suppression for V0M is significantly greater than for SPD.

Figure~\ref{6ax} (right) shows 13 TeV data/TCM spectrum ratios (dashed) for class II V0M/TCM and class VII SPD/TCM, data/data ratio SPD/V0M (solid) that appears in Fig.~4 of Ref.~\cite{alicenewspec} and the corresponding ratio of TCM spectra (dash-dotted).  Commenting on the SPD/V0M spectrum ratio in its Fig.~4 (identical to the solid curve here) Ref.~\cite{alicenewspec} states that ``For transverse momenta within 0.5-3 GeV/c the spectra [sic] for the [V0M] multiplicity class II is harder than that for the [SPD] multiplicity class VII$'$.'' However, the 5\% difference in $\bar \rho_0$ for V0M (20.5) and SPD (19.5) plays a significant role as indicated by the TCM ratio (dash-dotted). Since TCM model functions $\hat S_0$ and $\hat H_0$ are fixed, charge density $\bar \rho_0$ determines the TCM spectrum shape. The TCM ratio demonstrates the effect of the V0M vs SPD multiplicity difference: at least half of the SPD/V0M ratio deviation from unity arises from the difference in $\bar \rho_0$. Absent detailed comparisons with a model the term ``harder'' may mean a modified jet fragment distribution or simply more or less jet production according to $\bar \rho_h \approx \alpha \bar \rho_s^2$.

One should also note that class VII is comparatively low \nch\ for SPD whereas class II is relatively high \nch\ for V0M. Given the trends in Fig.~\ref{comparex} (b,d) one should then expect a substantial difference in {\em bias} for the two cases, as observed. Concerning the short-wavelength structure, the peaks near 5 GeV/c in the data/TCM ratios certainly correspond to the bipolar structure near $y_t = 4.4$ in Fig.~\ref{logderiv} (left) that is common to all \nch\ classes and therefore most probably results from the inefficiency correction. That structure then cancels in the SPD/V0M data ratio even though the spectra are from quite different event classes.

\subsection{Spectrum running integrals} \label{running}

In Sec. IV of Ref.~\cite{ppprd} running integrals of \yt\ spectra were the basis for discovery of the two-component structure of 200 GeV \pp\ \pt\ spectra without {\em a priori} assumptions. It was observed that spectra normalized not with total charge density $\bar \rho_0$ but with a ``soft component'' $\bar \rho_s$ defined as the root of $\bar \rho_0 = \bar \rho_s + \alpha \bar \rho_s^2$ with $\alpha \approx O(0.01)$ coincided below $y_t = 2$ ($p_t \approx 0.5$ GeV/c) within data uncertainties and that the endpoints of spectrum running integrals also followed a trend consistent with the above quadratic equation. The same approach is applied here to 13 TeV \pp\ spectrum data. One purpose is demonstration that the \pt\ spectrum TCM is {\em required} by \pp\ data for any energy, is not imposed {\em a priori}.

\begin{figure}[h]
	\includegraphics[width=1.65in,height=1.6in]{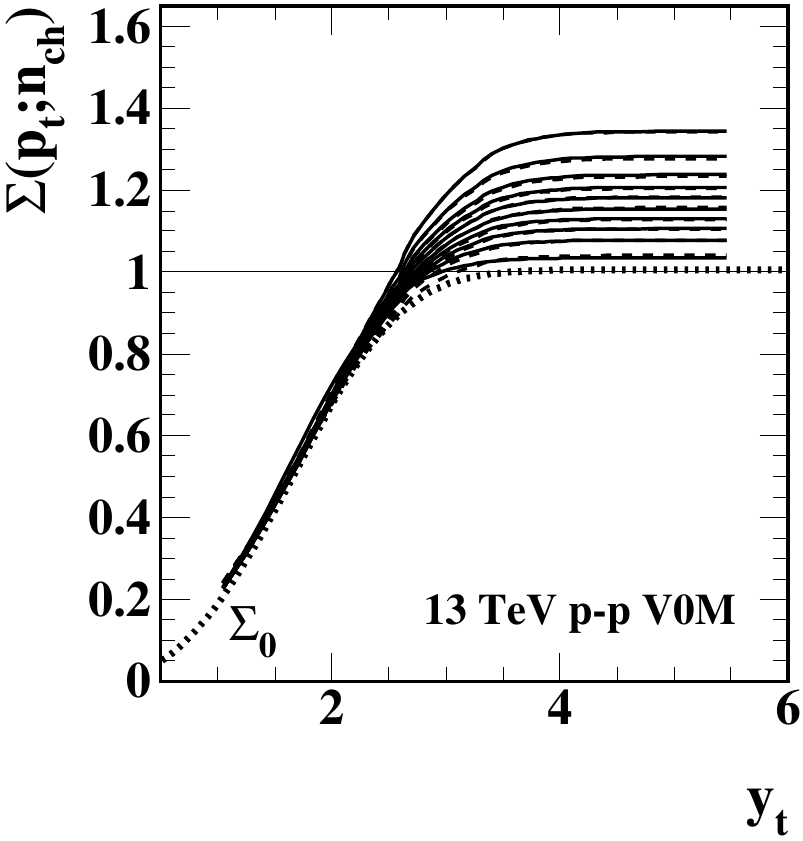}
	\includegraphics[width=1.65in,height=1.6in]{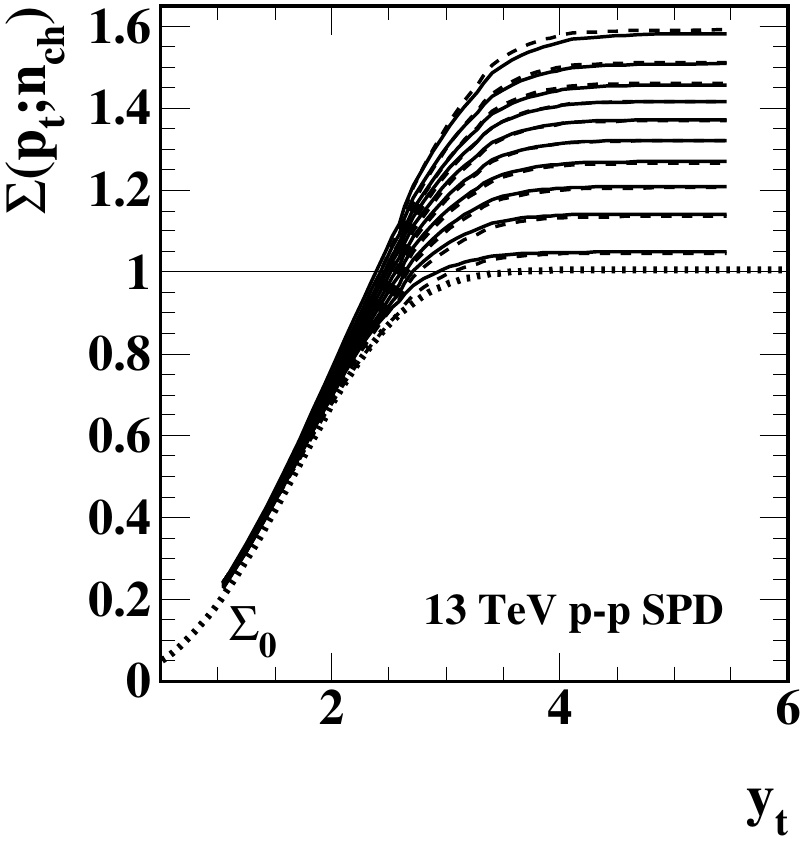}
	\caption{\label{runint}
		Running integrals for data (solid) and TCM (dashed) \pt\ spectra from 13 TeV \pp\ collisions and for V0M (left) and SPD (right) event selection. Running integrals have been corrected for incomplete \pt\ acceptance as described in the text. The bold dotted curve corresponds to model $\hat S_0(y_t)$.
	} 
\end{figure}

Figure~\ref{runint} shows running integrals of \yt\ spectra for ten \nch\ classes of 13 TeV \pp\ collisions each for V0M (left) and SPD (right) event selection criteria derived from data (solid) and TCM (dashed) spectra. Spectra are normalized by $\bar \rho_s$ inferred from $\bar \rho_0$ reported in Ref.~\cite{alicenewspec} using the quadratic equation defined above with $\alpha = 0.017$ for 13 TeV, 12\% higher than reported in Ref.~\cite{alicetomspec}. The uncorrected spectrum running integrals are then defined as 
\bea
\Sigma'(p_t;n_{ch},p_{t,cut}) &=& \frac{1}{\bar \rho_s} \int_{p_{t,cut}}^{p_t} dp_t' p_t' \bar \rho_0(p_t';n_{ch}).
\eea
Note that a factor $p_t'$ in the integrand is required in order to be consistent with the definition of $ \bar \rho_0(y_t;n_{ch})$ in Eq.~(\ref{rhotcm}).
Running integrals on data \pt\ values within the ALICE \pt\ acceptance can be simply corrected for incomplete \pt\ acceptance. The correction is addition of estimated $1 - \xi$ corresponding to $p_{t,cut} \approx 0.15$ GeV/c (see Fig.~\ref{2fe}, left and related text).

The corrected data running integrals can be expressed in TCM form on \yt\ as
\bea
\Sigma(y_t) &=& \Sigma_s(y_t) + \Sigma_h(y_t).
\eea
As in Ref.~\cite{ppprd} the corrected data running integrals coincide below $y_t = 2$ ($p_t \approx 0.5$ GeV/c) within data uncertainties then separate and achieve saturation above $y_t = 4.0$ ($p_t \approx 3.8$ GeV/c). Running integral $\Sigma_0(y_t)$ (bold dotted) of TCM soft component $\hat S_0(y_t)$ is {\em defined} as the limit of data running integrals as $n_{ch} \rightarrow 0$ where $\Sigma_h(y_t) \rightarrow 0$. The functional form of $\hat S_0(y_t)$ given by Eqs.~(\ref{s00}) and (\ref{s0m}) is then observed to generate the required limiting form of  $\Sigma_0(y_t)$ within data uncertainties. $\Sigma_0(y_t)$ saturates at 1 by definition. The same form is used for V0M and SPD data.

The data trends in Fig.~\ref{runint} demonstrate the following: (a) The shape of data soft component $\Sigma_s(y_t)$ does not vary significantly with \nch, is consistent with $\Sigma_0(p_t)$. (b) The complementary data hard components $\Sigma_h(y_t;n_{ch})$ are consistent with an erf$(y_t)$ function as running integral, demonstrating that data {\em spectrum} hard components are similarly-shaped peaked distributions with mode near \yt\ = 2.7 ($p_t \approx 1$ GeV/c) as demonstrated in Ref.~\cite{ppprd}. It is then of interest to examine the \nch\ trend of the data running-integral endpoints for TCM consistency.

\begin{figure}[h]
	\includegraphics[width=1.65in,height=1.6in]{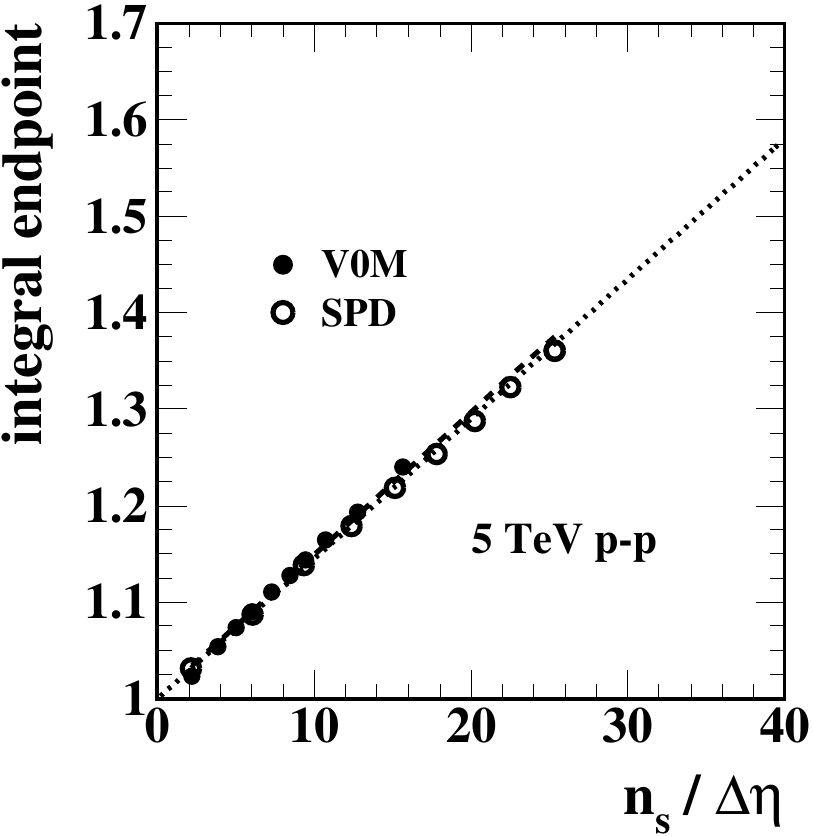}
	\includegraphics[width=1.65in,height=1.6in]{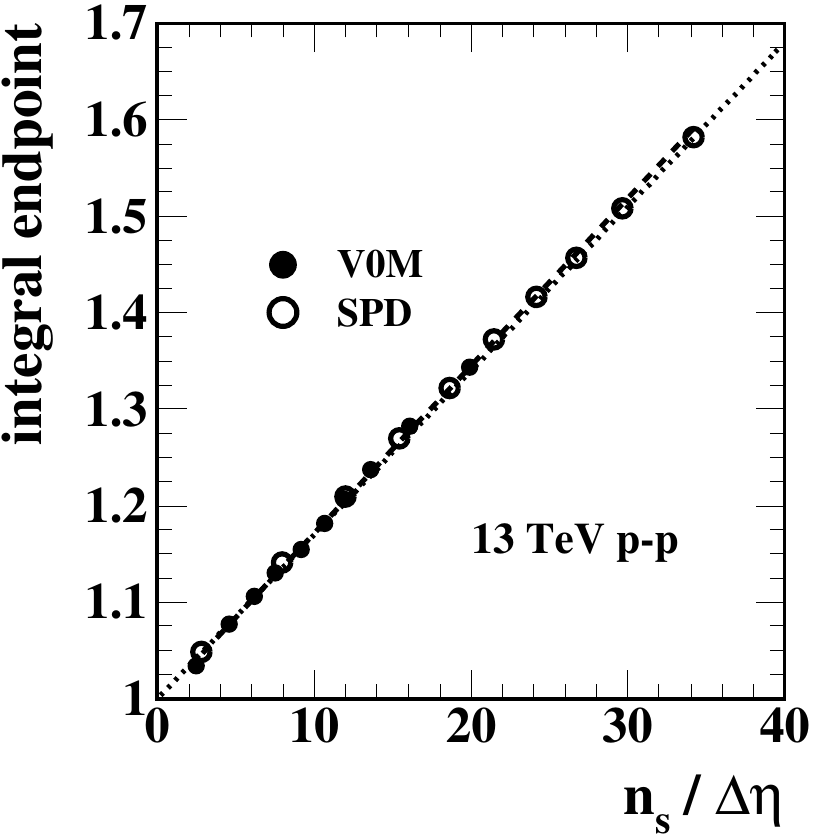}
	\caption{\label{endpoint}
		Endpoints of data running integrals in Fig.~\ref{runint} for V0M (solid dots) and SPD (open circles) and for 5 TeV (left) and 13 TeV (right) \pp\ collisions. The dashed lines are TCM endpoints on SPD \pt\ values. The dotted lines are TCM trends $1 + x(n_s) \rightarrow 1 + \alpha \bar \rho_s$ with $\alpha = 0.017$ (0.0145) for 13 (5) TeV.
	} 
\end{figure}

Figure~\ref{endpoint} shows running-integral endpoints vs $\bar \rho_s$ for 5 TeV (left) and 13 TeV (right) derived from data spectra for V0M (solid dots) and SPD (open circles) event selection. The result for uncorrected spectra is $\xi + x(n_s)$. The inefficiency correction is then $1 - \xi \approx 0.16$ for the TCM defined on SPD \pt\ values (dashed lines) as noted. 


Those results demonstrate {\em experimentally} the precise quadratic relation between hard and soft data components: (a) All spectra coincide precisely for $p_t < 0.5$ GeV/c given normalization via $\bar \rho_s$ as computed from measured $\bar \rho_0$ and (b) corrected-running-integral endpoints fall along straight-line loci $1 + \alpha \bar \rho_s$ (dotted lines) with $\alpha \bar \rho_s \approx x(n_s)$. The endpoint trends in turn accurately indicate the fractional contribution $x/ (1+x)$ of jet fragments to total spectra. For the highest multiplicity classes $x \approx 0.5$ and 33\% of hadrons are jet fragments.

In terms of TCM model functions, spectrum data demonstrate that one spectrum component scales linearly with $\bar \rho_s$ at lower \pt\ and that a second component scales {\em quadratically} with $\bar \rho_s$ at higher \pt. The \pt\ structures of the data components vary little with \nch.  Running integrals provide a less-sensitive way to probe spectrum {\em details} compared to Figs.~\ref{tcm5} and \ref{tcm13}. However, there are no assumptions about spectrum structure, per Ref.~\cite{ppprd}. It is notable that the straight-line trends in Fig.~\ref{endpoint} are followed down to the lowest event multiplicities although the data hard components are substantially biased there. The present study demonstrates that the \pt\ spectrum TCM, with its quadratic relation between soft and hard components, is {\em necessary} to describe 13 TeV \pt\ spectra (modulo biases described and interpreted in Sec.~\ref{biases}). 


\section{\titlepp\ ensemble \titlempt\ systematics} \label{meanpt}

Ensemble-mean \mmpt\ data are inferred from hadron spectra via integration. The \mmpt\ values in Ref.~\cite{alicenewspec} are obtained by integrating over  $p_t \in [0.15,10]$ GeV/c. Accurate values corresponding to ideal spectrum data could be obtained by extrapolating data spectra with a reliable spectrum model. Values obtained with unextrapolated spectra may be strongly biased, and correct interpretation of biased experimental  \mmpt\ values may be difficult. This section demonstrates how to relate a TCM reference to biased \mmpt\ values obtained from spectrum data. The resulting bias is here estimated and corrected. 

Reference~\cite{alicempt} reported a comprehensive analysis of \mmpt\ data vs event multiplicity for \pp, \ppb\ and \pbpb\ collision systems. The strong increase of \mmpt\ with \nch\ for \pp\ collisions was there interpreted in terms of color reconnection as modeled within the PYTHIA Monte Carlo. The data trends were said to ``pose a challenge to most of the existing models.'' A TCM analysis of the same data was presented in Ref.~\cite{alicetommpt}. The observed \mmpt\ vs \nch\ trends were found to be consistent with the jet (hard component) contribution to hadron production in all cases.

Reference~\cite{alicenewspec} introduces spherocity measure $S_0$ intended to select more- or less-``jetty'' events according to its value. Variation of jet production in a given event sample is expected to bias the \mmpt\ vs \nch\ trend. It is noted that as $S_0$ decreases \mmpt\ increases, possibly due to an increased jet contribution to spectra as anticipated. In addition to biases resulting from incomplete \pt\ acceptance and from V0M and SPD event selection methods, biases from event selection via spherocity $S_0$ are considered. 

\subsection{Ensemble \titlempt\ TCM for $\bf p$-$\bf p$ collisions}  \label{ppmptapp}

The TCM for ensemble-mean \mmpt\ data from \pp\ collisions is summarized. It is assumed that due to incomplete \pt\ acceptance (lower bound $p_{t,cut} \approx 0.15$ GeV/c) only a fraction $\xi < 1$ of the \pt\ spectrum soft component is accepted. It is also evident that for such a low cutoff value the entire spectrum hard component is accepted. The \mmpt\ values obtained from TCM model functions as defined by Table~\ref{engparamy} are as follows: For models defined on the continuum ($\xi \approx 1$) the mean values are $\bar p_{ts} \approx 0.48$ GeV/c and $\bar p_{th} \approx 1.39$ GeV/c. For models defined on spectrum data \pt\ values the mean values are $\bar p_{ts} \approx 0.58$ GeV/c (corresponding to $\xi \approx 0.84$) and $\bar p_{th} \approx 1.38$ GeV/c.

The TCM for charge densities averaged over some angular acceptance $\Delta \eta$ (1.6 for $|\eta| < 0.8$ as in Ref.~\cite{alicenewspec}) is 
\bea \label{nchns}
\bar \rho_0 &=& \bar \rho_{s} + \bar \rho_{h}
\\ \nonumber
&=& \bar \rho_{s}[1+ x(n_s)],
\\ \nonumber
\frac{\bar \rho_0'}{ \bar \rho_s}&=& \frac{n_{ch}'}{n_s} ~=~ \xi+ x(n_s),
\eea
where $x(n_s) \equiv \bar \rho_{h} / \bar \rho_{s} \approx  \alpha \bar \rho_s$ is the \pp\ ratio of hard-component to soft-component yields~\cite{ppprd} and $\alpha(\sqrt{s})$ is defined in Ref.~\cite{alicetomspec}. $\bar \rho_0'$ is an {\em uncorrected} charge density corresponding to incomplete \pt\ acceptance with $\xi < 1$. 

The TCM for extensive ensemble-mean {\em total} $p_t$ integrated within some angular acceptance $2\pi$ and $\Delta \eta$ from \pp\ collisions for given $(n_{ch},\sqrt{s})$ can be expressed as
\bea \label{mptsimple}
\bar P_t &=& \bar P_{ts} + \bar P_{th}
\\ \nonumber
&=& n_s \bar p_{ts} + n_h \bar p_{th}.
\eea
The conventional {\em intensive} ratio of extensive quantities
\bea \label{ppmpttcm}
\frac{\bar P_t' }{n_{ch}'} \equiv \bar p_t' &\approx & \frac{\bar p_{ts} + x(n_s) \bar p_{th}(n_s)}{\xi + x(n_s)}
\eea
(assuming $\bar P_{ts}' \approx \bar P_{ts}$~\cite{tommpt})%
\footnote{Because of the additional factor $p_t'$ in $ \bar P_{ts}' = \int_{p_{t.cut}}^\infty dp'_t {p'_t}^2 \bar  \rho_s(p_t')$ the effect of the low-\pt\ cutoff is minimal and $\bar P_{ts}' \approx \bar P_{ts}$.}
 in effect partially cancels dijet manifestations represented by ratio $x(n_s)$ that may be of considerable interest.  The alternative ratio
\bea \label{niceeq}
\frac{n_{ch}'}{n_s} \bar p_t'   \approx \frac{ \bar P_t}{n_s} &= & \bar p_{ts} + x(n_s) \bar p_{th}(n_s)
\\ \nonumber
&\approx& \bar p_{ts} + \alpha(\sqrt{s})\, \bar \rho_s \, \bar p_{th}(n_s,\sqrt{s})
\eea
preserves the simplicity of Eq.~(\ref{mptsimple}) and provides a convenient basis for testing the TCM hypothesis precisely.

\subsection{Spherocity event selection}

Reference~\cite{alicenewspec} studies ensemble \mmpt\ trends vs spherocity 
\bea \label{spher}
S_0(\%) = 100 \times \frac{\pi^2}{4} \min_{\bf \hat n_s}\left( \frac{\sum_i |\vec p_{t,i} \times {\bf \hat n_s}|}{\sum_i \vec p_{t,i}} \right)^2 \in [0,100],
\eea
where $\bf \hat n_s$ is a unit vector varied to minimize $S_0$.
Limit 0\% would result for a single \pt\ vector. Limit 100\% would result for an infinite number of \pt\ vectors uniformly distributed on azimuth with {\em equal magnitudes}, in which case the quantity in parenthesis becomes $(1/\pi) \int_0^\pi d\phi \sin(\phi)$ = $2/\pi$. Events are sorted into ten spherocity classes with the intent to identify events based on the particle fraction arising from jets, i.e. the event ``jettiness.''

\subsection{ALICE ensemble \titlempt\ results} \label{alicempt}

\begin{figure}[h]
	\includegraphics[width=3.3in]{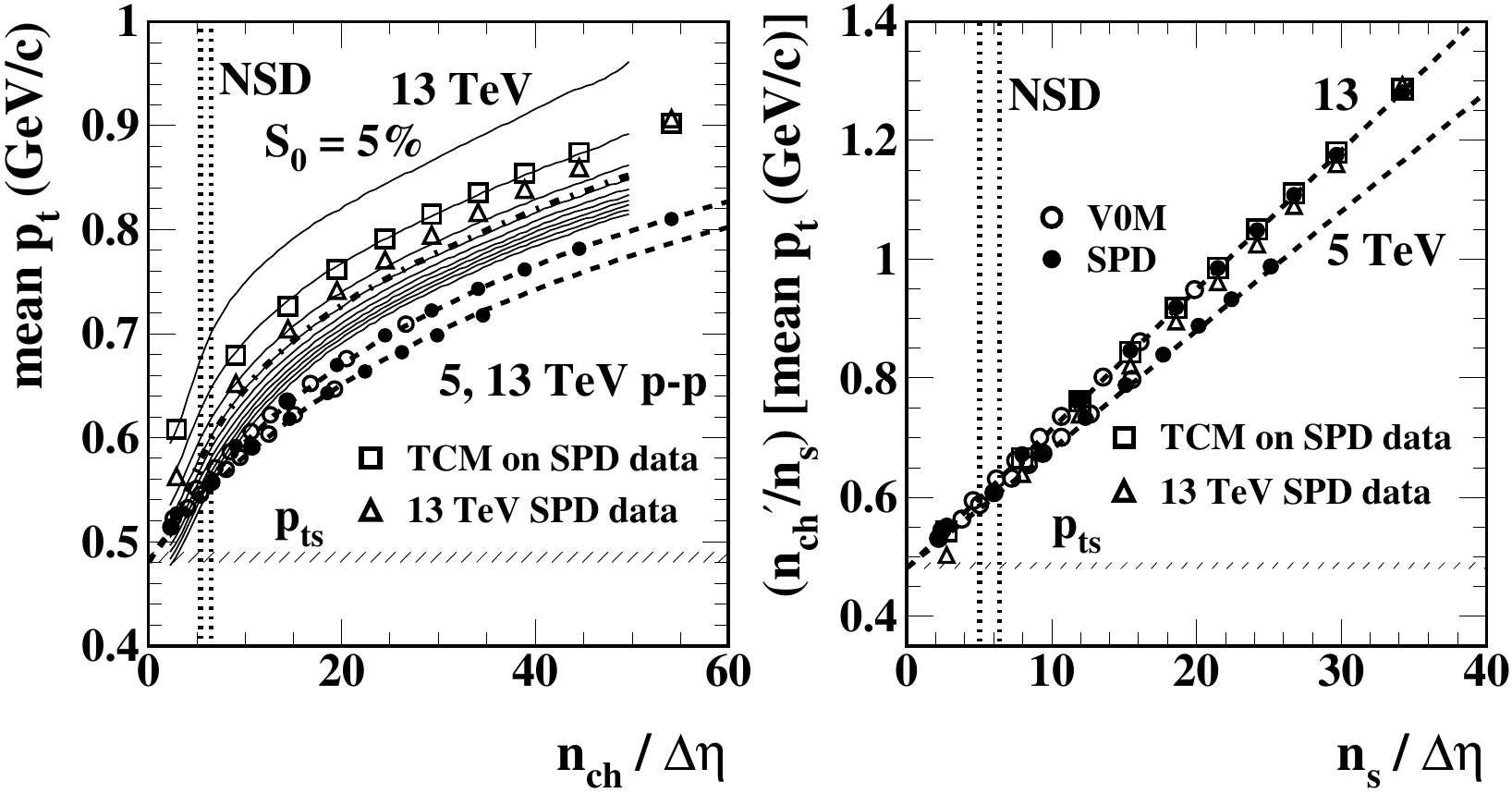}
	\caption{\label{2ax}
		Left: Ensemble-mean \mmpt\ vs \nch\ trends for 13 TeV events (thin solid) vs spherocity $S_0$. The dash-dotted curve is an unweighted mean of those curves. The open triangles are \mmpt\ from 13 TeV SPD data. The open squares are from uncorrected TCM spectra on data \pt\ values. The open and solid circles are the TCM ``on a continuum'' (see text). The dashed curves are TCM Eq.~(\ref{ppmpttcm}).
		Right: \mmpt\ vs \nch\ trends in the left panel converted per Eq.~(\ref{niceeq}) with $\xi = 1$ for dots and circles and 0.84 for open squares and open triangles (corrected). 
	}  
\end{figure}

\begin{figure}[h]
	\includegraphics[width=1.65in,height=1.6in]{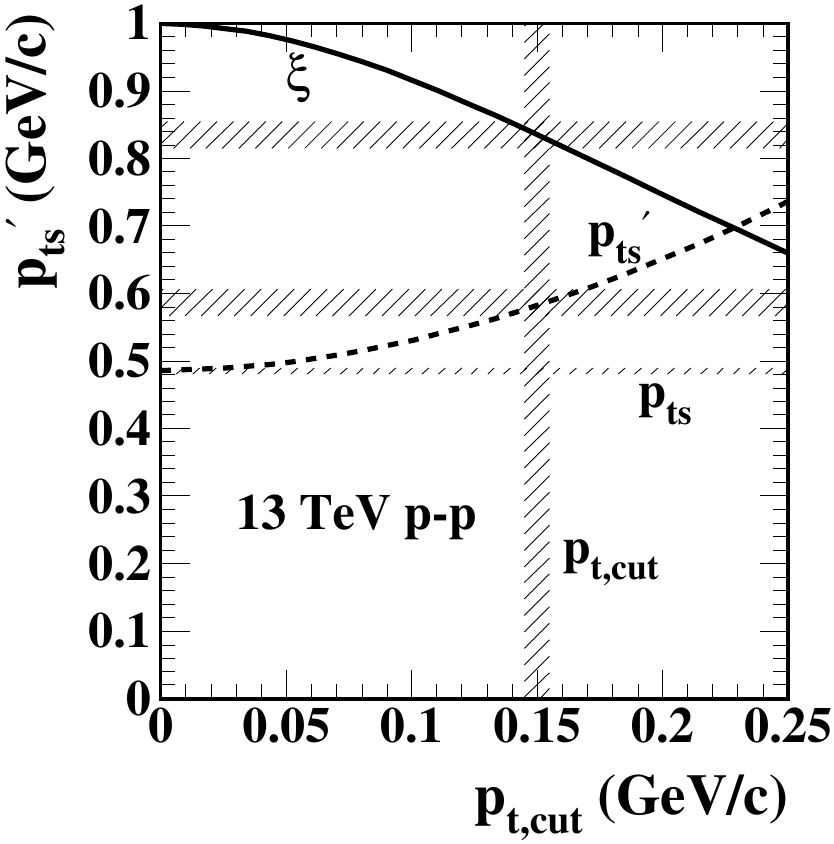}
	\includegraphics[width=1.65in,height=1.6in]{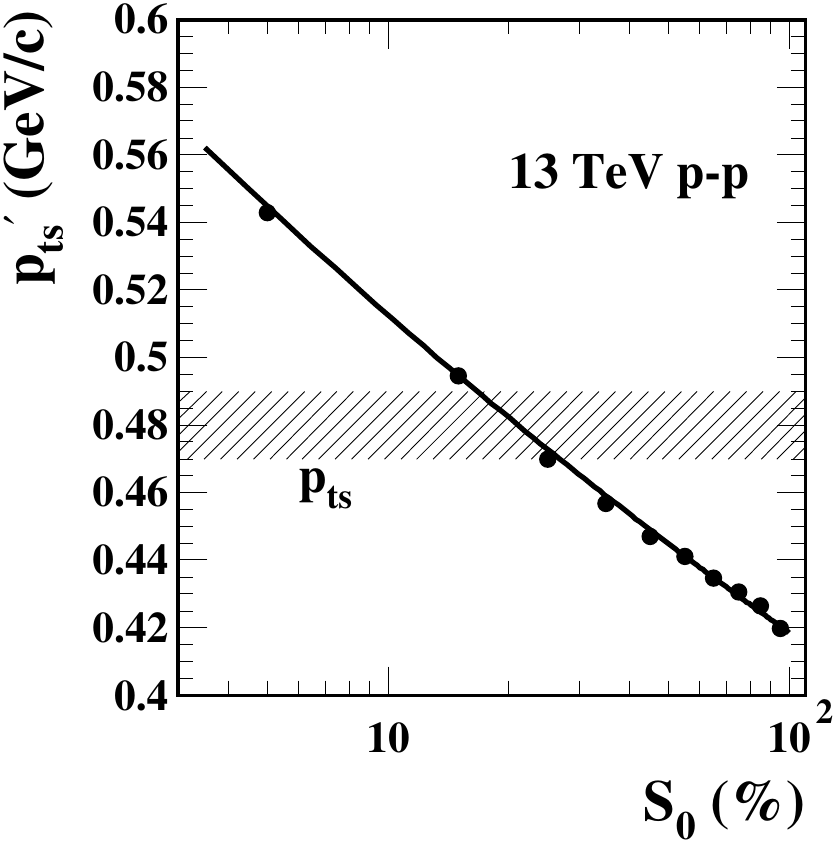}
	\caption{\label{2fe}
		Left: Fraction $\xi$ (solid) of TCM soft component $\hat S_0(y_t)$ integrated above cutoff $p_{t,cut}$ and biased $\bar p_{ts}' = \bar p_{ts}/\xi$ (dashed). For nominal acceptance cut $p_{t,cut} \approx 0.15$ GeV/c $\xi \approx 0.84$ and $\bar p_{ts}' \approx 0.58$ GeV/c (hatched bands).
		Right: Biased $\bar p_{ts}'$ vs spherocity $S_0$ (in percent) for corrected \mmpt\ vs \nch\ trends compared to $\bar p_{ts} \approx 0.48$ GeV/c (hatched band). Note that symbol $\bar p_{ts}'$ represents different bias sources in the two panels -- low-\pt\ acceptance cutoff vs spherocity $S_0$.
	}  
\end{figure}

Figure~\ref{2ax} (left) shows uncorrected (biased) ensemble $\bar p_t'$ values vs \nch\ (thin solid) from 13 TeV \pp\ collisions for  ten classes of spherocity $S_0$. According to Ref.~\cite{alicenewspec} uncorrected \mmpt\ is determined within acceptance $p_t \in [0.15,10]$ GeV/c but \nch\ is corrected by extrapolation to zero. The bold dash-dotted curve is a simple unweighted average of those spectra. As a result of the incomplete \pt\ acceptance the \mmpt\ values are biased upward as for $\bar p_t'$ in Eq.~(\ref{ppmpttcm}) with $\xi \approx 0.84$. The method of event selection on \nch \ is 
based on \nch\ appearing in angular acceptance $|\eta| < 0.8$ (i.e.\ SPD). The solid dots and open circles are \mmpt\ values obtained from TCM spectra defined on a ``continuum'' (100 points equally spaced on $y_t \in [0,6]$) for V0M and SPD \nch\ which are then unbiased ($\xi \approx 1$). The dashed curves are TCM trends for 5 and 13 TeV described by Eq.~(\ref{ppmpttcm}) with $\xi = 1$ and hard component $\bar p_{th}(n_s) = 1.39$ GeV/c  fixed. The open squares are the TCM defined on SPD data \pt\ values and integrated without extrapolation (i.e.\ are biased). The open triangles are from 13 TeV SPD \pt\ spectra reported in Ref.~\cite{alicenewspec} also integrated without extrapolation (biased). The \mmpt\ bias for uncorrected data is about 0.1 GeV/c (open squares vs upper dashed curve), consistent with Fig.~\ref{2fe} (left).

\begin{figure}[h]
	\includegraphics[width=3.3in]{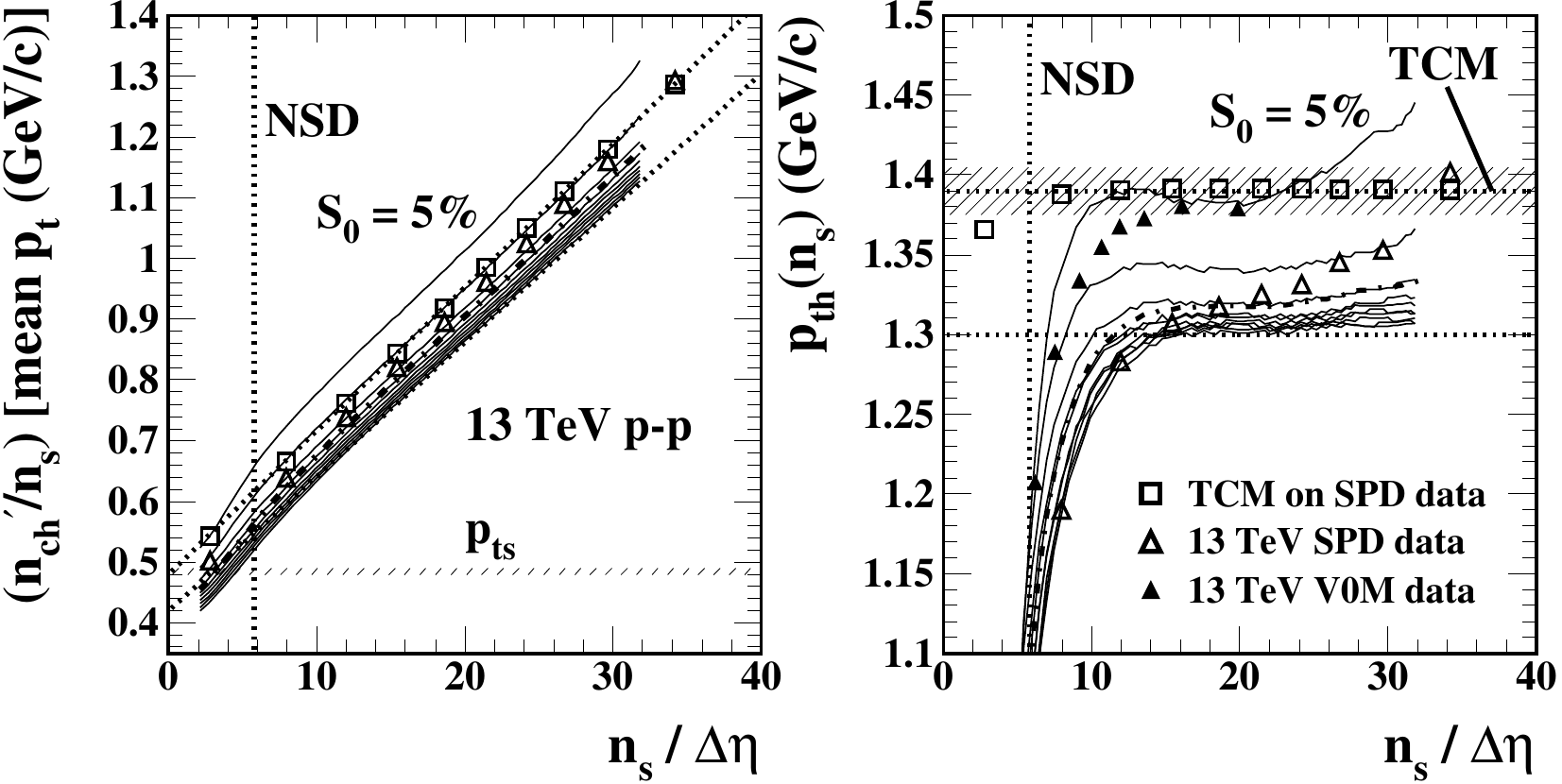}
	\caption{\label{2c}
		Left: 13 TeV \mmpt\ vs \nch\ data vs spherocity $S_0$ (thin solid, dash-dotted) from Fig.~\ref{2ax} (left) transformed and corrected via Eq.~(\ref{niceeq}) with $\xi \approx 0.84$. The open squares are the corrected 13 TeV TCM on data \pt\ values from Fig.~\ref{2ax} (right). $\bar p_{ts}$ is derived from the TCM $\hat S_0(y_t)$ model function.
		Right: \mmpt\ hard component $\bar p_{th}(n_s)$ derived from the contents of the left panel via Eq.~(\ref{niceeq}). For the $S_0$ data $\bar p_{ts} \rightarrow \bar p_{ts}'(S_0)$ values from Fig.~\ref{2fe} (right) are used in Eq.~(\ref{niceeq}). The dotted lines in left and right panels indicate the TCM trend (upper) and the lowest of the $S_0$ curves above $\bar \rho_s = 15$.
	} 
\end{figure}

Figure~\ref{2ax} (right) shows the same data and curves (with the exception of spherocity-related curves, see Fig.~\ref{2c}) in the form of Eq.~(\ref{niceeq}).  The TCM trends follow straight-line loci whose {\em slopes} are determined by the spectrum jet contribution, controlled in this case by parameter $\alpha$. The TCM defined on data points (open squares) and 13 TeV data SPD \mmpt\ values (open triangles) have been corrected according to Eq.~(\ref{niceeq}) with $\xi \approx 0.84$ in Eq.~(\ref{nchns}) (see below). The dashed lines are Eq.~(\ref{niceeq}) with $\bar p_{ts} = 0.48$ GeV/c and $\bar p_{th} = 1.39$ GeV/c as described in Sec.~\ref{ppmptapp}.

Figure~\ref{2fe} (left) shows the consequence of an incomplete \pt\ acceptance for calculation of \mmpt\ per Ref.~\cite{tommpt}. The solid curve is the running integral of $1 - \hat S_0(p_t)$  that determines the inefficiency parameter $\xi$ as a function of spectrum lower bound $p_{t,cut}$, the fraction of $\hat S_0(p_t)$ that survives the cut. The lowest hatched band indicates \mmpt\ soft component $\bar p_{ts} \approx 0.48$ GeV/c determined from TCM fixed spectrum soft-component $\hat S_0(y_t)$ with no cutoff. $p_{ts}' = \bar p_{ts}/\xi$ (dashed) is the biased \mmpt\ soft component resulting from the cutoff. For cutoff $p_{t,cut} \approx 0.15$ GeV/c  in Ref.~\cite{alicenewspec} $\xi \approx 0.84$ (as in the previous paragraph) and $\bar p_{ts}' \approx 0.58$ GeV/c, i.e.\ about 0.1 GeV/c higher  than the unbiased value. The right panel is explained below.

Figure~\ref{2c} (left) shows  \mmpt\ vs \nch\ data trends in Fig.~\ref{2ax} (left) as they vary with spherocity $S_0$ (thin solid), the unweighted ensemble average (bold dash-dotted) and the TCM on 13 TeV SPD data \pt\ values (open squares) transformed according to Eq.~(\ref{niceeq}) with $\xi \approx 0.84$ that can be compared with Fig.~\ref{2ax} (right). The open triangles are 13 TeV SPD data \mmpt\ values treated the same. In this plotting format it is evident that for most data spectra the primary variation with spherocity $S_0$ is soft-component mean $\bar p_{ts}$ ($y$-axis intercepts in this plot) since the jet contribution (i.e.\ $\bar p_{th}$), determining {\em slopes} in this format, shows little variation with $S_0$. The upper dotted line is $0.48 + 1.39 \alpha \bar \rho_s$ consistent with TCM $\bar p_{ts} = 0.48\pm 0.01$ GeV/c and $\bar p_{th} = 1.39 \pm 0.015$ GeV/c. The lower dotted line with $\bar p_{ts} = 0.42$ GeV/c and $\bar p_{th} = 1.30$ is consistent with  the lowest $S_0$ curve for $\bar \rho_s > 15$.

Figure~\ref{2fe} (right) shows empirically-determined biased soft-component values $\bar p_{ts}'$ for the individual $\bar p_t(n_{ch})$ trends in Fig.~\ref{2c} (left). The estimated values are simply described by the power-law expression $\bar p_{ts}'(S_0) \approx [0.048/S_0(\%)]^{0.088}$ GeV/c (straight line) over the entire $S_0$ range. The hatched band denotes the unbiased soft-component mean $\bar p_{ts} \approx 0.48$ GeV/c. The bias in this instance is not due to a \pt\ cutoff (which has been corrected), is instead the result of the imposed $S_0$ condition. The two points above the hatched band thus do not ``cause'' the two elevated trends in Fig.~\ref{2c} (right).

Figure~\ref{2c} (right) shows $\bar p_{th}(n_s)$ inferred from data in the left panel via Eq.~(\ref{niceeq}) (second line) with $\bar p_{ts}'(S_0)$ taken from Fig.~\ref{2fe} (right) for each spherocity class. The value 1.39 GeV/c (hatched band) is just the $\bar p_{th}$ corresponding to the 13 TeV TCM hard-component $\hat H_0(y_t)$ in Sec.~\ref{tcmmodel}. The adopted criterion for determining $\bar p_{ts}$ values in Fig.~\ref{2fe} (right) is a requirement that $\bar p_{th}(n_s)$ trends be as level as possible above $\bar \rho_s = 15$, consistent with a constant $\bar p_{th}(n_s)$ trend in Eq.~(\ref{niceeq}) reflecting the TCM. The horizontal dotted lines are the dotted lines at left transformed to obtain $\bar p_{th}(n_s)$ according to Eq.~(\ref{niceeq}).  

Deviations from the TCM reference in Fig.~\ref{2c} (right) could in general carry significant new information about collision dynamics. However, the data mainly indicate that while jet production increases about {\em 30-fold} over the interval $\bar \rho_s \in [6,33]$ it continues to follow the observed $\bar \rho_h \propto \bar \rho_s^2$ trend precisely no matter what the value of $S_0$. 

The comparison between V0M and SPD $\bar p_{th}$ values indicates that the SPD hard component mean \pt\ is biased down by almost 0.1 GeV/c {\em independent of} $S_0$. The effect is evident in Fig.~\ref{v0mspd} in that the data trend on \yt\ for higher \nch\ and for V0M is flat (i.e.\ agreeing with the TCM) whereas for SPD the hard component is shifted down on \yt\ (hence the tilt). Thus, for large SPD \nch\ the high-\pt\ tail is hardened while the main part of the hard component is softened. Trends for higher $S_0$ (thin solid) {\em agree} with the basic SPD data. Only the lowest two $S_0$ values show significant hardening of the hard component.

Information carried by \pt\ spectra sorted by $S_0$ thus has three contributions: (a) bias of $\bar p_{ts}$ as in Fig.~\ref{2fe} (right) indicates bias of the spectrum {\em soft} component in response to imposition of an azimuthal (a)symmetry condition via $S_0$, (b) bias of the spectrum hard component in the form of strong modification of the hard-component shape with low \nch\ condition as in Fig.~\ref{comparex} resulting in lower \mmpt\ {\em independent} of $S_0$, and differently for V0M and SPD, and (c) minor modification of the jet contribution to \mmpt\ varying with $S_0$ as in Fig.~\ref{2c} (right).
Note that for the SPD data used in the $S_0$ study the {\em absolute} jet production increases (relative to NSD \pp\ collisions) by $\approx $ 30-fold because of the \nch\ variation (see Fig.~7 of Ref.~\cite{alicenewspec}) while the largest effect of $S_0$ variation on the jet-related \mmpt\ hard component is less than 10\% (see Fig.~\ref{2c}, right).

\subsection{ALICE ensemble \titlempt\ summary}

Results in this section suggest several conclusions: 
(a) The basic $S_0$-related \mmpt\ data  in Fig.~\ref{2ax} (left) (thin solid) are strongly biased by the incomplete \pt\ acceptance (i.e.\ lower bound $p_{t,cut}$). The TCM results in the same panel illustrate the consequence of determining \mmpt\ by integration down to zero \pt. In contrast, integrating TCM spectra on data \pt\ values (open squares, with cutoff) is close to (but not equal to) integrating spectra from Ref.~\cite{alicenewspec} without extrapolation (open triangles).

(b) The $p_{t,cut}$ bias can be removed via Eq.~(\ref{niceeq}) as shown in Fig.~\ref{2ax} (right). The biased TCM data in the left panel (open squares) are then in good agreement with unbiased TCM data in the right panel. The bias correction relies on estimating efficiency $\xi \approx 0.84$ based on a TCM soft-component model that provides good data descriptions down to 0.15 GeV/c as demonstrated in Sec.~\ref{tcmspecdat}. 

(c) The conventional $\bar p_t$ ratio format of Eq.~(\ref{ppmpttcm}) and Fig.~\ref{2ax} (left) mixes two distinct physical mechanisms and is therefore difficult to interpret. Aside from bias corrections the data format corresponding to Eq.~(\ref{niceeq}) and Fig.~\ref{2ax} (right) provides a clear distinction between \mmpt\ data soft and hard components, especially some differential features of the isolated jet contribution $\bar p_{th}(n_s)$.

(d) Even with complete \pt\ acceptance the data spectra from Ref.~\cite{alicenewspec} are still strongly biased by the event selection methods, both SPD {\em and} V0M, as illustrated by comparison with fixed (not fitted) TCM spectra in Sec.~\ref{alicedata} (ratios in right panels). With demand for lower \nch, spectra for any $S_0$ are increasingly softened leading to decreased \mmpt\ relative to the TMC trend in Fig.~\ref{2ax} (left) (compare open squares and open triangles) and Fig.~\ref{2c} (left, note the strong drop-off near and below the NSD $\bar \rho_s$ value). The effect is most apparent in Fig.~\ref{2c} (right) because $\bar p_{th}(n_s)$ corresponds to slopes in the left panel.

(e) Event selection via spherocity $S_0$ was intended primarily to bias jet production~\cite{alicenewspec}. Data indicate that $S_0$ instead biases the spectrum soft component as  evident from the nearly-linear trends in Fig.~\ref{2c} (left) where the slopes (jet contribution) vary little while the $y$-axis intercepts (soft component) vary strongly according to a simple power-law trend shown in Fig.~\ref{2fe} (right). The origin of the bias is simple to identify. Large values of $S_0$ favor uniform azimuth distributions of {\em nearly-equal \pt\ values}. That condition then biases against higher-\pt\ contributions to the soft component which reduces $\bar p_{ts}$. Strong $S_0$ bias of $\bar p_{ts}$ would occur even in the absence of jets. 
The inferred soft-component power-law $\bar p_{ts}'(S_0)$ trend in Fig.~\ref{2fe} (right) is consistent for all values of $S_0$. Figure~\ref{2c} (right) shows that if $S_0$ bias of  $\bar p_{ts}$ is corrected there is significant bias of $\bar p_{th}(n_s,S_0)$ beyond what is evident in Figs.~\ref{ptspec1} and \ref{ptspec2} (b,d), but the bias magnitude is small compared to variation of jet production with \nch.

(f) Just as for \pt\ spectra, the TCM provides a basic reference for understanding ensemble-mean \mmpt\ trends. The TCM is not fitted to individual data sets or collision systems. It is, in effect, a global representation of a large data volume including yields, spectra and two-particle correlations from multiple A-B collision systems over a broad range of collision energies. Deviations of properly-corrected particle data from the TCM then reveal {\em differential details of the information carried by those data.}

These conclusions can be compared with those reported in Ref.~\cite{alicenewspec}. In the introduction appears  ``The aim of this study is to investigate the importance of jets in high-multiplicity pp collisions and their contribution to charged-particle production at low $p_T$.'' The contribution of jets to hadron spectra at lower \pt\ in \pp\ collisions has been established in a series of papers over fifteen years~\cite{ppprd,eeprd,fragevo,hardspec,anomalous,jetspec,jetspec2}, none of which is cited in Ref.~\cite{alicenewspec}. The Ref.~\cite{alicenewspec} abstract states ``Within uncertainties, the functional form of $\langle p_t \rangle(n_{ch})$ is not affected by the spherocity selection,'' but that statement conflicts with actual data properties as revealed in Fig.~\ref{2c} (right).   $S_0$ bias relating to the jet contribution to spectra is small compared to that for soft component $\bar p_{ts}$. Imposition of an $S_0$ condition on spectra does result in systematic bias of the jet-related spectrum hard component which is detected only by means of a highly differential analysis technique based on the TCM. As spectra and ensemble \mmpt\ data are presented in Ref.~\cite{alicenewspec} that information is inaccessible.

\section{Systematic uncertainties} \label{syserr}

Estimation of data systematic uncertainties typically addresses the reliability of numerical values resulting from physical measurements. In addition to inevitable statistical fluctuations the reliability of instrument calibrations and resulting data corrections is estimated. That approach is adequate if the interpretation of numerical values is unambiguous. In the present case where data volumes of 60M or 105M events are the basis for analysis systematic errors may dwarf statistical errors, making their correct estimation all that more important.

In this analysis the two-component model, a formal procedure with elements inferred from lower-energy \pp\ spectrum data, is used to separate two disjoint contributions to measured 13 TeV \pp\ \pt\ spectra. For the analysis of systematic uncertainties in this study both the accuracy of the separation and physical interpretation of the results are in question.
This section addresses the following questions: (a) What is the overall accuracy of the TCM for \pp\ \pt\ spectrum data descriptions? (b) What is the significance and physical interpretation of selection-bias trends relative to the TCM? (c) What is the accuracy and interpretation of spectrum parametrizations related to jets? (d) What is the effect of spherocity as a basis for jet preference and the physical interpretation of results?

\subsection{Overall accuracy of the TCM per Sec.~\ref{pptcm}}

Based on Figs.~\ref{v0mspd} and \ref{comparex} it might be argued that the TCM shows highly significant deviations from \pp\ spectrum data and is therefore a poor model. But the TCM is a {\em predictive reference} based on a broad survey of yield, spectrum and correlation data~\cite{ppprd,ppquad,alicetomspec,anomalous,ppbpid,jetspec2,fragevo} suggesting that the concept of systematic uncertainty should be reconsidered. How should the TCM be required to describe data and how well does it do that? Do significant data-model deviations reveal a faulty model or unexpected information carried by particle data?

In the present study the TCM is applied as a fixed reference to a minimum-bias ensemble of \pp\ \pt\ spectra sorted into multiplicity classes via two methods. The fixed TCM then provides highly differential information on resulting spectrum bias. If the TCM were fitted to individual spectra the bias trends would then be represented by varying fit parameters whose physical interpretation might be difficult. By examining resulting data deviations from a fixed model in relation to statistical errors as in Fig.~\ref{comparex} a likely physical interpretation may be possible that was not intuitively obvious beforehand.

The fixed TCM thus provides a valuable reference that is highly constrained by the requirement to describe \pp\ \pt\ spectra for all energies from 17 GeV to 13 TeV via simple systematic variation of model parameters and agreement with predictions based on jet data as illustrated in App.~\ref{old13}. The model parameters that appear in Table~\ref{engparamy} are those appearing in Table~\ref{engparamx} with the following exceptions: 
The values for $q$ used in the present study are 4.0 and 3.8 respectively for 5 and 13 TeV whereas the values in Table~\ref{engparamx} are 3.85 and 3.65. The values for $\sigma_{y_t}$ used in the present study are 0.58 and 0.60 respectively for 5 and 13 TeV whereas the values in Table~\ref{engparamx} are 0.58 and 0.615. Those changes arise because the 13 TeV spectrum data employed in Ref.~\cite{alicetomspec} were in effect SPD data. In the present study it was decided to favor V0M data with the TCM. The motivation is evident in Fig.~\ref{logderiv} (right). The values for $\alpha$ estimated in the present study are 0.0145 and 0.0170 whereas  the values in Table~\ref{engparamx} are 0.013 and 0.015 based on the earlier 13 TeV data with its limited \nch\ reach (maximum $\bar \rho_0 \approx 12$ vs 54 for SPD in the later study) and 1M total events vs 60M for the later study.
With those changes the TCM provides a good description of V0M data for the six highest \nch\ classes as demonstrated in Fig.~\ref{comparex}. Deviations for lower \nch\ classes carry important new information about biases as noted.

The TCM may be further parametrized to describe \nch\ trends of spectrum data in detail as illustrated in App.~\ref{old13}. The \nch\ dependence of 13 TeV SPD spectra can then be described within data uncertainties mainly by accommodating hard-component trends. The soft component exhibits no significant \nch\ dependence in its shape. Within the TCM context variation of hard-component parameters with \nch\ may then be physically interpreted in terms of mean jet characteristics altered due to selection bias.

\subsection{Bias trends and ``fit'' quality per Sec.~IV}

For the Ref.~\cite{alicenewspec} analysis three issues are important for data interpretation: (a) spectrum normalization for each \nch\ class, (b) the jet contribution to spectra for each \nch\ class and (c) selection bias for each \nch\ class and each event selection method. Without a well-defined  reference model it is difficult to distinguish among those issues.

\begin{figure}[h]
	\includegraphics[width=1.65in,height=1.63in]{alispec1i}
	\includegraphics[width=1.65in,height=1.6in]{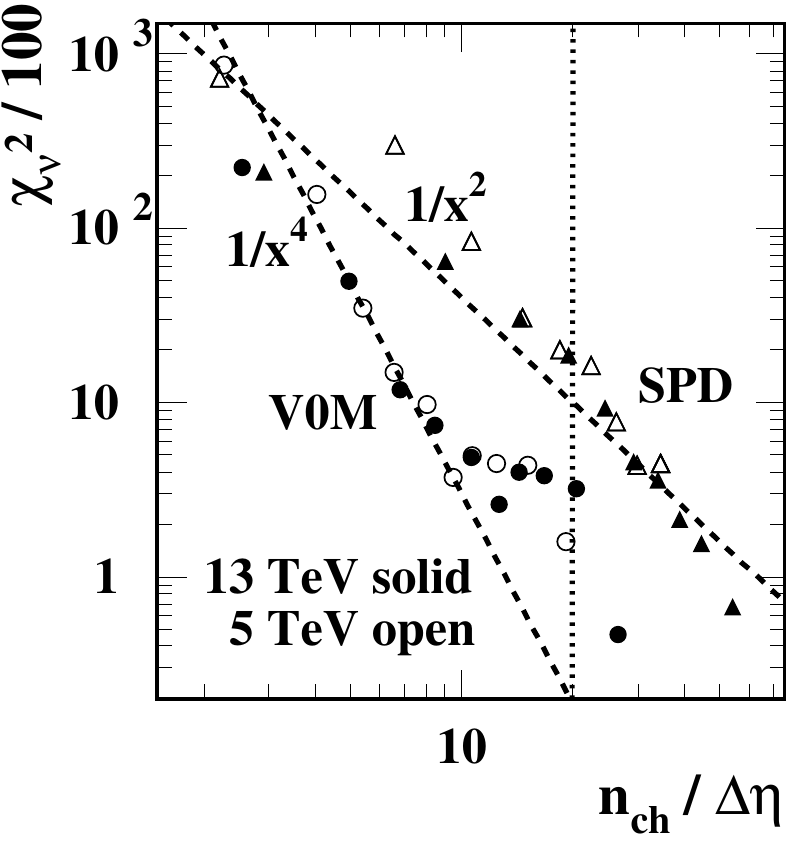}
	\caption{\label{6axx}
		Left: Repeat of Fig.~\ref{6ax} (right) for further discussion.
		Right:  $\chi_\nu^2$ values [second line of Eq.~(\ref{chinu})] for each curve in Fig.~\ref{comparex}. Data are observed to approximate power-law trends.
	} 
\end{figure}

Figure~\ref{6axx} (left) shows Fig.~\ref{6ax} (right) repeated for further consideration. The dashed curves for V0M/TCM and SPD/TCM spectrum ratios are those in Fig.~\ref{ptspec2} (b) and (d) for event classes II (V0M) and VII (SPD), each with $\bar \rho_0 \approx 20$. The data spectrum ratio SPD/V0M (solid) is consistent with the lower panel of Fig.~4 in Ref.~\cite{alicenewspec}. The statistical noise artifact common to V0M and SPD spectra (dashed curves) cancels in ratio as an example of common-mode noise rejection. As noted in Sec.~\ref{biasresp} the SPD/V0M trend is interpreted by Ref.~\cite{alicenewspec} to indicate that the V0M spectrum is ``harder'' than the SPD spectrum. However, comparison of the corresponding TCM reference spectra (dash-dotted) indicates that most of the deviation from unity is simply due to the difference in charge density $\bar \rho_0$ for the two event classes. 

The data-model comparisons of Fig.~\ref{comparex} illustrate the statistical significance of the data-model deviations. It is useful to compare the ``fit'' quality of the TCM compared to V0M and SPD data spectra, although the TCM is not fitted to individual spectra. The quality of the data description for a given model can be estimated by the {\em reduce}d $\chi^2$ statistic $\chi^2_\nu$
\bea  \label{chinu}
\chi^2_\nu &=& \frac{1}{\nu} \sum_{i=1}^N \frac{(O_i - E_i)^2}{\sigma_i^2}
\\ \nonumber
&\approx& \frac{1}{N}\sum_{i=1}^N Z_i^2,
\eea
where $O_i$ are observations, $E_i$ are model predictions and $\nu$ is the number of degrees of freedom -- the number of observations $N$ less the number of model parameters. In the present context a comparable model quality measure is approximated by the second line based on Z-scores as defined in Eq.~(\ref{zscore}). Since the TCM is not fitted to individual spectra and the resulting statistic is not compared to a $\chi^2$ probability distribution the second form is applied.

Figure~\ref{6axx} (right) shows $\chi_\nu^2$ values [second line of Eq.~(\ref{chinu})] for each curve in Fig.~\ref{comparex}. Note that $\chi^2$ is ordinarily employed as a measure of goodness of fit but in the present case is a measure of the amount of {\em new information} carried by data relative to the TCM reference. 
The V0M and SPD trends are separately equivalent when comparing 5 TeV and 13 TeV data. Both event selection types happen to approximate power-law trends on $\bar \rho_0$.

The vertical dotted line at $\bar \rho_0 = 20$ locates class II of V0M (20.5) and class VII of SPD (19.5) events for 13 TeV. There is a factor 5 difference between $\chi_\nu^2$ values. However, the difference would be much greater if not for the structures in Fig.~\ref{6axx} (left) that are common to all spectra as noted in connection with Fig.~\ref{logderiv} (left). Without that noise the higher \nch\ classes of V0M events are approximately statistically consistent with the TCM.

\subsection{Power-law fits: data vs TCM per Sec.~V}

This refers to Fig.~5 of Ref.~\cite{alicenewspec} vs Fig.~\ref{logderiv} of the present study and the substantial differences in the two results. One result arises from a power-law fit to a limited \pt\ interval; the other result is obtained with a log-derivative applied to spectrum data over a larger interval. The latter data are averaged over an interval chosen based on  differential data in Fig.~\ref{logderiv} (left) that directly indicates what \pt\ interval can be approximated by a power law.

Numerically, the power-law model fits result in variation for V0M from 6.0 down to 5.7 and for SPD from 6.2 down to 4.9 whereas the log-derivative produces a fixed value 5.6 for V0M and variation  from 5.7 down to 5.2 for SPD. The fixed log-derivative results for V0M are consistent with data/TCM ratios in Fig.~\ref{ptspec2} (d): i.e.\ no significant deviation from the TCM hard-component {\em shape} above \yt\ = 4. The same conclusion can be inferred from Fig.~\ref{tcm13} (b). The varying $n$ results for SPD are consistent with Eq.~(\ref{2qxx}) for $2/q$ in App.~\ref{old13}. Note that plotting estimates of $2/q$ rather than $q$ leads to a simple linear trend as in Fig.~\ref{checkx} (left) indicating that the effective high-\pt\ ``width'' of the spectrum hard component (i.e.\ $2/q$) increases linearly with SPD yield, an informative result.

It is unfortunate that relatively large artifacts appear in the data as sources of systematic uncertainty, as in Fig.~\ref{logderiv} (left). The statistical power of 60 million 13 TeV events is thereby degraded, and in the case of the log-derivative requires rejecting the lowest and highest \nch\ classes. Nevertheless, the consistency of both V0M and SPD exponent data with their respective smooth trends is indicative of the power of the log-derivative method.

\subsection{Ensemble $\bf \bar p_t$ per Sec.~VI}

Although Ref.~\cite{alicenewspec} estimates systematic uncertainties for ensemble \mmpt\ (quoted as 1-2\%) the substantial bias from incomplete \pt\ acceptance is not discussed. It is stated that ``The efficiency correction [to \mmpt]...is found to be $\sim 1$\%'', and ``The effect of track cuts on $\langle p_T \rangle$ was found to be...of the order of 1\%.'' The only mention of the low-\pt\ cutoff is ``This [primary particle composition] uncertainty takes into account the extrapolation of the spectra to low $p_T$....'' It is further stated that ``The transverse momentum spectra...are {\em fully corrected}...[emphasis added].'' As demonstrated in Sec.~\ref{alicempt} the \pt\ cutoff bias to $\bar p_{ts}'$ is about 0.1 GeV/c corresponding to 10-20\% of the \mmpt\ data values, compared to ALICE estimated systematic uncertainties  $O(1\%)$ as noted above. 

Given that the total excursion of uncorrected $\bar p_t'$ in Fig.~\ref{2ax} (left) is about 65\% of the soft-component value 0.48 GeV/c, {\em interpretation} of uncorrected data is problematic. The nominal goal of the ALICE spectrum study is determination of jet contributions to spectra, and \mmpt\ data are essential for achieving that goal. But one can contrast Fig.~\ref{2ax} (left), where  interpretation of the data is quite uncertain, with Fig.~\ref{2ax} (right) where separation of jet and non-jet contributions is accurately achieved.

Another important uncertainty relating to \mmpt\ data interpretation is the consequence of event selection based on spherocity $S_0$. It is expected that low spherocity will prefer ``jetty'' events and high spherocity will prefer ``isotropic'' events.  The structure of  Fig.~\ref{2ax} (left) obscures how the \mmpt\ vs \nch\ trend relates to jets whereas the corrected structure of Fig.~\ref{2ax} (right) brings clarity. Given that clarification  the simplicity of Fig.~\ref{2c} (left) leads to the correct interpretation of  ``...within uncertainties the overall shape of the [\mmpt\ vs \nch] correlation...is not spherocity-dependent.'' The correct interpretation is that while jet production varies approximately quadratically with \nch, and jet production for SPD events  increases $\approx 30$-fold relative to NSD \pp, spherocity has almost no effect on the jet contribution to spectra,

\section{Discussion} \label{disc}

In its introduction Ref.~\cite{alicenewspec} asserts that a \pt\ spectrum ``carries information of the dynamics of soft and hard interactions.'' As noted  ``The aim of [Ref.~\cite{alicenewspec}] is to investigate the importance of jets in high-multiplicity pp collisions and their contribution to charged-particle production at low $p_T$.''  It is proposed to ``disentangle the energy and multiplicity dependence'' of \pt\ spectra.'' The TCM has been applied to \pp\ \pt\ spectra over three orders of magnitude of \pp\ collision energy, and after fifteen years of development arguably extracts all available information from \pt\ spectra~\cite{ppprd,alicetomspec,ppbpid}. The energy and multiplicity dependences of \pt\ spectra are indeed factorizable, but the energy dependence requires a large energy interval to identify systematic variations accurately; the interval 5 to 13 TeV is too small to do so~\cite{alicetomspec,jetspec2}. Given those observations it is instructive to consider certain comments within Ref.~\cite{alicenewspec} relative to TCM results.

\subsection{Spectrum evolution with $\bf n_{ch}$}

This topic mainly concerns Figs.~2 and 3 of Ref.~\cite{alicenewspec} which present ratios of spectra for different \nch\ classes, two energies and two event selection criteria to minimum-bias INEL $> 0$ spectra. It is noted that ``the features of the spectra...are qualitatively the same for both energies'' and only the 13 TeV result is further discussed. That strategy can be compared with ``disentangle the energy and multiplicity dependence.'' If two things are ``qualitatively the same'' then they are quantitatively {\em dissimilar}, i.e.\ the difference is information carried by \pt\ spectra.

Commenting on Figs.~2 and 3 ``The [spectrum] ratios to the INEL $> 0$ $p_T$ distribution exhibit two distinct behavior [sic].'' In brief, at lower \pt\ the spectrum ratios exhibit small \pt\ dependence, but above 0.5 GeV/c the ratios are strongly dependent on \nch\ and \pt. Referring to Fig.~2 is the comment ``...the $p_T$ spectra become harder as the multiplicity increases, which contributes to the increase of the average transverse momentum with multiplicity.'' But the seemingly dramatic spectrum ``hardening'' in Fig.~2 does not dominate \mmpt(\nch) trends where the central issue is dijet production as a function of \nch\ described comprehensively via the TCM~\cite{alicetommpt,tommpt}. The term ``hardening'' is ambiguous between increased jet number with \nch\ and possible bias of jet fragment distributions~\cite{fragevo}.

In abstract and main text is the statement ``The high-$p_T$ ($> 4$ GeV/c) yields of charged particles increase faster than the charged-particle multiplicity, while the increase is smaller [i.e.\ less rapid than \nch] when we consider lower-$p_T$ particles.'' In relation to its Fig.~6 describing charge integrals within specific $\Delta p_t$ intervals Ref.~\cite{alicenewspec} observes ``Despite the large uncertainties, it is clear the data show a non-linear [i.e.\ faster than \nch] increase.'' But the TCM has provided an accurate {\em quantitative} picture of such trends for fifteen years as noted above. The low-\pt\ part of spectra ($< 0.5$ GeV/c or $y_t \approx 2$) increases $\propto \bar \rho_s \approx \bar \rho_0 - \alpha \bar \rho_s^2$, i.e.\ slower than $\bar \rho_0 = n_{ch} / \Delta \eta$, whereas the higher-\pt\ part of spectra (i.e.\ $> 4$ GeV/c or $y_t \approx 4$) increases {\em precisely} $\propto \bar \rho_s^2$, i.e.\ approximately {\em quadratically} with \nch. 
Again as noted, a characteristic aspect of certain observations in Ref.~\cite{alicenewspec} is confusion among several issues: (a) spectrum normalization (or not), (b) jet production vs \nch, and (c) various selection-bias effects. 

\subsection{Manifestations of jets in \titlept\ spectra}

Given a primary goal of Ref.~\cite{alicenewspec} -- determination of the jet contribution to particle production at lower \pt\ -- it is difficult to find any responding result in the paper. 
Regarding spectra in relation to pQCD ``...the high-$p_T$ ($p_T > 10$ GeV/c) particle production
is quantitatively well described by perturbative QCD (pQCD) calculations....'' But no experimental evidence is presented relating jet contributions at higher \pt, or jet production in general, to hadron production at lower \pt. In contrast, the TCM quantitatively isolates  minimum-bias jet contributions to hadron production over the entire \pt\ acceptance as in Sec.~\ref{tcmspecdat}, and spectrum hard components have been quantitatively related to pQCD via {\em measured} jet energy spectra and fragmentation functions~\cite{fragevo,jetspec2}.

Reference~\cite{alicenewspec} does acknowledge information derived from model fits to spectra: ``Commonly, the particle production is characterized by quantities like integrated yields or any fit parameter of the curve extracted from fits to the data, for example, the so-called inverse slope parameter [$T$]...'' and then emphasizes power-law exponent $n$ as relating to jet production. For description of spectra at higher \pt\ ``the natural choice is fitting a power-law function...to the invariant yield [\pt\ spectrum?] and studying the multiplicity dependence of the exponent ($n$) extracted from the fit.'' One conclusion -- ``the results [the $n(n_{ch})$ trend] using the two multiplicity estimators [V0M and SPD] are consistent within the overlapping multiplicity interval'' -- is problematic as argued below. 

A mechanism for \nch\ dependence of exponent $n$ is conjectured as follows: ``In PYTHIA 8, it has been shown that the number of high-$p_T$ jets increases with event multiplicity.'' In fact, the exact multiplicity dependence of jet production in inelastic \pp\ collisions is reported in Ref.~\cite{jetspec2} based on event-wise-reconstructed jet {\em measurements}, not a Monte Carlo. The conjecture continues: ``...based on PYTHIA 8 studies, the reduction of the power-law exponent [n] with increasing multiplicity [\nch] can be attributed to an increasing number of high-$p_T$ jets.'' But for V0M event selection in Fig.~\ref{logderiv} (right) the exponent $n$ trend on \nch\ is consistent with a {\em constant} (13 TeV) or even slight {\em increase} (5 TeV). The same jet population is accessible to either selection criterion and jet {\em number} is strictly dependent on \nch\ as demonstrated by Fig.~\ref{endpoint}. All that differs is the spectrum bias at high \pt\ induced by the selection method. Such arguments confuse basic QCD jet production (with its well-established systematics) and  manifestations of event selection bias.

The dijet production rate for NSD \pp\ collisions can be predicted from measured jet cross sections, and such predictions can be compared quantitatively with jet contributions to spectra and two-particle correlations identified by their unique \nch\ dependence (e.g.\ as represented by the TCM)~\cite{fragevo}. For 200 GeV NSD collisions 3\% of events include a dijet per unit of eta~\cite{ppprd}. For $\approx 10$ TeV NSD collisions 15\% of events include a dijet per unit of eta~\cite{jetspec2}, where the quoted percentages are $100(1/\sigma_{\text {NSD}})d\sigma_{\text {jet}}/d\eta$. For a typical range of \pp\ charge densities $\bar \rho_0 \in [\bar \rho_{0\text {NSD}},10\bar \rho_{0\text {NSD}}]$ the dijet yield should increase by factor 100 due to the {\em observed} trend $\bar \rho_h \propto \bar \rho_s^2$. 13 TeV \pp\ collisions with SPD $\bar \rho_0 \approx 54$ should include 11 dijets per unit pseudorapidity on average. ``Jets'' here assumes a minimum-bias jet spectrum wherein most jets appear near 3 GeV (the effective jet spectrum mode).

\subsection{Selection biases and QCD}

The same minimum-bias INEL $> 0$ event ensemble is partitioned into ten multiplicity classes according to two selection criteria -- V0M and SPD. The V0M or ``forward [on $\eta$] multiplicity estimator'' is said to ``minimize the possible autocorrelations induced by the use of the midpseudorapidity estimator.'' The statement suggests that the V0M criterion should produce substantially less bias. ``The comparison of results obtained with these [V0M and SPD] estimators allows to understand potential biases from measuring the multiplicity and $p_T$ distributions in overlapping $\eta$ regions.'' As with statements about jet contributions to lower \pt\ it is difficult to find within Ref.~\cite{alicenewspec} any such understanding. In fact ``bias'' as in that sentence does not appear again in the paper.

For spectrum data plotted as ratios, as in Figs.~\ref{ptspec1} and \ref{ptspec2} (b,d), selection bias from V0M and SPD criteria appear quite different. However, the apparent large high-\pt\ bias for SPD involves a tiny fraction of all particles and even a small fraction of jet fragments. When the two criteria are compared based on significance as in Fig.~\ref{comparex} the SPD high-\pt\ bias does not dominate the structure. Ironically, V0M bias is {\em at least as} significant as SPD bias and the \pt\ dependence is similar, contradicting the expectation that ``autocorrelations'' are minimized by disjoint $\eta$ intervals. 

Particle production and QCD dynamics must be strongly correlated between V0M and SPD acceptances. Hadron production at midrapidity depends in part on a parton splitting cascade within each projectile proton that also contributes hadrons at larger $\eta$. Thus, fluctuations in V0M must be strongly correlated with fluctuations in SPD, albeit SPD has additional contributions from parton scattering and fragmentation to jets. Figure~\ref{comparex} reveals that fluctuations {\em correlated} between V0M and SPD are statistically dominant, whereas fluctuations in low-energy jet formation play a less-significant role.

\subsection{Ensemble $\bf \bar p_t$ vs spherocity}

As part of a strategy to identify jet contributions at lower \pt\ Ref.~\cite{alicenewspec} introduces spherocity $S_0$ (effectively an azimuthal asymmetry measure):  ``The present paper reports a novel multi-differential analysis aimed at understanding charged-particle production associated to partonic scatterings with large momentum transfer and their possible correlations with soft particle production.'' But most of the jet-related hard component arises from lower-energy partons~\cite{jetspec2,fragevo}, and most fragments from any jet appear at low \pt~\cite{eeprd}.  ``Transverse spherocity...has been proven to be a valuable tool to discriminate between jet-like and isotropic events....'' That statement is based on material presented in Ref.~\cite{ortizspher} which is a study of various event-shape measures applied to PYTHIA simulations. It is not clear from that study what spherocity contributes to understanding real \pp\ collisions.

``Studying observables as a function of spherocity reveals interesting features.'' Relative to the \mmpt\ vs \nch\ trend for the INEL $> 0$ ensemble average as a reference the trend for ``isotropic'' events (high $S_0$) ``stays systematically below'' the reference, whereas  ``...for jet-like events'' (low $S_0$) the \mmpt\ trend is consistently higher. The following observation is interesting: ``Moreover,...the overall shape of the correlation [\mmpt(\nch)], i.e.\ a steep linear rise below $dN_{ch}/d\eta = 10$ followed by a less steep but still linear rise above, is {\em not spherocity-dependent} [emphasis added].''
  
That qualitative description relates to a confusing data trend arising from questionable analysis variables. The data trend for uncorrected $\bar p_t'(n_{ch})$ includes the misleading dependence expressed by Eq.~(\ref{ppmpttcm}) and hard-component suppression at lower \nch\ exhibited in Fig.~\ref{comparex}.   In contrast, the approximate linear rise of corrected \mmpt\ vs $\bar \rho_s$ in Fig.~\ref{2ax} (right) and Fig.~\ref{2c} (left) is equivalent to the linear rise of integrated yields with $\bar \rho_s$ in Fig.~\ref{endpoint}. In other words, the TCM structure of Eq.~(\ref{niceeq}) is equivalent to the TCM structure of Eq.~(\ref{nchns}). Jets control the {\em slope} of the linear trend over a large \nch\ interval: in  the absence of jets there would be no linear rise, i.e.\ the slope of the trend is a measure of jet production. By observing that \mmpt\ vs \nch\ trends in Fig.~\ref{2ax} (left), equivalent to trends in Fig.~\ref{2c} (left), are ``not spherocity-dependent'' Ref.~\cite{alicenewspec} admits that $S_0$ does little to control jet production in \pp\ collisions. That the high-$S_0$ and low-$S_0$ trends appear respectively systematically below and above the reference in Fig.~7 of Ref.~\cite{alicenewspec} is a consequence of $S_0$ biasing the nonjet spectrum {\em soft} component as noted in Sec.~\ref{alicempt}.

\section{summary} \label{summ}

This article reports a study of high-statistics \pt\ spectra from 5 TeV and 13 TeV \pp\ collisions at the large hadron collider. This study is based on the two-component (soft + hard) model (TCM) of hadron production near midrapidity. The spectrum data and additional spectrum analysis were reported by the ALICE collaboration with the nominal goal to estimate jet contributions to \pt\ spectra, especially at lower \pt.  Part of the motivation was a response to recent claims of ``collective'' behavior (flows) in small (asymmetric?) collision systems. The principal data presentation was \pt\ spectra for various event classes in ratio to an ensemble average over all collision events.

The TCM has been demonstrated over a number of years to be a valuable tool for isolating distinct hadron production mechanisms in A-B collisions. In particular, hadron contributions to spectra from jet-related production (hard component) are accurately distinguished (at the percent level) from nonjet production (soft component). The signature element of the TCM for \pp\ collisions is a {\em quadratic relation} between the hard component and soft component as inferred from spectrum data.

A major goal for the ALICE study was characterization of {\em selection bias} resulting from sorting collision events into classes according to particle yields in two different pseudorapidity $\eta$ intervals denoted by SPD (midrapidity) and V0M (forward rapidity). If event classes are simply compared to an ensemble average the present study finds that three issues are consequently confused: (a) spectrum normalization, (b) the jet contribution to \pp\ spectra as it varies {\em in known ways} with multiplicity \nch\ and (c) the selection bias in question. This study demonstrates that ratios of data spectra to TCM spectra remove issues (a) and (b) and expose selection bias (c) to direct study.

One aspect of the present study is introduction of the {\em Z-score} statistical measure: Whereas spectrum {\em ratios} tend to visually exaggerate deviations at higher \pt\ and suppress those at lower \pt\ Z-scores provide a measure of the {\em statistical significance} of data-model deviations. Data-model deviations at high \pt\ that seem to dominate ratios are actually only modestly significant whereas deviations at lower \pt\ are much more significant and {\em quite similar} for V0M and SPD event classifications.

The present analysis reveals that selection bias has two main manifestations depending on the event selection criterion. The spectrum soft component appears unaffected by selection method. 
The V0M and SPD selection methods both bias the lower-\pt\ parts of the jet-related hard component (fragment distribution) similarly. For lower charge multiplicities \nch\ the peaked hard component is shifted to lower \pt, resulting in spectrum suppression at higher \pt\ (above the peak mode) and enhancement at lower \pt\ (below the mode). However, the high-\pt\ tails of spectra are negligibly affected by V0M selection but strongly affected by SPD selection: with increasing \nch\ SPD spectrum tails become increasingly harder (smaller power-law exponent). The combined effects suggest that selection bias responds to fluctuations in jet production in different ways depending on the relevant $\eta$ acceptance.

Ensemble-mean \mmpt\ data as a function of \nch\ were obtained for different values of {\em spherocity} $S_0$, a measure of the azimuth asymmetry of the vector $\vec p_t(\phi)$ distribution. The intent was to bias events according to their ``jettiness,'' a lower $S_0$ value expected to prefer events with more or more-frequent jets. The published \mmpt\ data are biased because of the incomplete detector \pt\ acceptance (lower limit at 0.15 GeV/c) making data interpretation more difficult. In the present analysis the \mmpt\ data are corrected and transformed to a TCM configuration with fixed soft-component contribution and hard component varying with \nch\ in a manner that should reveal the jet contribution to \mmpt. This study concludes that the main affect of $S_0$ selection on \mmpt\ vs \nch\ trends is bias of the $\bar p_{ts}$ {\em soft} component. The effect on the hard component (i.e.\ jet contribution) is barely detectable. Given $S_0$ as an azimuth asymmetry measure, its lower values tend to suppress events where the soft component includes more high-\pt\ particles, thus reducing its ensemble-mean $\bar p_{ts}$. 

In summary, the TCM is observed to be a necessary and sufficient description of \pp\ \pt\ spectra and arguably represents all information carried by spectrum data for unidentified hadrons. The TCM hard component has been quantitatively related to the properties of event-wise reconstructed jets. Determination of an accurate TCM for {isolated} spectra (rather than ratios) over a broad range of event multiplicities and and event selection criteria as in the present study establishes an accurate and efficient representation of a large volume of spectrum data. The contribution of jets to \pp\ \pt\ spectra is accurately determined at the percent level and the consequences of several forms of event selection are isolated.

\begin{appendix}

\section{Previous 13 $\bf TeV$ data TCM} \label{old13}

A previous TCM analysis of 13 TeV \pp\ spectra was reported in Ref.~\cite{alicetomspec}. That study emphasized multiplicity and energy dependence of the spectrum hard component in the context of QCD theory and jet measurements. The 13 TeV \pp\ spectrum data as reported in Ref.~\cite{alicespec} were quite limited as to range of event multiplicity -- see the 13 TeV solid triangles and open circles in Fig.~\ref{checkx} (left) -- and the presentation format was based on ratios of spectra from three \nch\ classes to a minimum-bias INEL $> 0$ reference. It was required therefore to develop new techniques to isolate the hard and soft components for the three \nch\ classes based on analysis of 200 GeV spectrum data for which the TCM structure is well-established~\cite{ppprd}. In this appendix relevant results from Ref.~\cite{alicetomspec} are reviewed to provide context for the present study of a much more extensive sample of 13 TeV \pp\ spectrum data. 

\subsection{Spectrum TCM multiplicity dependence} \label{ppmult}

In Sec.~\ref{pptcm} TCM model functions are held fixed for all \nch\ classes and for both V0M and SPD. Systematic data biases are then revealed in a precise way. However, in Ref.~\cite{alicetomspec} the TCM hard component was parametrized to accommodate the \nch\ dependence of the hard-component shape for 13 TeV SPD spectra. In this subsection those results (for 200 GeV as well as 13 TeV) are reviewed and the quality of the data description is evaluated. The 200 GeV data are equivalent to Ref.~\cite{alicenewspec} SPD with $|\eta| < 1.0$

\begin{figure}[h]
	\includegraphics[width=1.62in]{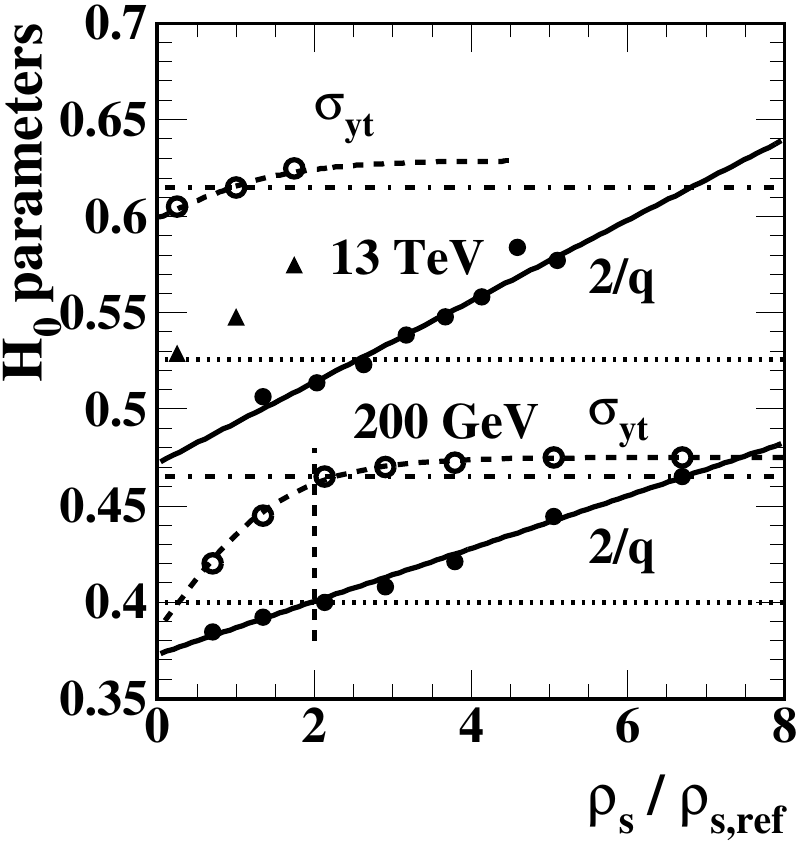}
	\includegraphics[width=1.66in]{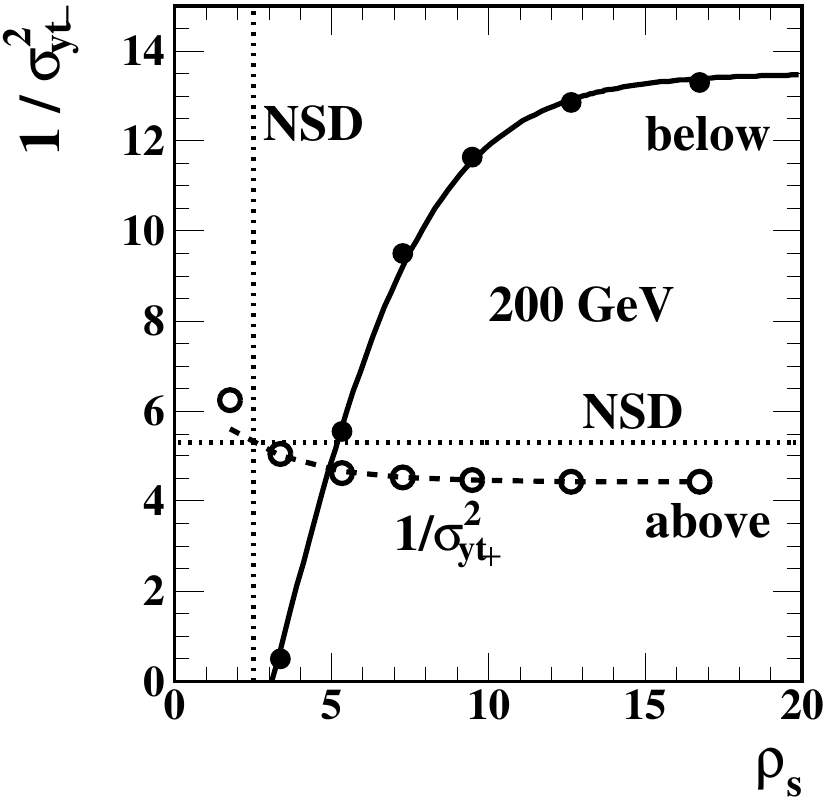}
	\caption{\label{checkx}
		Left: TCM hard-component parameters varying with $n_{ch}'$ or $\bar \rho_s$ for a revised TCM. 200 GeV solid and dashed curves through parameter data are defined by Eqs.~(\ref{parameqx}). The 13 TeV open circles and solid triangles are as reported in Ref.~\cite{alicetomspec}. The upper solid dots and solid line are updates determined in the present analysis, derived from more extensive \pp\ data. $\bar \rho_{s,ref} = 2.45$ for 200 GeV NSD \pp\ collisions and 5.8 for 13 TeV NSD collisions. The factor 2 in $2/q$ permits greater plot sensitivity.  The mean-value energy trend for $\sigma_{y_t}$ is shown in Fig.~\ref{enrat3x} (left) and for $1/q$ is shown in Fig.~\ref{softcompx} (left).
		Right:  Variation of the Gaussian width below the hard-component mode $\sigma_{y_t-}$ (solid points) for $n = 2$-7 that accommodates data in that \yt\ interval. The Gaussian width above the mode $\sigma_{y_t+}$ (open points) is included for comparison. The curves are defined in the text. The correlation of two trends is notable.
	}  
\end{figure}

Figure~\ref{checkx} (left) shows variation of two 200 GeV $\hat H_0(y_t)$ parameters with $\bar \rho_s$ (lower solid and dashed curves) that provides accurate description of spectrum ratios above the hard-component mode for all multiplicity classes. Optimized TCM parameters follow simple $\bar \rho_s$ trends
\bea \label{parameqx}
2/q &=& 0.373 + 0.0137 (\bar \rho_s / \bar \rho_{s,ref})~~\text{(lower solid)}
\\ \nonumber
\sigma_{y_t} &=& 0.385 + 0.09 \tanh(\bar \rho_s/4)~~~\text{(lower dashed)}.
\eea
The nominal parameter values for the 200 GeV {\em fixed} $\hat H_0(y_t)$ model are represented by the dotted and dash-dotted lines (corresponding to parameter values for $\bar \rho_s / \bar \rho_{s,ref} \approx 2$). Variation of two parameters in combination serves to broaden the hard-component model {\em above} the mode toward higher \yt. The saturation of $\sigma_{y_t}$ at larger $\bar \rho_s$ is a consequence of increasing $2/q$. The transition on $\hat H_0(y_t)$ from Gaussian to exponential form then moves back toward the mode and the exponential/power-law tail increasingly dominates the higher-\pt\ structure. $\bar \rho_{s,ref}$ values are noted in the Fig.~\ref{checkx} caption.

The corresponding parameter trends for 13 TeV inferred from spectrum {\em ratios} in the earlier analysis of Ref.~\cite{alicetomspec} are shown by the upper solid triangles and open circles.  New values for $2/q$ (upper solid dots) are inferred in the present study via log derivative applied directly to  13 TeV SPD \yt\ spectra via Eq.~(\ref{logg}) (upper line). Those results correspond accurately with $n$ values plotted in Fig.~\ref{logderiv} (right) obtained with Eq.~(\ref{logg}) (lower line) demonstrating the correspondence $n \approx q + 1.8$. The $2/q$ data are well described by
\bea \label{2qxx}
2/q &=& 0.472 + 0.021 (\bar \rho_s / \bar \rho_{s,ref})~\text{(upper solid)}.
\eea
Previous 13 TeV $2/q$ values derived from spectrum ratios in Ref.~\cite{alicetomspec} (triangles) differ substantially from those in the present study and demonstrate the difficulty of deriving accurate results from limited spectrum ratios.

Figure~\ref{checkx} (right)  shows variation of the 200 GeV hard-component width required to accommodate data {\em below} the mode in the form $1/\sigma_{y_t-}^2$ (solid points). The solid curve through points is $13.5 \tanh[(\bar \rho_s - 3.1)/5]$.  Also included is the trend for the width above the mode from the left panel and Eq.~(\ref{parameqx}) (lower) plotted as $1/\sigma_{y_t+}^2$ (open points and dashed curve respectively) demonstrating correlation of the two trends. The  two widths become equal near $\bar \rho_s \approx 5$, or  $\bar \rho_s / \bar \rho_{s,ref} \approx 2$ in Fig.~\ref{checkx} (left) where the hard-component  model is then, near its mode, approximately symmetric as in Refs.~\cite{ppprd,ppquad}.

\begin{figure}[h]
	\includegraphics[width=1.65in,height=1.6in]{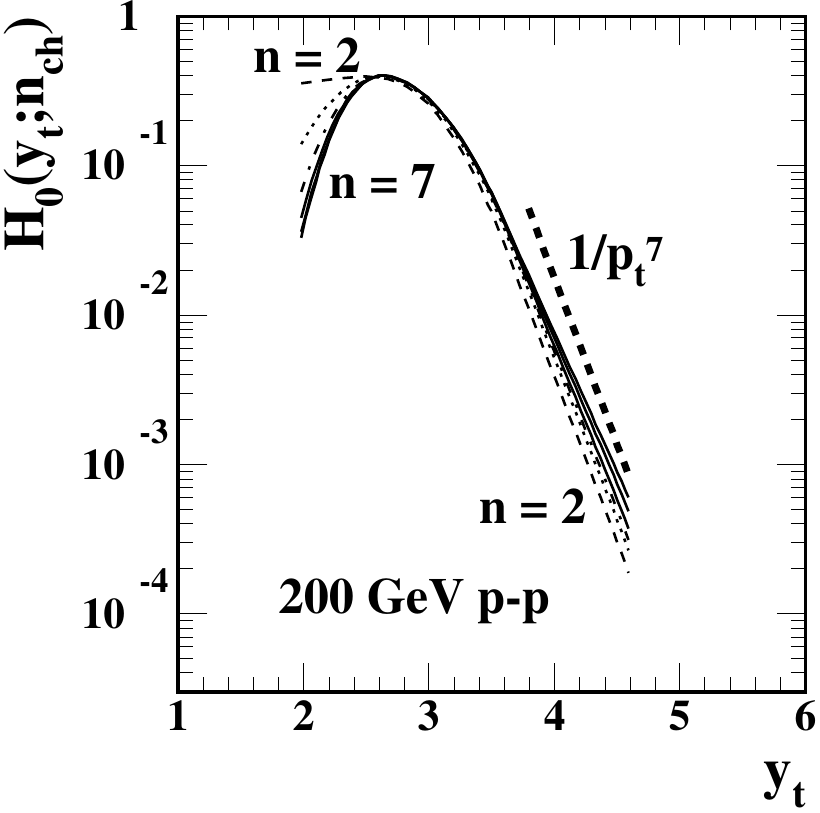}
	\includegraphics[width=1.65in,height=1.6in]{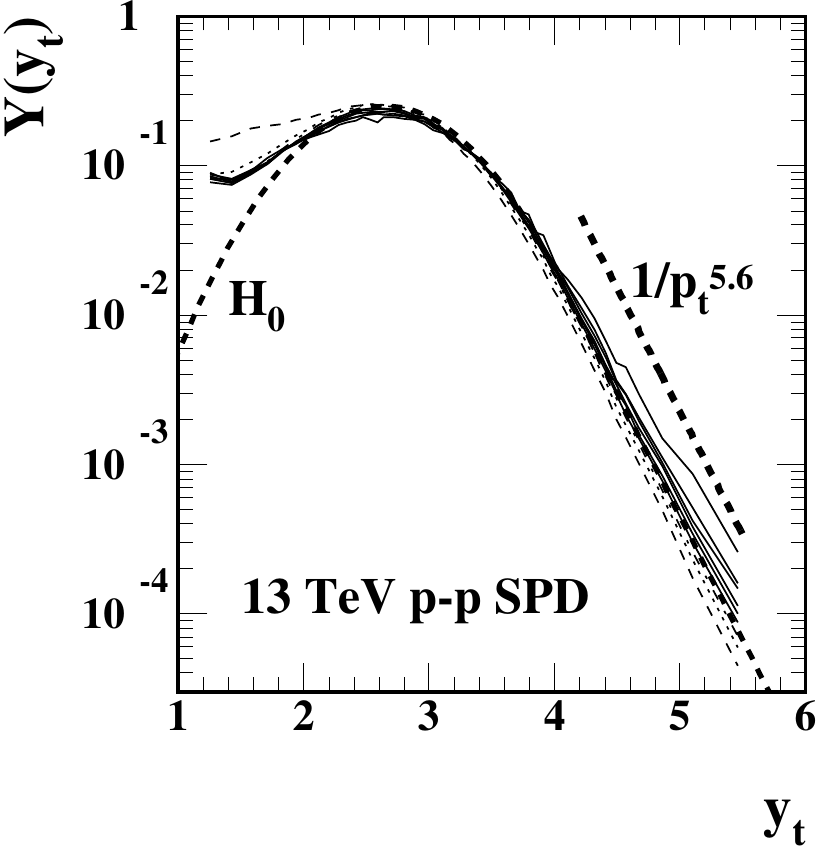}
	\caption{\label{chi2xx}
		Left: Evolution of the 200 GeV hard-component model over six multiplicity classes that exhausts all information in high-statistics spectrum data from Ref.~\cite{ppquad}.
		Right:  Spectrum hard components for SPD event selection and for nine multiplicity classes of 13 TeV \pp\ collisions. The order of line styles is reversed compared to other figures in this article to match the order in the left panel, and the order of index $n$ is also reversed from the present study.
	} 
\end{figure}

Figure~\ref{chi2xx}  (left)  summarizes the revised 200 GeV TCM hard-component model for six multiplicity classes. The hard component for the lowest \pp\ multiplicity class $n = 1$ is typically severely distorted due to selection bias (see Sec.~\ref{tcmspecdat} for example).

Figure~\ref{chi2xx}  (right)  shows 13 TeV SPD spectrum hard components (curves) appearing in Fig.~\ref{tcm13} (d). SPD event selection based on particle yields within $|\eta| < 0.8$ is consistent with event selection for 200 GeV data based on yields within $|\eta| < 1$. Line types for this panel are reversed in order compared to other figures in this study to match the convention in the left panel. As noted, the hard component for the lowest multiplicity class (here $n = 1$, not 10 as in the present study) is not shown since there is very little jet contribution to those events due to strong selection bias. Given those minor differences the event-selection biases for 200 GeV and 13 TeV \pp\ \pt\ spectra are remarkably similar. The similarity at higher \pt\ is consistent with the solid lines in Fig.~\ref{checkx} (left).

\subsection{Spectrum TCM collision-energy dependence}

\begin{figure}[h]
	\includegraphics[width=1.65in,height=1.6in]{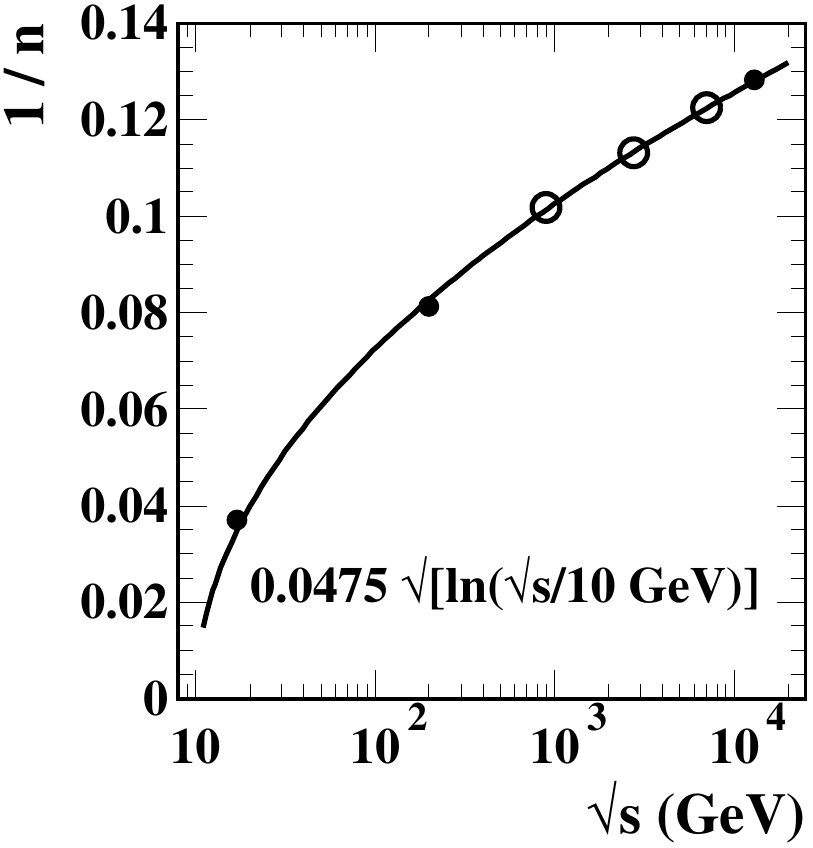}
	\includegraphics[width=1.67in,height=1.6in]{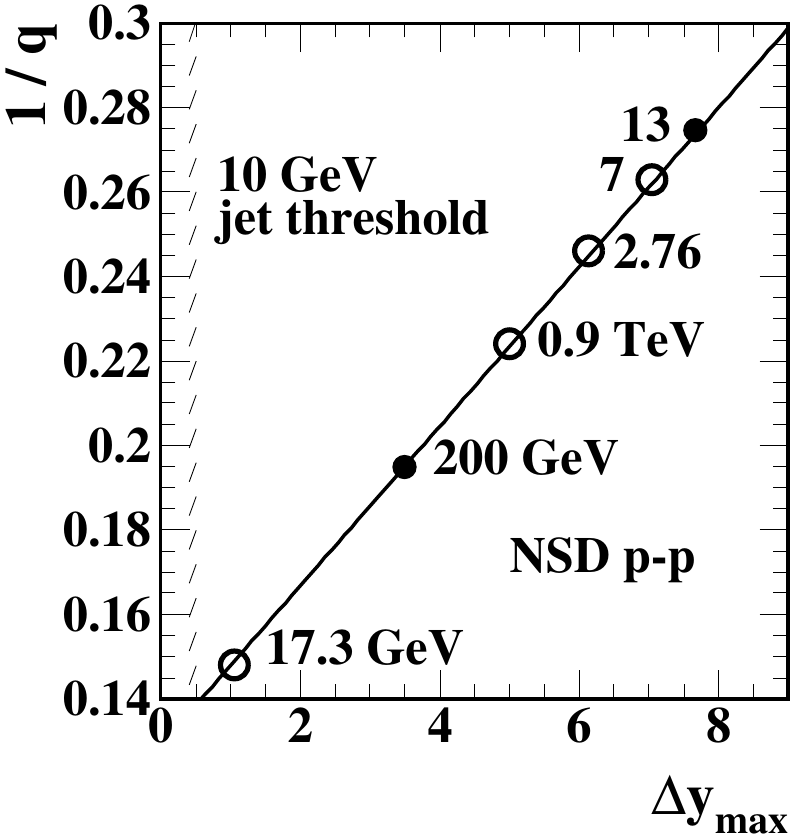}
	\caption{\label{softcompx}
		Left:  Measured L\'evy exponents for three collision energies (solid points). The curve is a fit by eye of the function $A \sqrt{\ln(\sqrt{s} / \text{10 GeV})}$ (with $A = 0.0475$) motivated by the possibility of Gribov diffusion controlling the growth of transverse momentum for low-$x$ partons (gluons)~\cite{gribov}.  Open symbols are interpolations at energies relevant to the study in Ref.~\cite{alicetomspec}.
		Right: Hard-component exponents $q$ determined by analysis of \pp\ spectrum data (solid points) from Ref.~\cite{ppquad} and Ref.~\cite{alicetomspec}. The solid curve is based on a jet-spectrum parametrization in Ref.~\cite{jetspec2} that also describes ensemble-mean-\pt\ hard-component energy variation~\cite{alicetommpt}. Open points are interpolations and extrapolation relevant to Ref.~\cite{alicetomspec}.
	} 
\end{figure}

Figure~\ref{softcompx} (left) shows soft-component exponents in the form $1/n$ inferred from spectrum data for three collision energies (solid points) at the SPS, RHIC and LHC. The solid curve is an algebraic hypothesis based on variation of the soft component due to conjectured Gribov diffusion~\cite{gribov}. Low-$x$ gluons result from a virtual parton splitting cascade within projectile nucleons whose mean depth on $x$ is determined by the collision energy. Each step of the cascade adds transverse-momentum components in a random-walk process. The depth of the cascade is proportional to $\ln(s/s_0)$, and $\sqrt{s_0} \approx 10$ GeV is inferred from dijet systematics~\cite{anomalous,jetspec2}. Given the properties of a random walk and with $1/n$ as a measure of transverse-momentum excursions~\cite{wilk} its trend is estimated as $\propto \sqrt{\ln(\sqrt{s} / \text{10 GeV})}$ (solid curve).  The open circles at 0.9, 2.76 and 7 TeV are interpolations of the L\'evy exponent to $n = 9.82$, 8.83 and $8.16$ respectively.

Figure~\ref{softcompx} (right) shows inverse values (solid points) of exponents $q = 5.15$ for 200 GeV and $q = 3.65$ for 13 TeV plotted vs quantity $\Delta y_{max} \equiv \ln(\sqrt{s} / \text{6 GeV})$ that is observed to describe the energy trend for jet spectrum widths $\propto \Delta y_{max}$ from NSD \pp\ collisions assuming a jet-spectrum low-energy cutoff near 3 GeV~\cite{jetspec2} (see Fig. 5 of Ref.~\cite{jetspec2} for a direct comparison with measured jet spectra). The inverse $1/q$ effectively measures the hard-component peak width at larger \yt. The relation $1/q \propto  \Delta y_{max}$ (solid line) is expected given that the \pp\ \pt-spectrum hard component can be expressed as the convolution of a fixed \pp\ fragmentation-function ensemble with a collision-energy-dependent jet spectrum~\cite{fragevo}, and the jet-spectrum width trend has the same dependence~\cite{jetspec2}. The vertical hatched band indicates an inferred cutoff to dijet production from low-$x$ gluon collisions near 10 GeV. That the same relation applies to the ensemble \mmpt\ hard component was established in Ref.~\cite{alicetommpt}. $1/q$ for $q = 3.80$ for 7 TeV, $q = 4.05$ for 2.76 TeV and $q = 4.45$ for 0.9 TeV (open circles) are interpolations. 

\begin{figure}[h]
	\includegraphics[width=1.65in]{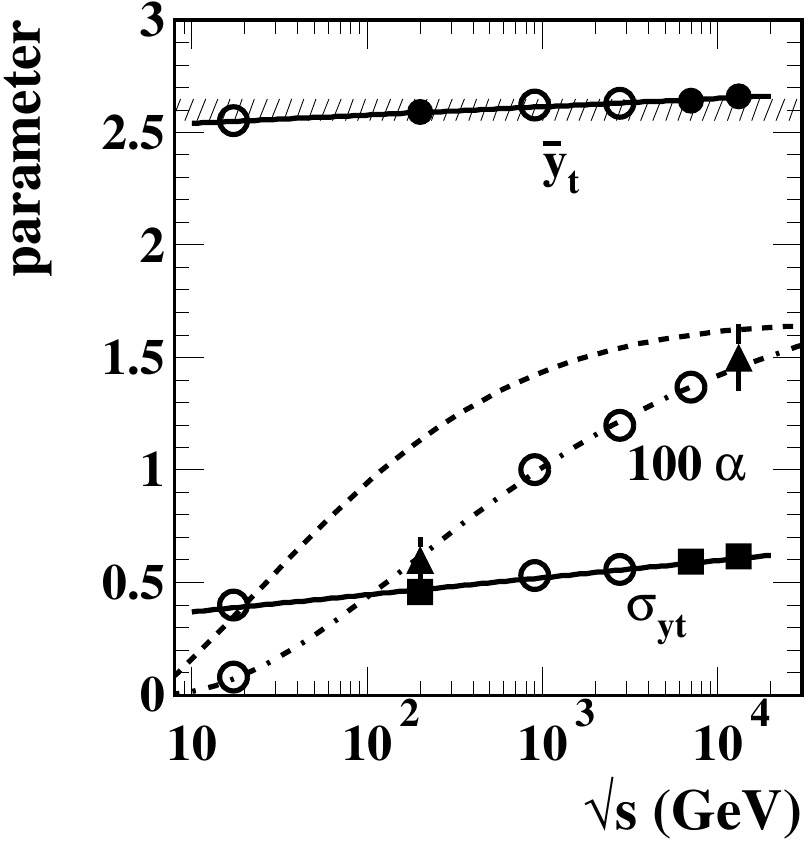}
	\includegraphics[width=1.65in]{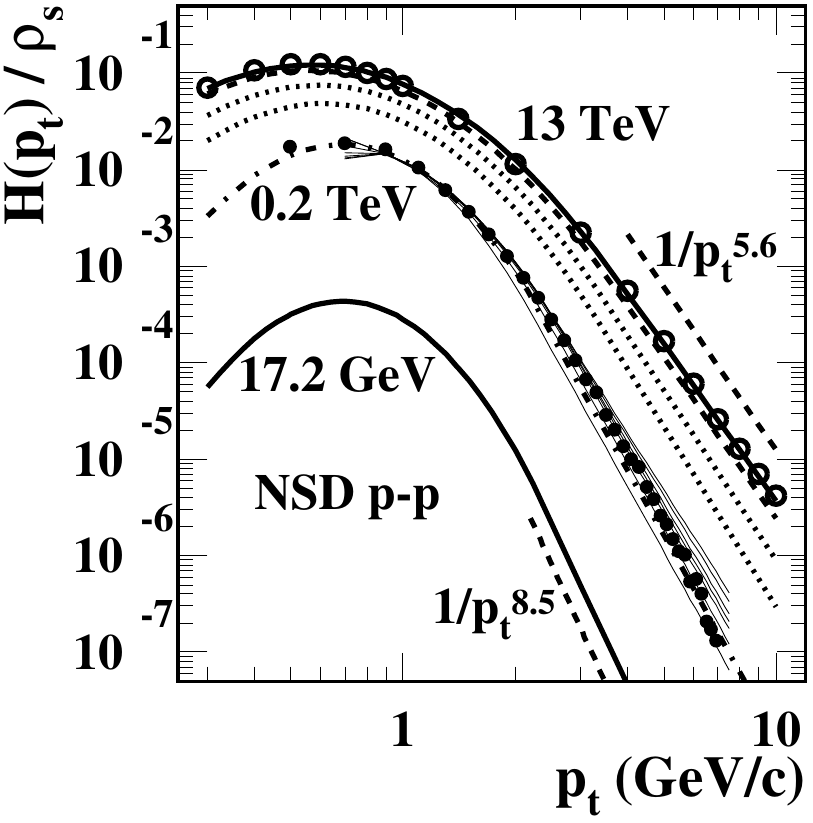}
	\caption{\label{enrat3x}
		Left: TCM NSD hard-component parameters derived from spectrum data (solid points). Open circles are interpolations or extrapolations relevant to Ref.~\cite{alicetomspec}. The solid lines are fits to data. The dashed and dash-dotted curves related to $\alpha(\sqrt{s})$ are described in Ref.~\cite{alicetomspec}.
		Right: Survey of spectrum hard components over the currently accessible energy range from threshold of dijet production (10 GeV) to LHC top energy (13 TeV). The curves are determined by parameters in Table~\ref{engparamx} except for the 200 GeV fine solid curves determined also by the $\sigma_{y_t}$ and $q$ trends in Fig.~\ref{checkx} (left). The points are from Refs.~\cite{ppquad} (200 GeV) and \cite{alicespec} (13 TeV).
	} 
\end{figure}

Figure~\ref{enrat3x} (left) shows NSD TCM hard-component model parameters $\bar y_t$ and $\sigma_{y_t}$ (points) vs collision energy. The solid points are derived from data. The open points are interpolations or extrapolations derived from the inferred or predicted trends in the figure (curves). The trends for $\bar y_t$ and $\sigma_{y_t}$ are consistent with straight lines. Whereas $\sigma_{y_t}$ increases by 50\% the upper limit on $\bar y_t$ variation is five percent (hatched band) and $\bar y_t$ may not actually vary significantly over  over three orders of magnitude of collision energy. The trend for  $\bar y_t$ is consistent with a fixed lower bound on the underlying jet spectrum near 3 GeV, also nearly independent of collision energy~\cite{fragevo,jetspec2}.

Figure~\ref{enrat3x} (right) shows the TCM for quantity $H(p_t;E) / \bar \rho_s(E) \approx \alpha(E) \bar \rho_s(E) \hat H_0(p_t;E)$ measuring the spectrum hard component  {\em per soft-component yield} corresponding to dijet production per participant low-$x$ gluon. The two dotted curves are for 0.9 and 2.76 TeV and the dashed curve is for 7 TeV. Isolated hard components rather than spectrum ratios clarify spectrum energy evolution and its relation to dijet production. The hard-component mode on \pt\ is near 0.5 GeV/c ($y_t \approx 2$) whereas the mode on \yt\ is near 2.7 ($p_t \approx 1$ GeV/c).

The predictions for six collision energies (curves) derived from parameter values in Table~\ref{engparamx} have been compared to data from four energies (13, 7, 0.9 and 0.2 TeV )~\cite{alicetomspec}. The 17.2 GeV extrapolation indicates  no {\em significant} jet contribution to yields and spectra at that energy and explains why no excess \pt\ fluctuations were observed at the SPS~\cite{na49fluct,tomaliceptfluct}. However, evidence for SPS jets {\em is} visible in 17 GeV azimuth correlations as a more sensitive detection method~\cite{ceres}.
The 200 GeV summary (thin solid) includes parametric variation of $\hat H_0(y_t;q,\sigma_{y_t},n_s)$ for six multiplicity classes as described in Sec.~\ref{ppmult}. Corresponding data (solid points) represent NSD \pp\ collisions. The overall result is a comprehensive and accurate description of dijet contributions to \pt\ spectra vs \pp\ collision energy over three orders of magnitude.

\subsection{Spectrum TCM parameter summary} \label{parsum}

Table~\ref{engparamx} summarizes NSD \pp\ TCM parameters for a broad range of energies. The entries are grouped as soft-component parameters $(T,n)$, hard-component parameters $(\bar y_t,\sigma_{y_t},q)$, hard-soft relation parameter $\alpha$ and soft density $\bar \rho_s$.
Slope parameter $T = 145$ MeV is held fixed for all cases consistent with observations.  Its value is determined solely by a low-\yt\ interval where the hard component is negligible. The interpolated L\'evy exponent $n$ values are derived from Fig.~\ref{softcompx} (left) (open circles). Interpolated hard-component $q$ values are derived from Fig.~\ref{softcompx} (right) (open circles). $\bar \rho_s$ values are derived from the universal trend $\bar \rho_s \approx 0.81 \ln(\sqrt{s} / \text{10 GeV})$ inferred from correlation and yield data. All 0.9 and 2.76 TeV values are predicted via interpolation. All remaining (unstarred) numbers are obtained from spectrum data.

\begin{table}[h]
	\caption{Spectrum TCM parameters for NSD \pp\ collisions  at several energies from Ref.~\cite{alicetomspec}.
		Starred entries are estimates by interpolation or extrapolation. Unstarred entries are derived from yield, spectrum or spectrum-ratio data.
			}
	\label{engparamx}
	\begin{center}
		\begin{tabular}{|c|c|c|c|c|c|c|c|} \hline
			$\sqrt{s}$ (TeV) & T\. (MeV) & $n$ & $\bar y_t$ & $\sigma_{y_t}$ & $q$ & $100\alpha$ & $\bar \rho_s$ \\ \hline
			0.0172  & 145  & 27 & 2.55$^*$ & 0.40$^*$  & 6.75$^*$  & 0.07$^*$ & 0.45 \\ \hline
			0.2  & 145  & 12.5 & 2.59 & 0.435  & 5.15  & 0.6 & 2.45 \\ \hline
			0.9  & 145  & 9.82$^*$ & 2.62$^*$ & 0.53$^*$  &  4.45$^*$  & 1.0$^*$ & 3.65 \\ \hline
			2.76  & 145  & 8.83$^*$ & 2.63$^*$ & 0.56$^*$  &  4.05$^*$  & 1.2$^*$ & 4.55 \\ \hline
			5.0 & 145  & 8.47$^*$ & 2.63$^*$ & 0.58$^*$  &  3.85$^*$  & 1.3$^*$ & 5.00 \\ \hline
			7.0  & 145  & 8.16$^*$ & 2.64 & 0.595  & 3.8$^*$  & 1.4$^*$ & 5.30  \\ \hline
			13.0  & 145  & 7.80 & 2.66 & 0.615  & 3.65  & 1.5 & 5.80 \\ \hline
		\end{tabular}
	\end{center}
\end{table}

\end{appendix}


\end{document}